\documentclass[preprint,superscriptaddress,nofootinbib,preprintnumbers,amsmath,amssymb]{revtex4-1}

\usepackage{bbm}
\usepackage{stmaryrd}
\usepackage[bb=boondox]{mathalfa}

\usepackage{amsfonts}
\usepackage{mathrsfs}
\usepackage{leftidx}
\usepackage{amssymb}
\usepackage{placeins}
\usepackage{relsize}
\usepackage{slashed}
\usepackage{multirow}
\usepackage[dvipsnames]{xcolor}

\usepackage[colorlinks]{hyperref}
\usepackage{epsfig}
\usepackage{graphicx}               
\usepackage{url}
\usepackage{float}

\makeatletter
\newcommand*{\ovA}[1]{%
  \m@th\overline{\mbox{$#1$}\raisebox{2.25mm}{}}%
}
\newcommand*{\ovB}[1]{%
  \m@th\overline{\mbox{$#1$}\raisebox{2.5mm}{}}%
}
\newcommand{\overleftrightsmallarrow}{\mathpalette{\overarrowsmall@\leftrightarrowfill@}}
\newcommand{\overrightsmallarrow}{\mathpalette{\overarrowsmall@\rightarrowfill@}}
\newcommand{\overleftsmallarrow}{\mathpalette{\overarrowsmall@\leftarrowfill@}}
\newcommand{\overarrowsmall@}[3]{%
  \vbox{%
    \ialign{%
      ##\crcr
      #1{\smaller@style{#2}}\crcr
      \noalign{\nointerlineskip}%
      $\m@th\hfil#2#3\hfil$\crcr
    }%
  }%
}
\def\smaller@style#1{%
  \ifx#1\displaystyle\scriptstyle\else
    \ifx#1\textstyle\scriptstyle\else
      \scriptscriptstyle
    \fi
  \fi
}

\makeatother
\newcommand{\te}[1]{\overleftrightsmallarrow{#1}}

\DeclareMathOperator{\tr}{tr}
\def\pr{\prime}

\def\CL{{\cal L}}

\def\CL{{\cal L}}

\def\BR{{\rm BR}}

\def\be{\begin{equation}}
\def\ee{\end{equation}}
\def\bea{\begin{eqnarray}}
\def\eea{\end{eqnarray}}

\def\pr{\prime}

\newcommand{\tppp}{t^{\prime \prime \prime}}
\newcommand{\tpp}{t^{\prime \prime}}
\newcommand{\tp}{t^{\prime}}

\newcommand{\bpp}{b^{\prime \prime}}
\newcommand{\bp}{b^{\prime}}
\newcommand{\pb}{p^{\prime}}
\newcommand{\tpbar}{\bar{t}^{\prime}}
\newcommand{\pbbar}{\bar{p}^{\prime}}
\newcommand{\tphi}{ \tilde{\phi} }
\newcommand{\tcapphi}{ \widetilde{\Phi} }
\newcommand{\nppp}{n^{\prime \prime \prime}}
\newcommand{\npp}{n^{\prime \prime}}
\newcommand{\np}{n^{\prime}}

\newcommand{\xp}{x_{+}}
\newcommand{\xpbar}{\bar{x}_{+}}
\newcommand{\ym}{y_{-}}
\newcommand{\ymbar}{\bar{y}_{-}}
\newcommand{\cm}{c_{-}}
\newcommand{\cmbar}{\bar{c}_{-}}
\newcommand{\fbderiv}{ \te{\partial} }

\newcommand\beq{\begin{eqnarray}}
\newcommand\eeq{\end{eqnarray}}

\begin{document}


\title{Composite Higgs Models with a Hidden Sector}
\author{Ann E. Nelson}
\email{aenelson@uw.edu}
\affiliation{Department of Physics, University of Washington, Seattle, WA}  
\author{Michael Park}
\email{q1park@uw.edu}
\affiliation{Department of Physics, University of Washington, Seattle, WA}  
\affiliation{Department of Physics, University of Washington, Bothell, WA}       
\author{Devin G. E. Walker}
\email{devin.g.walker@dartmouth.edu} 
\affiliation{Department of Physics and Astronomy, Dartmouth College, Hanover, NH}      

\begin{abstract}

We discuss the phenomenology of composite Higgs models that naturally produce a Standard Model-like Higgs boson with a mass of $126$ GeV. The effective theory below the compositeness scale is weakly coupled in these models, and the goldstone sector acts as a portal between the third generation of quarks and a hidden gauge sector. The addition of hidden-sector fermions gives rise to a calculable effective scalar potential with a naturally light scalar resonance. The generic prediction of these theories is the existence of additional pseudo-Nambu Goldstone bosons with electroweak-scale cross sections and masses. In this paper we analyze the collider signatures for some simple concrete realizations of this framework. We find that despite the existence of additional weakly and strongly coupled particles that are kinematically within reach of current experiments,  the generic signatures are  difficult to resolve at the LHC, and could remain well-hidden in the absence of an $e^+ e^-$ Higgs factory such as the CEPC or a surface detector such as MATHUSLA.
\end{abstract}

\maketitle
\tableofcontents

\section{Introduction}

Despite its success in discovering a Standard Model-like Higgs boson, the Large Hadron Collider (LHC) has yet to provide a satisfying explanation for the mechanism of electroweak symmetry breaking (EWSB). To date there is no discovery leading the way to new physics, and many of the popular explanatory frameworks are becoming constrained into finely tuned regions of their parameter spaces. Theoretical development over the last several decades has largely been motivated by criteria of {\it naturalness} and {\it parsimony}~\cite{Craig:2013cxa}. While there is a strong logical and historical motivation for this notion of naturalness, there is also an arguably comparable motivation for cautious skepticism in our conceptions about parsimony. In this paper we study a class of effective theories that generically give rise to a composite Higgs boson with Standard Model (SM)-like properties, and analyze their collider signatures. A light Higgs with a large quartic is achieved via the introduction of a hidden-sector that couples to the SM through the Higgs portal, thus preserving naturalness at the expense of parsimony. This work adds to a growing abundance of effective models for electroweak symmetry breaking with highly subtle hadron-collider signatures, and motivates several concrete analyses for further scrutiny.

The composite Higgs hypothesis \cite{Kaplan:1983fs} offers an attractive and radically conservative \cite{Wilczek:1988fw} possibility for the origin of a naturally macroscopic electroweak scale. In this framework the compositeness scale  is naturally higher than the electroweak symmetry breaking scale,  as  the Higgs arises as a pseudo-Nambu-Goldstone boson (pNGB) \cite{Kaplan:1983sm,Georgi:1984af,Georgi:1984ef,Dugan:1984hq,ArkaniHamed:2001nc,ArkaniHamed:2002qy,ArkaniHamed:2002qx,Low:2002ws,Kaplan:2003uc,Chang:2003un,Skiba:2003yf,Chang:2003zn,Piai:2004yb} from the spontaneous breaking of an approximate global symmetry by the compositeness dynamics. These models retain a weakly coupled description of the EWSB sector above the chiral symmetry breaking scale $f$, with a full UV completion appearing at a higher scale associated with the new strong dynamics $\Lambda$ \cite{Weinberg:1975gm, Susskind:1978ms}. The most parsimonious constructions of composite Higgs theories have historically found tension with measurements of precision electro-weak (PEW) observables due to their generic assumption of additional gauge structure above the scale $f$. The Intermediate Higgs (IH) or ``natural composite'' scenario \cite{Katz:2003sn,Katz:2005au,Contino:2006qr,Gripaios:2009pe} relaxes this assumption, which results in a quadratic sensitivity to the compositeness scale $\Lambda$ from loops of gauge bosons. However this contribution is numerically small enough to avoid serious fine tuning for a new physics scale as high as $\Lambda \sim 10$ TeV. The large radiative corrections to the Higgs mass from the top sector are canceled by new vector-like quarks with TeV-scale masses, utilizing the mechanism of collective symmetry breaking, and resulting in partially composite third generation~\cite{Kaplan:1991dc}. 

The recently measured mass of the SM Higgs boson $m_h = 126$ GeV~\cite{Chatrchyan:2012xdj,Aad:2012tfa} provides a significant piece of additional data for the composite Higgs hypothesis. In the IH scenario, the scalar potential is highly constrained by the explicit symmetry breaking pattern of the approximate global symmetries. Again the most parsimonious constructions have historically found difficulty generating a sufficiently light Higgs mass or an otherwise viable Higgs potential  \cite{Low:2004xc,Hubisz:2004ft,Bazzocchi:2005gs,Pappadopulo:2010jx,Vecchi:2013bja,Bellazzini:2014yua,Low:2015nqa}. In this paper we consider the effects of extending the composite sector to include additional hidden-sector interactions via Yukawa couplings between the pNGB matrix and a new multiplet of hidden-sector fermions. We find that the radiative corrections to the Higgs potential from these states at $f$ generically produce a pattern of EWSB that is consistent with current experimental data. The invariant prediction of these theories is the existence of new vector-like quarks with masses of $\mathcal{O}(f)$ and new weakly interacting scalars with masses of $\mathcal{O} (g f)$. All of the pNGB scalars remain uneaten in these models and manifest as physical resonances, however their weak scale cross sections are typically below threshold for discovery via direct production at the LHC. Furthermore the extended Yukawa sector can generically lead to non-standard decays of the new quarks that significantly weakens the limits from LHC searches for simplified top-partner models \cite{Khachatryan:2015gza,Sirunyan:2017usq}. The variety of possible coset spaces thus provides a wide range of theoretically motivated collider signatures that should now be considered more seriously. 

In this paper we select two contrasting examples for detailed analysis. Section \ref{sec:LL} discusses a model based on the $SU(5)/SO(5)$ coset space \cite{Georgi:1984af}, which generates a rich variety of new scalar resonances. Here we focus on the low-energy phenomenology as well as the conditions for a realistic pattern of EWSB. Large couplings between the pNGBs and the new vector-like quarks in this model lead to an enhanced production of high-multiplicity tops and bottoms, resulting in cascade decays with many leptons and $b$-quarks in the final state. The hidden-sector interactions result in a fairly generic hidden-valley phenomenology which we review in the context of these models. Section \ref{sec:SUSP} discusses the $SU(4)/Sp(4)$ coset space \cite{Georgi:1984af}, which generates only one new pNGB resonance beyond the Higgs multiplet. We briefly discuss the tension between the fermionic sector of this model and electroweak precision constraints on the bottom quark interactions and  how to alleviate this. We then focus on the difficulties in resolving the goldstone sector of this theory, which could remain well hidden until future Higgs factories come online. In Section \ref{sec:con} we conclude with a summary of these results.

\newpage

\section{$SU(5)/SO(5)$ Intermediate Model}
\label{sec:LL}

The symmetry breaking pattern $SU(5)/SO(5)$ produces $24-10=14$ pseudo Nambu-Goldstone Bosons (pNGBs). The sigma field transforms under $SU(5)$ as $V \Sigma V^T$ where $V$ is an $SU(5)$ matrix. It is convenient to specify an $SO(5)$ symmetric background field $\Sigma_0$, under which the unbroken $SU(5)$ generators $T$ and the broken generators $X$ satisfy

\be
\begin{matrix} 
  T \Sigma_0+ \Sigma_0 T^T=0 \\[2ex]
  X\Sigma_0-\Sigma_0 X^T=0 \\ 
\end{matrix}
\qquad \qquad \qquad
\Sigma_0 = 
\begin{pmatrix} 
  \quad & \quad & \mathbbm{1} \\
  \quad  & 1 & \quad \\
  \mathbbm{1}  & \quad & \quad \\ 
\end{pmatrix}.
\ee

\noindent The $SO(5)$ symmetry contains the global custodial $SO(4)_c = SU(2)_L \otimes SU(2)_R$ subgroup, where $SU(2)_R$ is approximate and contains $U(1)_Y$. The generators of this global symmetry $L^a$ and $R^a$ can be expressed as

\begin{align}
  L^a &=
  \frac{1}{2}\begin{pmatrix}
    \sigma^a & \quad & \quad \\
    \quad & $0$ & \quad \\
    \quad & \quad & -\sigma^{a *}
  \end{pmatrix} \\[2ex]
  R^a &=
  \frac{1}{2} \left \{ \begin{pmatrix}
    \quad & \quad & -i \sigma^{2} \\
    \quad & $0$ & \quad \\
    i \sigma^{2} & \quad & \quad
  \end{pmatrix} , \begin{pmatrix}
    \quad & \quad & -\sigma^{2} \\
    \quad & $0$ & \quad \\
    -\sigma^{2} & \quad & \quad
  \end{pmatrix} , \begin{pmatrix}
    \mathbbm{1} & \quad & \quad  \\
    \quad & $0$ & \quad \\
    \quad & \quad & -\mathbbm{1}
  \end{pmatrix} \right \}
  \label{eq:LLglobal}
\end{align}

\noindent To describe the quantum numbers, we embed the $SU(2)_L\otimes U(1)_Y$ gauge symmetry of the Standard Model into the global $SU(2)_L \otimes SU(2)_R$ with $Y=R^3$. These generators remain unbroken in the reference vacuum $\Sigma=\Sigma_0$ and the pNGB's are fluctuations about this background in the direction of the broken generators, $\Pi \equiv \pi^a X^a$. Under under the gauged $SU(2)_L \otimes U(1)_Y$ they transform as $1_0 \oplus 2_{\pm 1/2} \oplus 3_0 \oplus 3_{\pm 1}$ and may be parameterized as

\beq
 \begin{matrix} 
   \Sigma(x) = e^{2 i \Pi/f} \Sigma_0 \\ 
 \end{matrix}
 \qquad \qquad
\Pi  &=& \frac{1}{2} \begin{pmatrix} 
  \frac{1}{\sqrt{10}} \eta + \Phi & H^T & \tcapphi \\
  H^*  & -\frac{4}{\sqrt{10}} \eta ~\mathbbm{1} & H^c \\ 
  \tcapphi^\dagger & H^{c \dagger}  & -\frac{1}{\sqrt{10}} \eta + \Phi^* \\ 
\end{pmatrix}
\eeq

\noindent All of the goldstone modes in this theory remain uneaten and appear as physical fluctuations. These include the SM Higgs doublet $H = (H_+ ~H_0)$, a parity-odd electroweak singlet $\eta$, and a set of Georgi-Machacek scalars $\Phi$ and $\tcapphi$ \cite{Georgi:1985nv,Hartling:2014zca} which may be written as

\begin{equation}
  \tcapphi =
  \frac{1}{\sqrt{2}}\begin{pmatrix}
    \sqrt{2} \, \tphi_{++} & ~\tphi_{+} \\
    \tphi_{+} & ~\sqrt{2} \, \tphi_{0}\\
  \end{pmatrix} \qquad \qquad \Phi =
  \frac{1}{\sqrt{2}}\begin{pmatrix}
    \phi_{0}~ & ~\phi_{+} \\
    \phi_{+}^* \, & \phi_{0}\\
  \end{pmatrix} 
\end{equation}

\noindent Under the $SU(2)_L \otimes SU(2)_R$ global symmetry the pNGBs transform as $(2,2) \oplus (3,3) \oplus (1,1)$. The SM Higgs field transforms as a bi-doublet, while the real and complex triplets $\Phi$ and $\tcapphi$ transform together as a bi-triplet. The fermionic sector of this theory gives a large negative contribution to the mass of the Higgs, while the remaining goldstone masses are dominated by a positive contribution from the one-loop gauge interactions. The resulting vacuum breaks the gauge symmetry down to $U(1)_{\rm EM}$ and can be parameterized by a single angle $\theta$.

\beq
\theta^2 \equiv \frac{1}{4} \frac{H^\dagger H}{f^2} \qquad \qquad \langle \Sigma \rangle =
\begin{pmatrix} 0 & 0 & ~~0~~ & ~~~1~~~ & 0 \\
  0 & -\sin^2 \theta & ~~\frac{i}{\sqrt{2}} \sin 2 \theta~~ & ~~~0~~~ & \cos^2 \theta \\
  \frac{i}{\sqrt{2}} \sin 2 \theta & 0 & ~~\cos 2 \theta~~ & ~~~0~~~ & \frac{i}{\sqrt{2}} \sin 2 \theta \\
  1 & 0 & ~~0~~ & ~~~0~~~ & 0 \\
  0 & \cos^2 \theta & ~~\frac{i}{\sqrt{2}} \sin 2 \theta~~ & ~~~0~~~ & -\sin^2 \theta
\end{pmatrix}
\label{eq:hvev}
\eeq

\noindent The $SU(5)$ global symmetry is explicitly broken by the $SU(2)_L \times U(1)_Y$ gauge couplings. The covariant derivative can be expanded to second order in the vacuum giving tree level masses to the weak gauge bosons in terms of the $SU(2)_L \times U(1)_Y$ couplings $g, g^\prime$. The vacuum expectation value of the Higgs bi-doublet thus gives mass to the $W$ and $Z$ bosons which preserves the remnant custodial $SU(2)_c$, thus guaranteeing tree level relation $m_W/m_Z=\cos\theta_w$.

\begin{equation}
  D_\mu \Sigma = \partial_\mu \Sigma -  \left[ i g W_\mu^a (L^a \Sigma + \Sigma Q^{aT})  + i g' B_\mu (Y \Sigma + \Sigma Y^T) \right]
  \label{eq:cd}
\end{equation}
\begin{align}
m_W^2 = \frac{g^2}{2}\,f^2 \sin^2 2 \theta && m_Z^2 = \frac{g^2 + g'^2}{2}\, f^2 \sin^2 2 \theta.
\end{align}

\subsection{Fermion Sector}

If the $SU(5)$ symmetry arises accidentally from the dynamics of a strongly coupled theory then the Yukawa sector could contain interactions between the $\Sigma$ field and composite fermions in $SU(5)$ multiplets. Yukawa couplings that softly break the global symmetry can result in potentially large contributions to the scalar potential. However these interactions can naturally be constructed with a collective symmetry breaking property that guarantees the absence of quadratic divergences to the Higgs mass from fermion loops. Despite the absence of such quadratic divergences, constructing soft-symmetry breaking interactions that reproduce the measured properties of the Higgs boson remains non-trivial. We find that this goal can be achieved in a relatively simple way by extending the fermion sector of this theory to include two vector-like multiplets of fermions in the fundamental representation of $SU(5)$. One of the multiplets $(\psi, \,\bar{\psi})$ is color-charged and mixes with the SM quarks resulting in a partially composite third generation. The other multiplet $(\chi, \, \bar{\chi})$ is assumed to be charged under a hidden-sector gauge group $G$ that confines at some scale $\widetilde{\Lambda} < f$.

\subsubsection{Color-charged Fermions}

The most serious quadratic divergence to the Higgs potential from the top quark can be eliminated by extending the fermion sector of the Standard Model to include new vector-like quarks $(\psi, \,\bar{\psi})$ in the fundamental representation of $SU(5)$. The states in this multiplet mix with the massless chiral third generation $q$ and $(\bar{t}, \bar{b})$ to produce partially composite top and bottom mass eigenstates. The gauge quantum numbers of the vector-like quarks are fixed by the requirement of partial compositeness and the proposed embedding of $SU(2)_L \times U(1)_Y \subset SU(5)$. The corresponding embedding of the gauge eigenstate components in $(\psi, \,\bar{\psi})$ is given in Equation \ref{eq:llvlq}, and their transformation properties are given in Table \ref{tab:llqnumbers}
\be
\psi = \left( Q  ~~T ~~P \right) \qquad \qquad \bar{\psi} = \left( \overline{P} ~~\overline{T} ~~\overline{Q} \right)^{T}
\label{eq:llvlq}
\ee
\begin{center}
  \begin{table}[ht]
    \begin{tabular}{| c | c c c | c c c |}
      \hline
      & ~$\left( Q ~\overline{Q} \right)$~ & ~$\left( T ~\overline{T} \right)$~ & ~$\left( P ~\overline{P} \right)$~ & ~$q$~ & ~$\bar{t}$~ & ~$\bar{b}$~ \\ [0.5ex] 
      \hline
      $SU(3)_c$ & $(3 ~\overline{3})$ & $(3 ~\overline{3})$ & $(3 ~\overline{3})$ & $3$ & $\overline{3}$ & $\overline{3}$ \\
      $SU(2)_L$ & $(2 ~2)$ & $(\mathbbm{1} ~\mathbbm{1})$ & $(2 ~2)$ & $2$ & $\mathbbm{1}$ & $\mathbbm{1}$ \\
      $U(1)_Y$ & $\left( \frac{1}{6} ~\text{-}\frac{1}{6} \right)$ & $\left( \frac{2}{3} ~\text{-}\frac{2}{3} \right)$ & $\left( \frac{7}{6} ~\text{-}\frac{7}{6} \right)$ & $\frac{1}{6}$ & $\text{-}\frac{2}{3}$ & $\frac{1}{3}$ \\
      \hline
      $G$ & $(\mathbbm{1} ~\mathbbm{1})$ & $(\mathbbm{1} ~\mathbbm{1})$ & $(\mathbbm{1} ~\mathbbm{1})$ & $\mathbbm{1}$ & $\mathbbm{1}$ & $\mathbbm{1}$ \\
      \hline
    \end{tabular}
    \caption{Gauge quantum numbers for the new vector-like quarks. These include two new $SU(2)_L$ doublets $Q = (Q_T, \, Q_B)$ and $P = (P_T, \, P_B)$, in addition to a vector-like singlet $T$. The new vector-like states mix with the SM chiral doublet $q = (q_T, \, q_B)$ and singlets $(\bar{t}, \, \bar{b})$.}
    \label{tab:llqnumbers}
  \end{table}
\end{center}
The most general gauge invariant fermion interactions include terms that softly break the $SU(5)$ global symmetry. For simplicity we assume that the SM bottom quark mass term arises from the Yukawa interactions of an incomplete $SU(5)$ multiplet. This introduces an explicit source of $SU(5)$ breaking and a quadratic contribution to the Higgs potential from loops of bottom quarks. However due to the relatively small value of the bottom quark Yukawa coupling, these effects are numerically negligible compared to the leading logarithmic contribution from the top sector. We thus express the Yukawa interactions for the color-charged fermions as

\be
\CL_{\psi} =  \, y_1\,f\,\psi \,\Sigma\,\bar{\psi} + y_2 f \, q \,\overline{Q} + y_3\,f\,T\,\bar{t}\, + y_4\,f\,s_{2 \theta} \,q_B\,\bar{b} + {\rm h.c.}
\label{eq:lltopyuk}
\ee

\begin{center}
  \begin{table}[ht]
    \begin{tabular}{| c | c c c c |}
      \hline
      & $~~\overline{T}~~$ & $~~\overline{Q}_T~~$ & $~~\overline{P}_T~~$ & $~~\bar{t}~~$ \\ [0.5ex] 
      \hline
      $T$  & $y_1 f c_{2 \theta}$ & $\frac{i}{\sqrt{2}} y_1 f s_{2 \theta}$ & $\frac{i}{\sqrt{2}} y_1 f s_{2 \theta}$ & $y_3 f$ \\
      $Q_T$  & $\frac{i}{\sqrt{2}} y_1 f s_{2 \theta}$ & $y_1 f c^2_{\theta}$ & $-y_1 f s^2_{\theta}$ & $0$ \\
      $P_T$  & $\frac{i}{\sqrt{2}} y_1 f s_{2 \theta}$ & $-y_1 f s^2_{\theta}$ & $y_1 f c^2_{\theta}$ & $0$ \\
      $q_T$  & $0$ & $y_2 f$ & $0$ & $0$ \\
      \hline
    \end{tabular} \qquad \qquad   \begin{tabular}{| c | c c c |}
      \hline
      & $~~\overline{Q}_B~~$ & $~~\overline{P}_B~~$ & $~~\bar{b}~~$ \\ [0.5ex] 
      \hline
      $Q_B$ & $y_1 f c^2_{\theta}$ & $-y_1 f s^2_{\theta}$ & 0 \\
      $P_B$ & $-y_1 f s^2_{\theta}$ & $y_1 f c^2_{\theta}$ & $0$ \\
      $q_B$ & $y_2 f$ & $0$ & $y_4 s_{2 \theta}$ \\
      \hline
    \end{tabular}
    \caption{Mass matrices for the top-like quarks $M_T$ (left) and the bottom-like quarks $M_B$ (right) in the gauge eigenbasis.}
    \label{tab:llqmass}
  \end{table}
\end{center}
\vspace{-1cm}

\noindent The top and bottom quark mass matrices $M_T$, $M_B$ can be expressed in terms of the Higgs vacuum expectation value $\theta$ defined in Equation (\ref{eq:hvev}), and the UV-insensitivity of the Yukawa interactions is guaranteed by the following identity

\begin{align}
 \frac{\partial}{\partial \theta} \tr M_T^\dagger M_T &= \frac{\partial}{\partial \theta} \tr (M_T^\dagger M_T)^2 = 0
 \label{eq:llnodiv}. 
\end{align}

\noindent Diagonalizing $M_T$ and $M_B$ gives seven mass eigenstates. The up-type sector contains four charge $\pm 2/3$ states which we label in order of descending mass as $\tppp$, $\tpp$, $\tp$, $t$, and the lightest of these states corresponds to the SM top quark. The down-type sector contains three mass eigenstates. Two of these $\bp$, $b$ have charge $\mp 1/3$ and include the SM bottom quark, and there is additionally a ``peculiar'' charge $\pm 5/3$ quark $\pb$, which can lead to interesting phenomenological signatures. The heavy mass eigenstates are independent of the Higgs VEV at leading order, and the numerically diagonalized mass spectrum is shown in Figure (\ref{fig:llfmasses}). Their analytic expressions have been computed in Appendix \ref{app:llscalar} to leading order in $\sin^2 2 \theta$ and are approximated by the following scalings.

\be
\begin{aligned}[c]
m_t^2 &\sim \frac{f^2}{2} y_t^2 \sin^2 2 \theta \\
m_{\tp}^2 &\sim f^2 y_1^2 \\
m_{\tpp}^2 &\sim f^2 \left(y_1^2 + y_2^2 \right) \\m_{\tppp}^2 &\sim f^2 \left( y_1^2 + y_3^2 \right) 
\end{aligned}
\qquad\qquad\qquad
\begin{aligned}[c]
m_b^2 &\sim \frac{f^2}{2} y_b^2 \sin^2 2 \theta\\
m_{\pb}^2 &\sim f^2 y_1^2 \\
m_{\bp}^2 &\sim f^2 \left(y_1^2 + y_2^2 \right) \\
& 
\end{aligned}
\ee

\noindent The top-sector mass matrix $M_T$ also has the property that $\det M_T^\dagger M_T \propto \sin^2 2 \theta$, thus producing one massless state when $\theta = n \pi / 2$ for $n \in \mathbb{Z}$. The states corresponding to the SM top and bottom quarks $t$ and $b$ remain massless at zeroth order in $\theta$, and couple to the Higgs via the following Yukawa couplings

\be
y_t=\frac{y_1 y_2 y_3}{\sqrt{y_1^2 + y_2^2} \sqrt{y_1^2 + y_3^2}} \qquad \qquad
y_b=\frac{y_1 y_4}{\sqrt{y_1^2 + y_2^2}}
\label{eq:llSMyuk}
\ee

\subsubsection{Hidden-Sector Fermions}

A realistic pattern of EWSB via radiative effects can be achieved throughout this parameter space by introducing an additional multiplet of vector-like fermions $(\chi, \, \bar{\chi})$. In this analysis we will assume that $(\chi, \, \bar{\chi})$ are charged under a hidden-sector gauge group $G$, transforming in the $(\Box,5) \oplus (\ovB{\Box},5)$ representation of $G \times SU(5)$. If the hidden-sector gauge group $G$ confines at some scale $\widetilde{\Lambda} < f$, then the low energy spectrum will be comprised of hidden-sector mesons, baryons, and glueballs. Due to a conserved $U(1)_B$ in the hidden-sector, the baryons can be stable on cosmological time scales and may thus be an attractive candidate for dark matter if the lightest baryon is electrically neutral. The direct and indirect detection signatures for these scenarios have been well studied in the context of asymmetric dark matter~\cite{Kaplan:2009ag,Laha:2013gva,Laha:2015yoa,Mitridate:2017oky,Forestell:2017wov,Elor:2018xku,Braaten:2018xuw} and will not be discussed further. Here we choose a hypercharge assignment for the hidden-sector fermions that can naturally accomodate a number of low-lying neutral mass eigenstates. The hidden-sector baryons in this framework thus become a plausible candidate for dark matter, although in general this will depend on the discrete $\mathbb{Z}_N$ subgroup at the center of $G$ and the mass splitting interactions of the hidden baryon spectrum. The embedding of the gauge eigenstate components in $(\chi, \, \bar{\chi})$ is given in Equation \ref{eq:llvlqdark} and the gauge quantum numbers of these new states are given in Table \ref{tab:llqnumbersdark}. 

\be
\chi = \left( X  ~~N ~~Y \right) \qquad \qquad \bar{\chi} = \left( \overline{Y} ~~\overline{N} ~~\overline{X} \right)^{T}
\label{eq:llvlqdark}
\ee
\begin{center}
  \begin{table}[ht]
    \begin{tabular}{| c | c c c | c |}
      \hline
      & ~$\left( X ~\overline{X} \right)$~ & ~$\left( N ~\overline{N} \right)$~ & ~$\left( Y ~\overline{Y} \right)$~ & ~$\left( n ~\bar{n} \right)$~  \\ [0.5ex] 
      \hline
      $SU(3)_c$ & $(\mathbbm{1} ~\mathbbm{1})$ & $(\mathbbm{1} ~\mathbbm{1})$ & $(\mathbbm{1} ~\mathbbm{1})$ & $(\mathbbm{1} ~\mathbbm{1})$  \\
      $SU(2)_L$ & $(2 ~2)$ & $(\mathbbm{1} ~\mathbbm{1})$ & $(2 ~2)$ & $(\mathbbm{1} ~\mathbbm{1})$ \\
      $U(1)_Y$ & $\left( \frac{1}{2} ~\text{-}\frac{1}{2} \right)$ & $\left( 0 ~0 \right)$ & $\left( \text{-}\frac{1}{2} ~\frac{1}{2} \right)$ & $\left( 0 ~0 \right)$  \\
      \hline
      $G$ & $(\Box ~\ovA{\Box})$ & $(\Box ~\ovA{\Box})$ & $(\Box ~\ovA{\Box})$ & $(\Box ~\ovA{\Box})$  \\
      \hline
    \end{tabular}
    \caption{Gauge quantum numbers for the new vector-like hidden-sector fermions. These include two $SU(2)_L$ doublets $X = (X_+ ~~X_0)$ and $Y = (Y_- ~~Y_0)$, as well as two singlets $N$ and $n$.}
    \label{tab:llqnumbersdark}
  \end{table}
\end{center}

A realistic Higgs potential from one-loop fermion corrections can be obtained by introducing symmetry breaking interactions between the hidden-sector $SU(5)$ multiplet $(\chi, \, \bar{\chi})$ and an additional massless vector-like pair of hidden-fermions $(n, \, \bar{n})$. The Yukawa interactions thus include an $SU(5)$ symmetric coupling to the $\Sigma$-field in addition to terms that softly break the global symmetry

\be
  \CL_{\chi} = \tilde{y}_1\,f\,\chi \,\Sigma\,\bar{\chi} + \tilde{y}_2 f n \,\overline{N} + \tilde{y}_3 f N \,\overline{n}  + {\rm h.c.}
  \label{eq:lltopyukdark}
\ee

\begin{center}
  \begin{table}[ht]
    \begin{tabular}{| c | c c c c |}
      \hline
      & $~~\overline{N}~~$ & $~~\overline{X}_0~~$ & $~~\overline{Y}_0~~$ & $~~\bar{n}~~$ \\ [0.5ex] 
      \hline
      $N$  & $\tilde{y}_1 f c_{2 \theta}$ & $\frac{i}{\sqrt{2}} \tilde{y}_1 f s_{2 \theta}$ & $\frac{i}{\sqrt{2}} \tilde{y}_1 f s_{2 \theta}$ & $\tilde{y}_3 f$ \\
      $X_0$  & $\frac{i}{\sqrt{2}} \tilde{y}_1 f s_{2 \theta}$ & $\tilde{y}_1 f c^2_{\theta}$ & $-\tilde{y}_1 f s^2_{\theta}$ & 0 \\
      $Y_0$  & $\frac{i}{\sqrt{2}} \tilde{y}_1 f s_{2 \theta}$ & $-\tilde{y}_1 f s^2_{\theta}$ & $\tilde{y}_1 f c^2_{\theta}$ & $0$ \\
      $n$ & $\tilde{y}_2 f$ & $0$ & $0$ & $0$ \\
      \hline
    \end{tabular} \qquad \qquad
    \begin{tabular}{| c | c c |}
      \hline
      & $~~\overline{X}_+~~$ & $~~\overline{Y}_-~~$ \\ [0.5ex] 
      \hline
      $X_+$  & $\tilde{y}_1 f c^2_{\theta}$ & $-\tilde{y}_1 f s^2_{\theta}$ \\
      $Y_-$  & $-\tilde{y}_1 f s^2_{\theta}$ & $\tilde{y}_1 f c^2_{\theta}$ \\
      \hline
    \end{tabular}
    \caption{Mass matrix for the neutral hidden-sector fermions $M_N$ (left) and the charged hidden-sector fermions $M_C$ (right) in the gauge eigenbasis.}
    \label{tab:llqmassdark}
  \end{table}
\end{center}
\vspace{-1cm}

\noindent In this scenario the mass matrix for the neutral hidden-sector states $M_N$ has the property $\det M_N^\dagger M_N \propto \cos^2 2 \theta$, resulting in one massless particle when $\theta = (n+1/2) \pi / 2$ for $n \in \mathbb{Z}$. The absence of a quadratic senstivity to the Higgs mass from loops of hidden-sector fermions is guaranteed by the identity

\begin{align}
 \frac{\partial}{\partial \theta} \tr M_N^\dagger M_N &= \frac{\partial}{\partial \theta} \tr (M_N^\dagger M_N)^2 = 0
 \label{eq:llnodivdark}
\end{align}

\noindent The hidden-sector mass matrix for $M_N$ produces four electrically neutral mass eigenstates which we label in in order of descending mass as $\nppp$, $\npp$, $\np$, $n$. The lightest of these states is generically comparable in mass to the heaviest goldstone modes, and thus stable against pair production via resonant scalar decay. The down-type hidden-sector mass eigenstates include a charge $\pm 1$ vector-like pair $(\xp, \, \xpbar)$ and a charge $\mp 1$ vector-like pair $(\ym, \, \ymbar)$. Since these hidden-sector states are color-neutral, their relatively low production cross sections make all but the lightest state $n$ inconsequential for collider phenomenology. For the remainder of this section we thus consider a simplified parameter space for this sector in which $\tilde{y}_3 = \tilde{y}_2 = \tilde{y}_1$. Analytic expressions for general couplings have been derived in Appendix \ref{app:llscalar} to leading order in $\sin^2 2 \theta$. With this simplified parameter space the hidden-sector fermion masses scale approximately as

\be
\begin{aligned}[c]
m_n^2 &\sim \frac{f^2}{4} \tilde{y}_1^2 \cos^2 2 \theta \\[1ex]
m_{\np}^2 &\sim f^2 \tilde{y}_1^2 \\[1ex]
m_{\npp}^2 &\sim m_{\nppp}^2 \sim 2 f^2 \tilde{y}_1^2 \\
\end{aligned}
\qquad\qquad\qquad
\begin{aligned}[c]
& \\
m_{\ym}^2 &\sim m_{\xp}^2 \sim f^2 \tilde{y}_1^2 \\
& 
\end{aligned}
\ee

\subsubsection{Theory Space}

\begin{figure}
\begin{minipage}{.45\textwidth}
\includegraphics[trim={0cm 0cm 0cm 0cm},clip,width=1.0\linewidth]{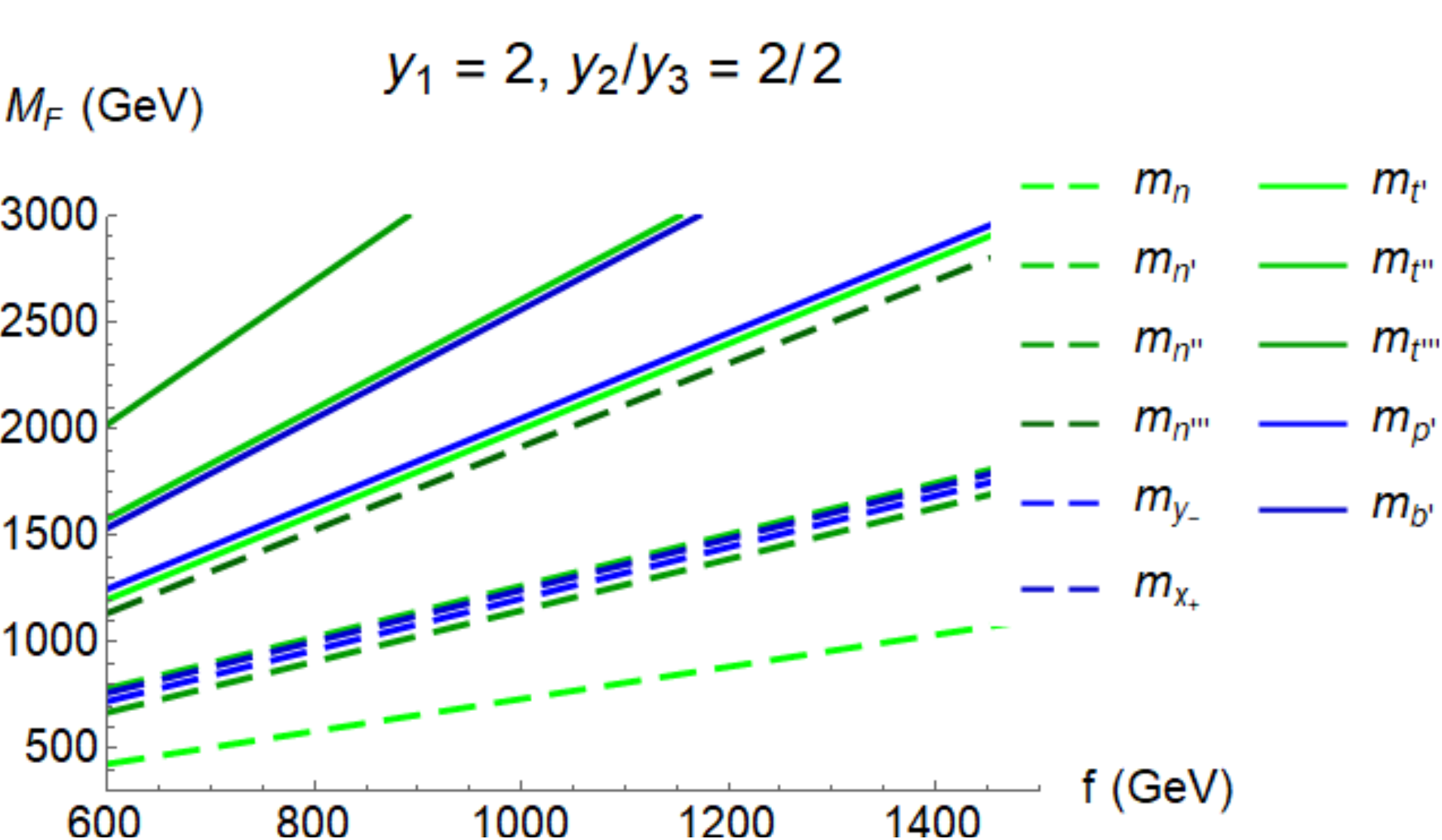}
\end{minipage}
\hspace{1cm}
\begin{minipage}{.45\textwidth}
\includegraphics[trim={0cm 0cm 0cm 0cm},clip,width=1.0\linewidth]{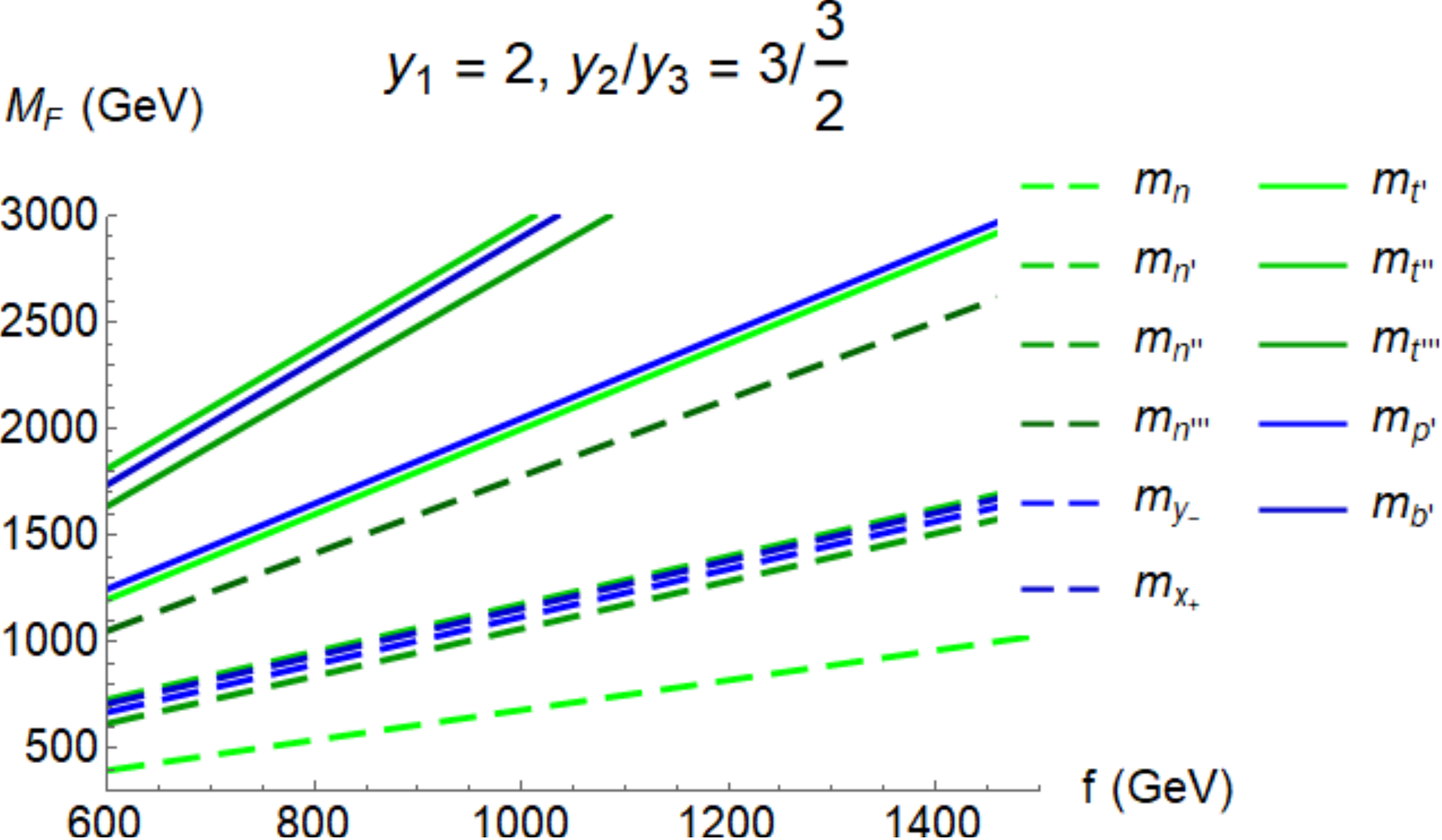}
\end{minipage}
\caption{Mass spectrum for the new fermions at two benchmark points. One corresponding to $y_2 / y_3= 2/2$ (left) and $y_2 / y_3 = 3/\frac{3}{2}$ GeV (right). The hidden-fermion masses are fixed relative to the color-charged states by the Higgs mass.}
\label{fig:llfmasses}
\end{figure}

The fermion interactions introduce a number of new parameters to the Standard Model. The new color-charged interactions introduce four Yukawa couplings $y_i$, two of which are fixed by the Standard Model top and bottom quark masses as implied by Equation \ref{eq:llSMyuk}. This leaves two free parameters $y_1$ and $y_2$ which are bound from below by the top-quark Yukawa coupling $y_t < y_1 y_2 / \sqrt{y_1^2 + y_2^2}$. The large observed top quark mass in this framework thus descends from the $\mathcal{O}(1)$ values of the composite sector Yukawa couplings. The new hidden-sector interactions introduce three new Yukawa couplings which we have reduced to a single parameter by assuming $\tilde{y}_1 = \tilde{y}_2 = \tilde{y}_3$. The fermion content of this theory has been chosen to allow for a pattern of electroweak symmetry breaking that can be tuned to agree with the measured properties of the Higgs boson. In this simplified parameterization, for each point in the two-dimensional parameter space of $(y_1, \, y_2)$ there exists a unique value for $\tilde{y}_1$ which gives a Higgs mass and quartic that is consistent with Standard Model predictions. The conditions of EWSB thus generically imply a mass scale for the new hidden-fermions that is $\mathcal{O}({\rm few})$ times lower than the masses of the new color-charged states. The masses of the new fermions as a function of the chiral symmetry breaking scale $f$ are illustrated for two benchmark points in Figure \ref{fig:llfmasses}.

\begin{figure}
\begin{minipage}{.31\textwidth}
\includegraphics[trim={0cm 0cm 0cm 0cm},clip,width=1.0\linewidth]{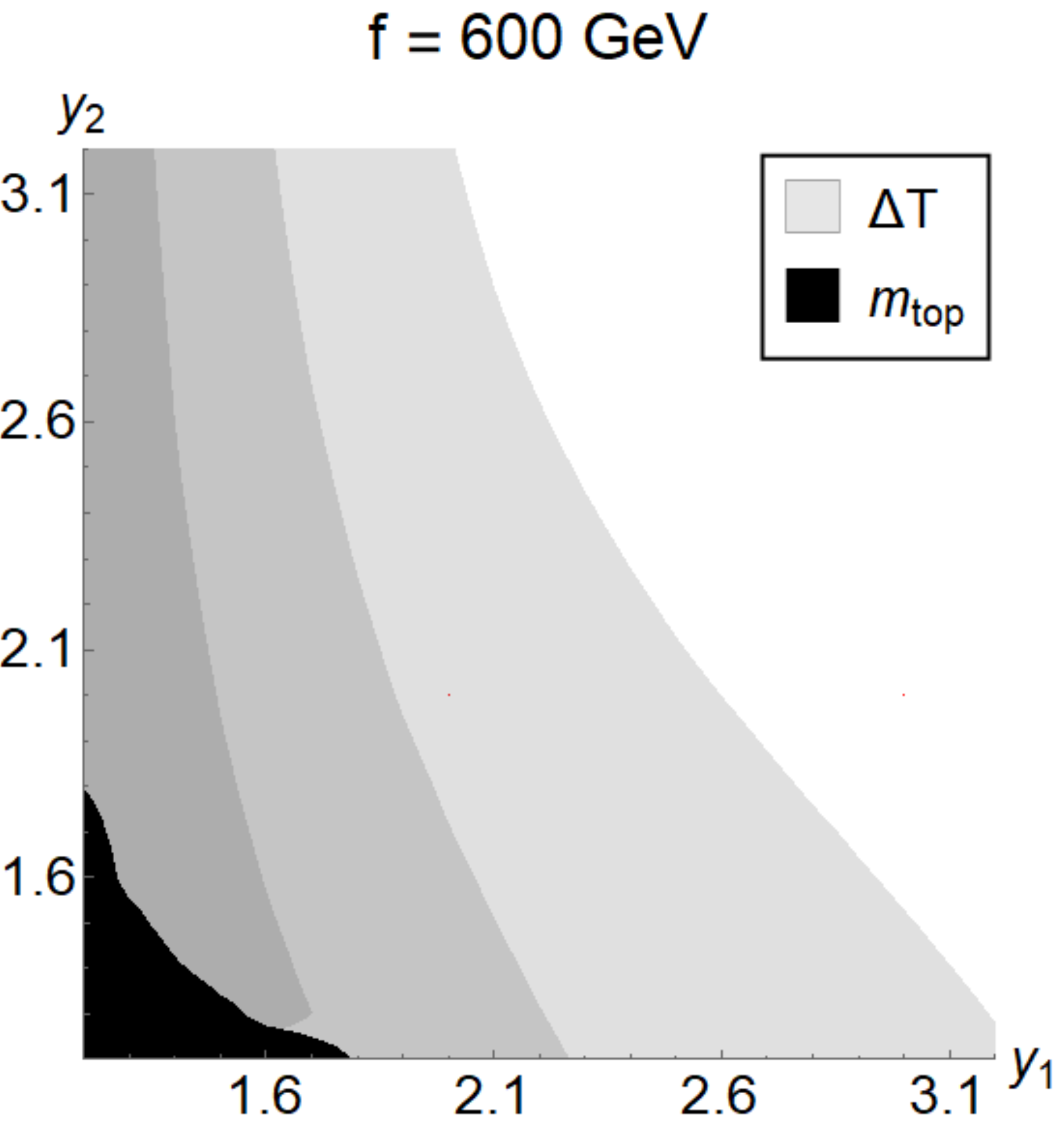}
\end{minipage}
\hspace{0.25cm}
\begin{minipage}{.31\textwidth}
\includegraphics[trim={0cm 0cm 0cm 0cm},clip,width=1.0\linewidth]{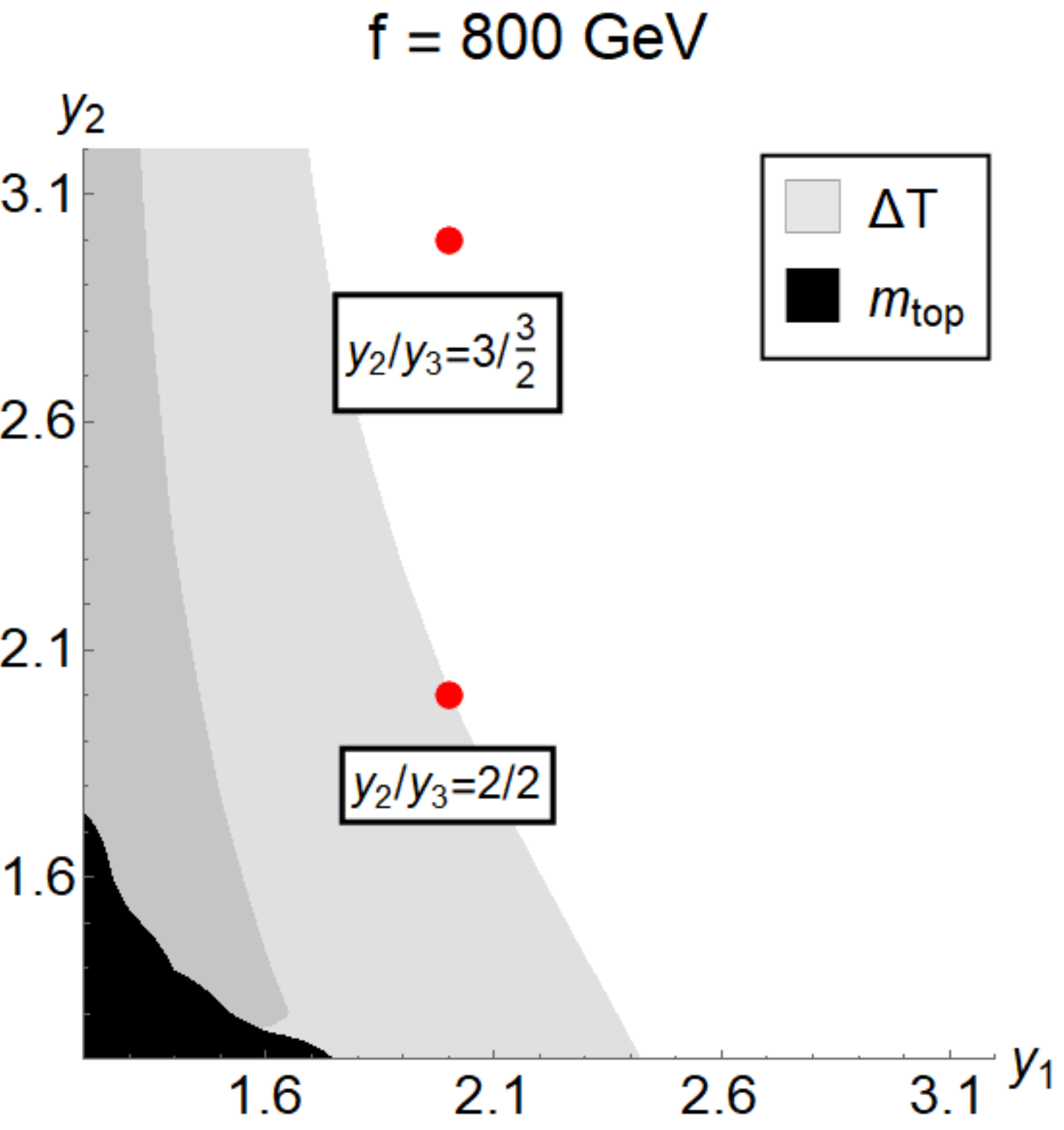}
\end{minipage}
\hspace{0.25cm}
\begin{minipage}{.31\textwidth}
\includegraphics[trim={0cm 0cm 0cm 0cm},clip,width=1.0\linewidth]{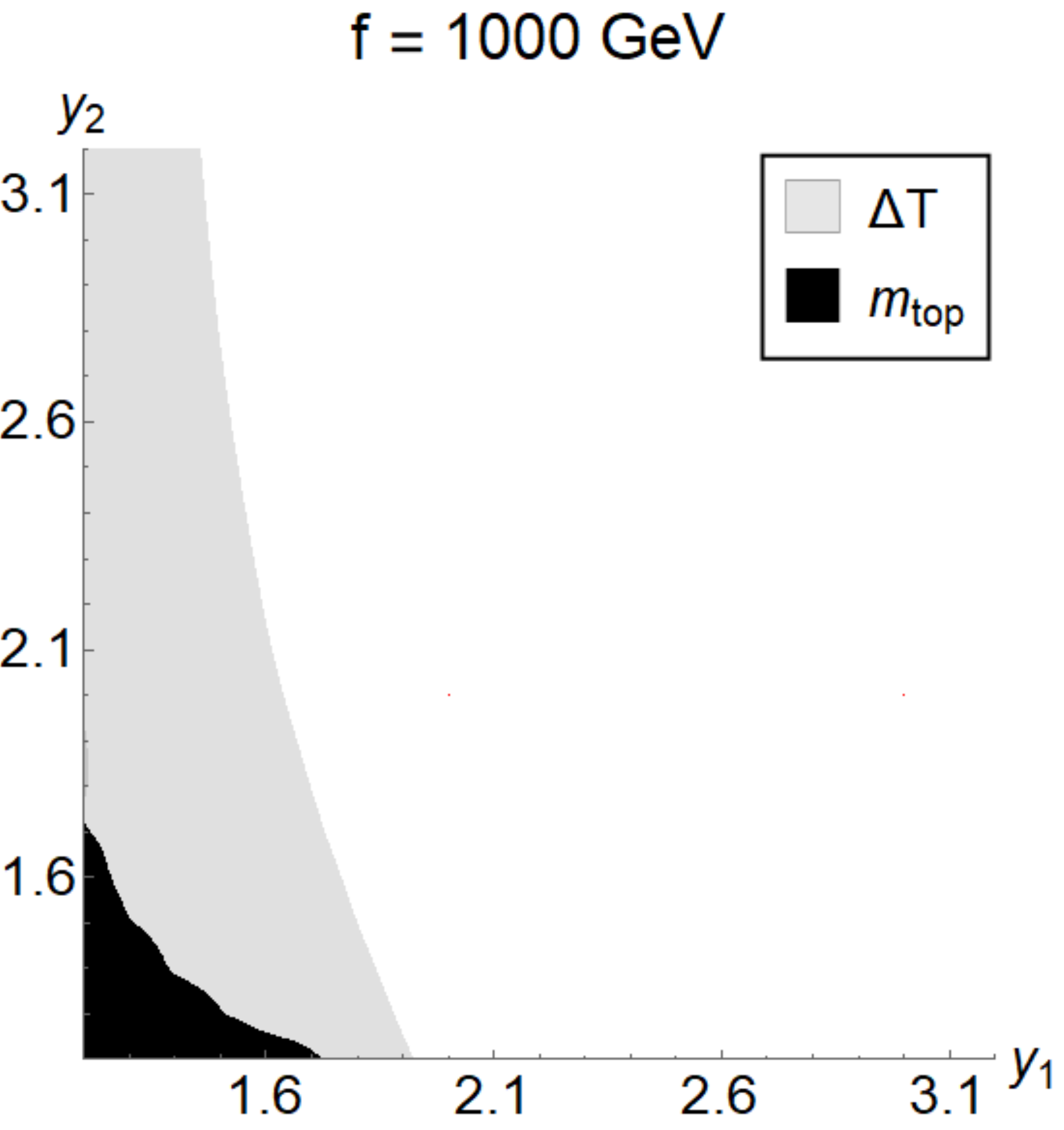}
\end{minipage}
\caption{Constraints from the $T$-parameter at $f = 600$ GeV (left), $800$ GeV (middle), and $1000$ GeV (right). The light/medium/dark gray correspond to the $1 \sigma$/$2 \sigma$/$3 \sigma$ deviations from experimentally measured values. The black corresponds to regions inconsistent with the top quark mass.}
\label{fig:PEWT}
\end{figure}

These Yukawa interactions also introduce additional sources of $SU(2)_c$ violation which can generate large contributions to the $T$-parameter at loop-level. These contributions will vanish as the chiral symmetry breaking scale $f$ increases due to the decoupling of these vector-like states, but can become unacceptably large at low values of $f$. The $T$-parameter may be computed from the fermionic contributions to the $W$ and $Z$-boson self-energies $\Pi_{WW}$ and $\Pi_{ZZ}$ as defined in Equation \ref{eq:llPEW}. We find that these $SU(2)_c$ violating interactions provide unacceptable contributions to the $T$-parameter when $f$ becomes on the order of $600$  GeV, as illustrated in Figure \ref{fig:PEWT}.

\begin{align}
  \alpha(m_Z) \, T &= \frac{\Pi_{WW} (0)}{m_W^2} - \frac{\Pi_{ZZ} (0)}{m_Z^2}
  \label{eq:llPEW}
\end{align}

\subsection{Electroweak Symmetry Breaking}

The scalar effective potential in this theory is sculpted by the radiative corrections from $SU(5)$ breaking interactions at $f$, which may be extracted from the quadratically and logarithmically sensitive terms in the Coleman Weinberg potential. The absence of additional gauge structure above the chiral symmetry breaking scale implies that the contribution from $W$ and $Z$ loops is quadratically sensitive to the compositeness scale $f$, which we take to be $\Lambda \sim 4 \pi f$. The form of this gauge contribution $\delta V_G$ can be extracted from the covariant derivative and expressed in terms of the gauge boson mass matrix $M_V$
\begin{align}
  \delta V_G (c, \theta) &= f^2 \tr M_V^2 (\Sigma) \\
  &= f^4 c \left( g^2 \tr \left[ L^a \Sigma (L^a \Sigma)^*\right] + {g'}^2 \tr \left[ Y \Sigma (Y \Sigma)^*\right] \right) \\
  &= - f^2 f_G^2 \, \cos 2 \theta \\[5pt]
 f_G^2 &= \frac{f^2 c}{2} \left( 3 g^2 + g^{\prime 2} \right)
  \label{eq:llgaugecw}
\end{align}
Here  $c$ is a UV sensitive constant which parameterizes the leading gauge contribution to the effective potential for the pNGBs. Dimensional analysis gives $c$ of order 1, while a technicolor-like UV completion of the model requires $c$ to be positive, that is, the gauge contribution to vacuum alignment prefers vanishing gauge boson masses. The analogous term in the  effective low energy description of QCD gives the (positive) mass squared splitting between the $\pi^+$ and $\pi^0$ ~\cite{Piai:2004yb}.

The absence of quadratically divergent one-loop contributions from the Yukawa interactions follows from the properties of the color-charged and hidden-sector fermion mass matrices in Equation \ref{eq:llnodiv} and Equation \ref{eq:llnodivdark}. The most significant contribution to the scalar potential from fermions $\delta V_F$ in this effective theory thus comes from the logarithmically sensitive term. The above relations again guarantee that the $\Lambda$ dependent terms cancel order by order in $\theta$. The result is thus sensitive only to scales of $\mathcal{O} (f)$, and can be written in terms of ratios of mass eigenstates.
\begin{align}
  \delta V_F (\theta) &= - \frac{3 }{16 \pi^2} \tr(M_F^\dagger M_F)^2 \log \frac{M_F^\dagger M_F}{\Lambda^2} \\[1.0ex]
  &= - \frac{3}{16 \pi^2} \displaystyle\sum_{F} |m_F^2 (\theta)|^2 \log |m_F^2(\theta)|
\end{align}
The collective symmetry breaking property of these mass matrices guarantees that these contributions to the effective potential are extremely well approximated by their leading order terms in $\det M_F^\dagger M_F$. As argued in Appendix \ref{app:llscalar}, higher order terms in $\det M_F^\dagger M_F$ are negligible due to corresponding suppression by powers of $(\tr M_F^\dagger M_F)^4$. The most significant logarithmic contributions from the color-charged top-like sector $\delta V_T$ and the neutral hidden-sector $\delta V_N$ thus enter with opposite sign, and the $\theta$ dependence is numerically well approximated by $\delta V_F \propto \pm \cos 4 \theta$. The dominant contribution to the Higgs potential from fermions may thus be parameterized by two scales $f_T$ and $f_N$, which are functions of the color-charged and hidden-sector Yukawa couplings respectively.
\begin{align}
  \delta V_F (\theta) &= \delta V_T (y_i, \theta) + \delta V_N (\tilde{y}_i, \theta) \\[2pt]
  &\approx f^2 \left( f_T^2 - f_N^2 \right) \cos 4 \theta
\end{align}
\be
f_T^2 = \frac{3 f^2}{32 \pi^2} \frac{y_1^2 y_2^2 y_3^2}{y_{2}^2 - y_{3}^2} \log \frac{y_1^2 + y_{2}^2}{y_1^2 + y_{3}^2} \qquad \qquad f_N^2 = \frac{3 f^2}{32 \pi^2} \tilde{y}_1^4
\ee

\noindent In this simplified parameter space, the hidden-sector fermion mass spectrum is fixed by the masses of the color-charged fermions and the conditions of electroweak symmetry breaking. The $\tilde{y}_i$ dependence of this potential for arbitrary hidden-sector Yukawa couplings are derived in Appendix \ref{app:llscalar}.

Since all of the Yukawa and gauge interactions are invariant under the axial $U(1)_a \subset SU(5)$, the goldstone excitation $\eta$ in this direction is left massless by the contributions considered thus far. In order to give a mass to this gauge-singlet large enough to avoid constraints from hadronic physics, we introduce an explicit source of $U(1)_a$ violation in the form of a spurion term that proportional to the reference vacuum $M_0 \equiv m_0 \Sigma_0$. Such a contribution would descend naturally from fermion mass terms in the UV completion.
\begin{align}
  \delta V_{0}(m_{0}, \theta) &= -f^3 \tr[ M_{0} \Sigma ] \\
  &= - f^2 f_0^2 \,\cos 2 \theta \\
 f_0^2 &= 2 \, f m_{0}
\label{eq:llspurion}
\end{align}
In the absence of additional symmetry breaking terms, the scalar potential is fully parameterized by the Yukawa coupings $y_i$ and $\tilde{y}_i$, the 1-loop gauge coefficient $c$, and spurion coefficient $m_0$. This potential has a generic EWSB vacuum when $f_T^2 - f_N^2 \gg f_0^2 + f_G^2 > 0$ and can be expressed as 
\begin{align}
  V_{\rm H} (\theta) &= - f^2 \left(f_0^2 + f_G^2 \right) \cos 2 \theta + f^2 \left( f_T^2 - f_N^2 \right) \cos 4 \theta \\[5pt]
  \langle \theta^2 \rangle &= \frac{\langle H^\dagger H \rangle}{4 f^2} = \frac{3}{2} \left[ \frac{4 ( f_T^2 - f_N^2 ) - \left( f_0^2 + f_G^2 \right)}{16 ( f_T^2 - f_N^2 ) - \left( f_0^2 + f_G^2 \right)} \right]
\label{eq:llscalarV}
\end{align}
The order parameter $\langle \theta^2 \rangle$ is assumed to develop a vacuum expectation value at $v^2 = 4 f^2 \langle \theta^2 \rangle = (246 \text{ GeV})^2$. The observed Higgs mass $m_h = 126$ GeV thus fixes the relation between the color-charged and hidden sector Yukawa couplings, as well as between the gauge renormalization coefficient $c$ and the axial-breaking spurion coefficient $m_0$.
\begin{align}
  V_{\rm H} ( \theta ) \xrightarrow{H \rightarrow v+h}& \quad \frac{m_h^2}{2} h^2 + \frac{\lambda}{4} h^4 + \mathcal{O}(h^6) \\[10pt]
  f_T^2 - f_N^2 &= \frac{m_h^2}{24} \left( \frac{3}{2} \frac{1}{\langle \theta^2 \rangle} - 1 \right) \\[5pt]
f_0^2 + f_G^2 &= \frac{2 m_h^2}{3} \left( \frac{3}{8} \frac{1}{\langle \theta^2 \rangle} - 1 \right)
\label{eq:llhiggs}
\end{align}

The mass spectrum of the non-SM scalars can also be extracted from these higher dimensional operators. The fermionic contributions to the potential of the scalars $\tcapphi$ and $\Phi$ are suppressed by $\mathcal{O}(\theta^2)$. Their masses are thus roughly independent of the Yukawa couplings $y_i$ and dominated by the positive contribution from gauge interactions.
\be
  m_{\tcapphi}^2 \sim 4 f^2 \left[ c \left( g^2 + \frac{g^{\prime 2}}{2} \right) + \frac{m_0}{f} \right] \qquad\qquad m_{\Phi}^2 \sim 4 f^2 \left[ c\, g^2 + \frac{m_0}{f} \right]
\ee
\be
  m_{\eta}^2 \sim 4 f m_0
\ee

\begin{figure}
\begin{minipage}{.31\textwidth}
\includegraphics[trim={0cm 0cm 0cm 0cm},clip,width=1.0\linewidth]{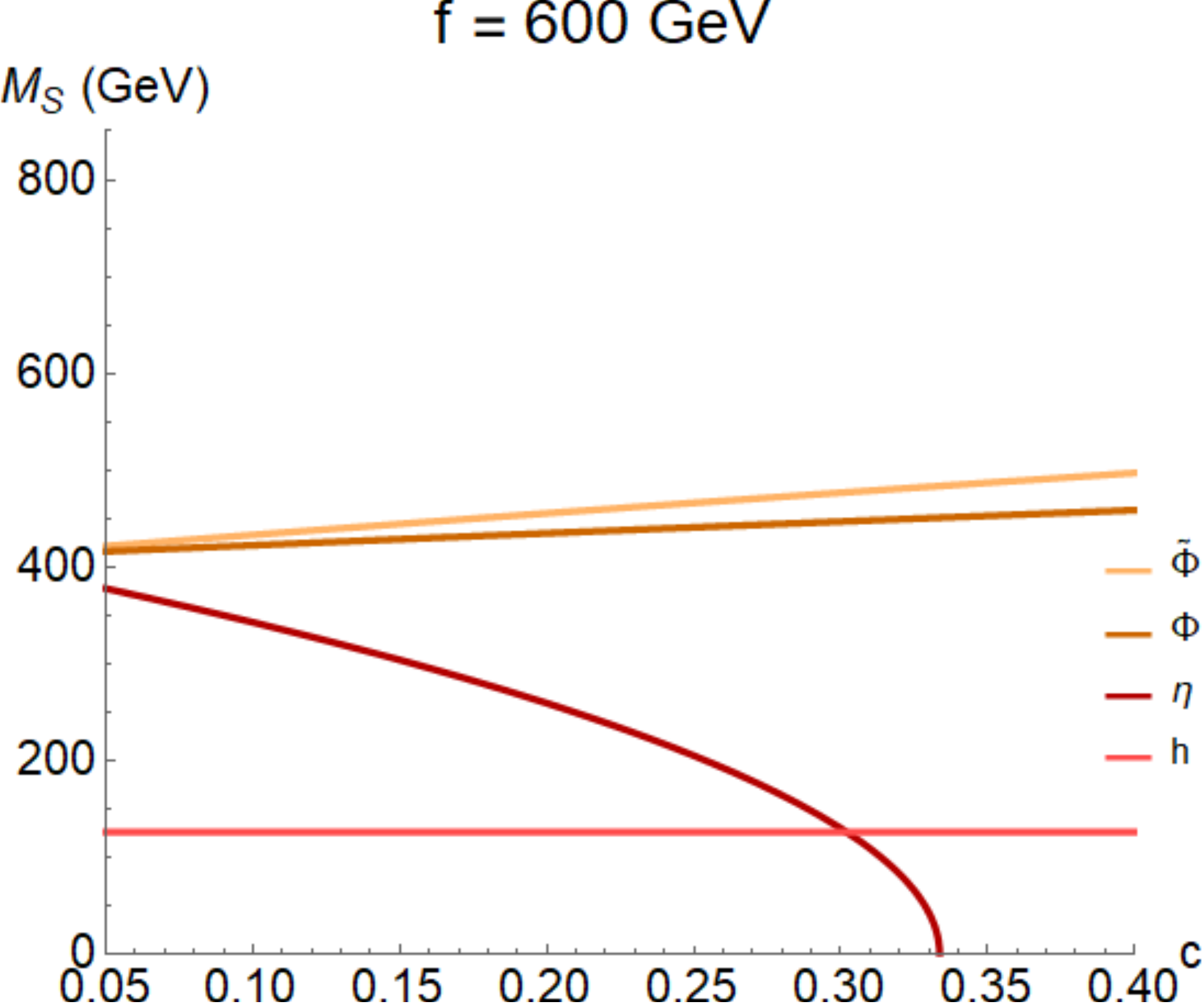}
\end{minipage}
\hspace{0.25cm}
\begin{minipage}{.31\textwidth}
\includegraphics[trim={0cm 0cm 0cm 0cm},clip,width=1.0\linewidth]{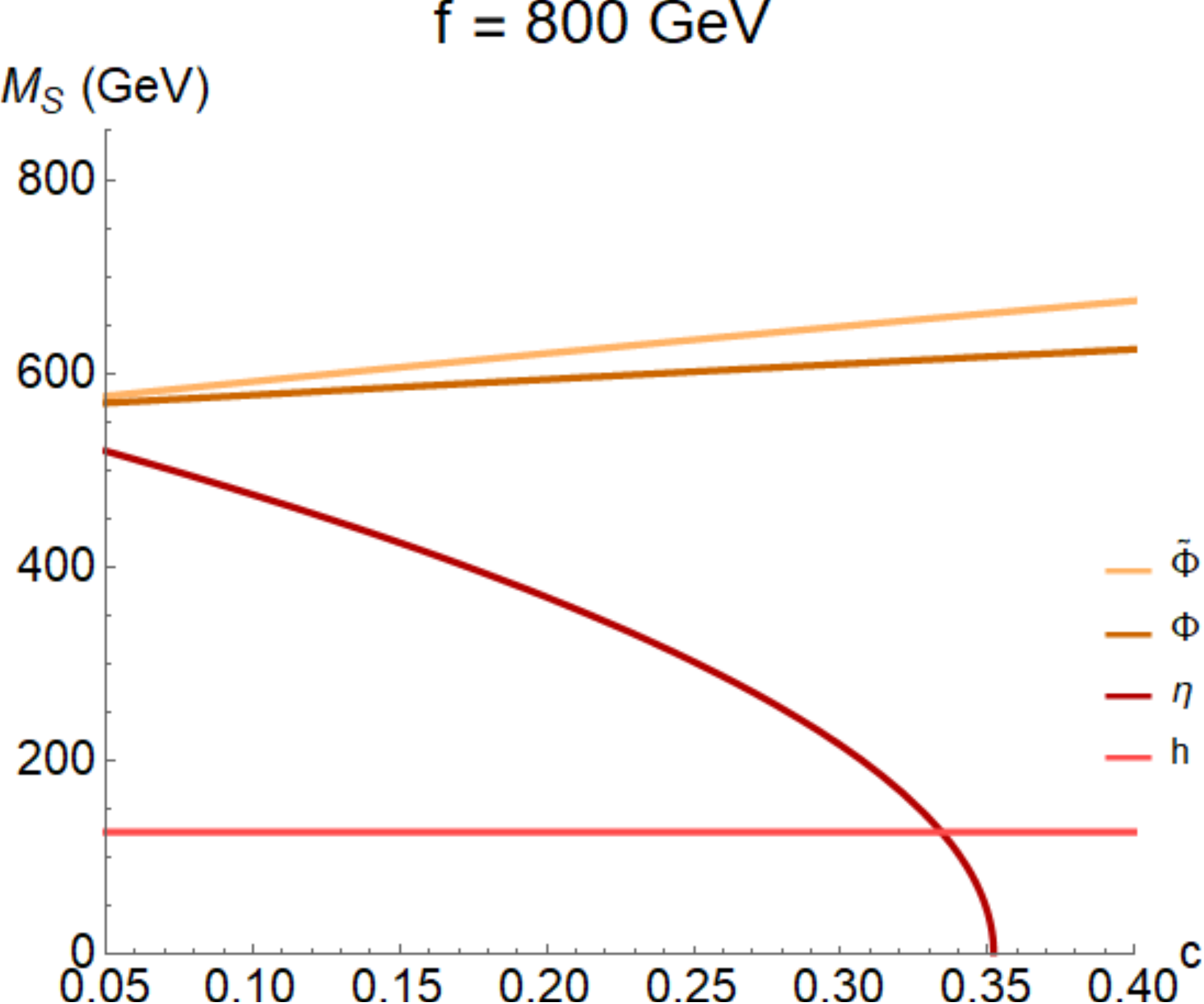}
\end{minipage}
\hspace{0.25cm}
\begin{minipage}{.31\textwidth}
\includegraphics[trim={0cm 0cm 0cm 0cm},clip,width=1.0\linewidth]{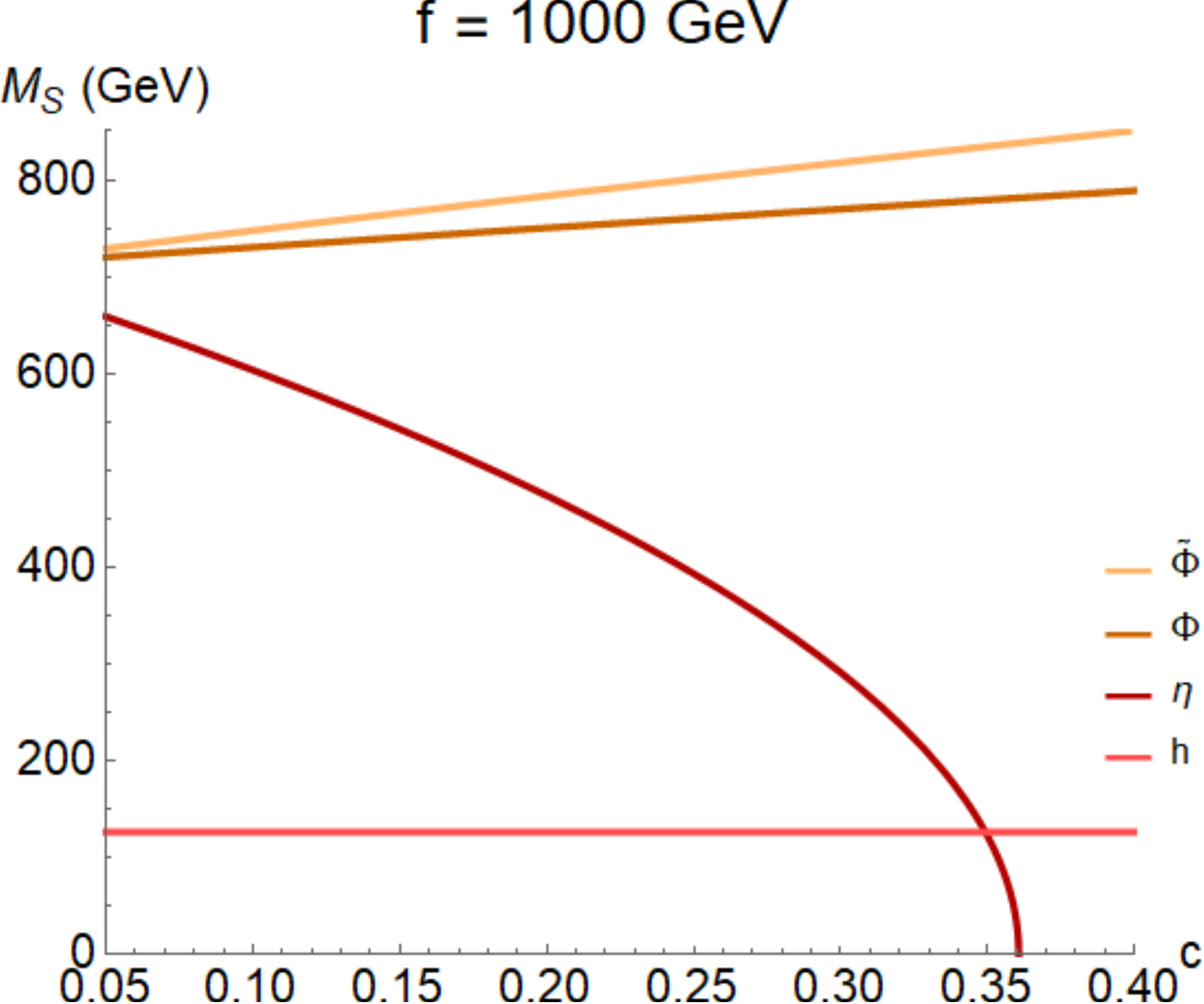}
\end{minipage}
\caption{Mass spectrum for the goldstone scalars as a function of the renormalization constant $c$ for $f=600$ GeV (left), $f=800$ GeV (middle), and $f=1000$ GeV  (right)}
\label{fig:llsmasses}
\end{figure}

The mass of the gauge-singlet $\eta$ is simply proportional to the coefficient of the spurion operator $m_0$, which is a function of the Higgs mass $m_h$ and the gauge renormalization coefficient $c$. The mass splitting between the Georgi-Machacek scalars $\tcapphi / \Phi$ and the gauge-singlet $\eta$ thus grows as a function of $c$, with $\eta$ developing a vacuum expectation value at $c \sim 0.35$. These features of the scalar mass spectrum are shown in Figure (\ref{fig:llsmasses}).

\subsection{Phenomenology}
\label{sec:LLexp}

\begin{figure}
\begin{minipage}{.45\textwidth}
\includegraphics[trim={0cm 0cm 0cm 0cm},clip,width=1.0\linewidth]{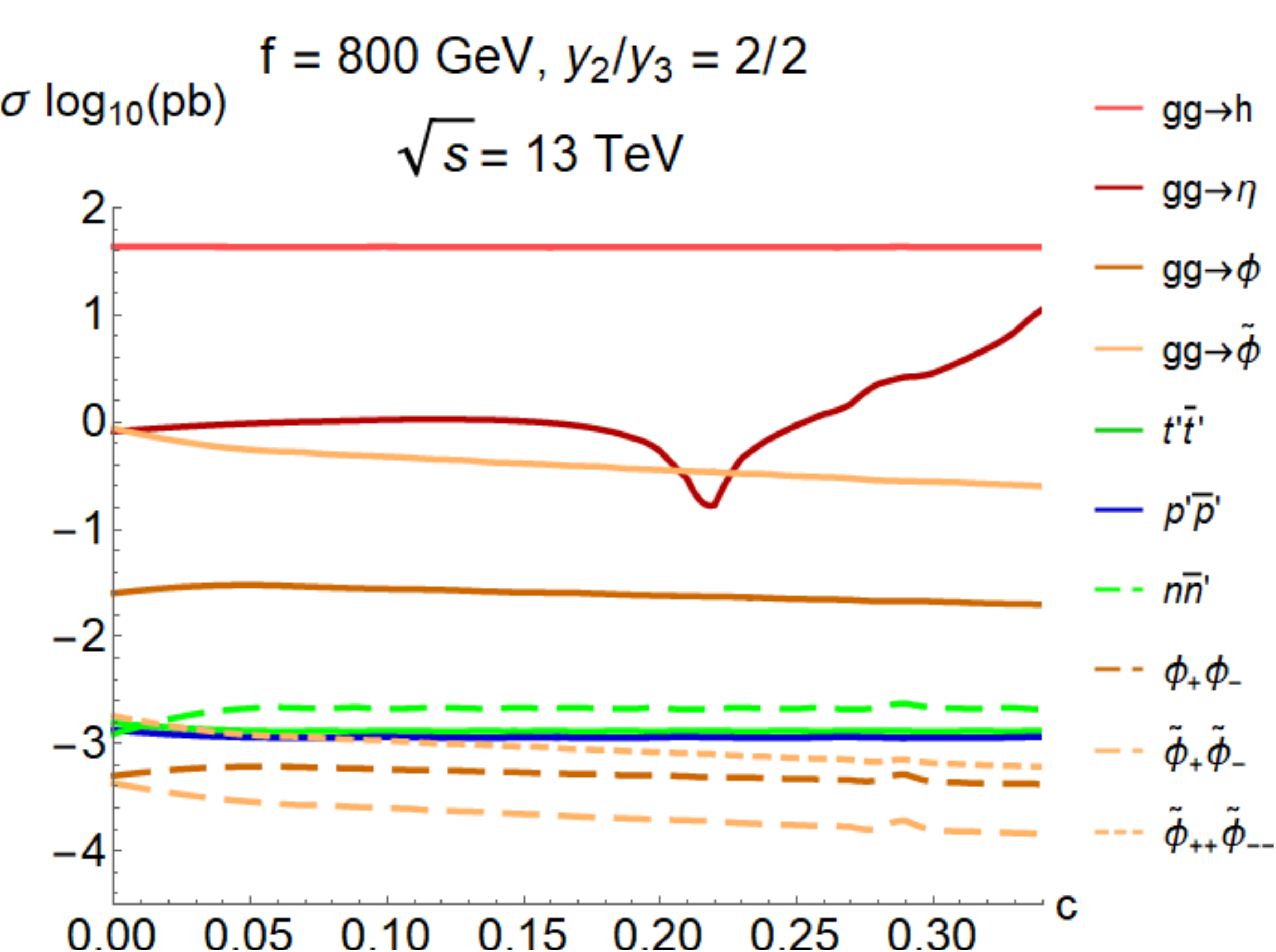}
\end{minipage}
\hspace{1cm}
\begin{minipage}{.45\textwidth}
\includegraphics[trim={0cm 0cm 0cm 0cm},clip,width=1.0\linewidth]{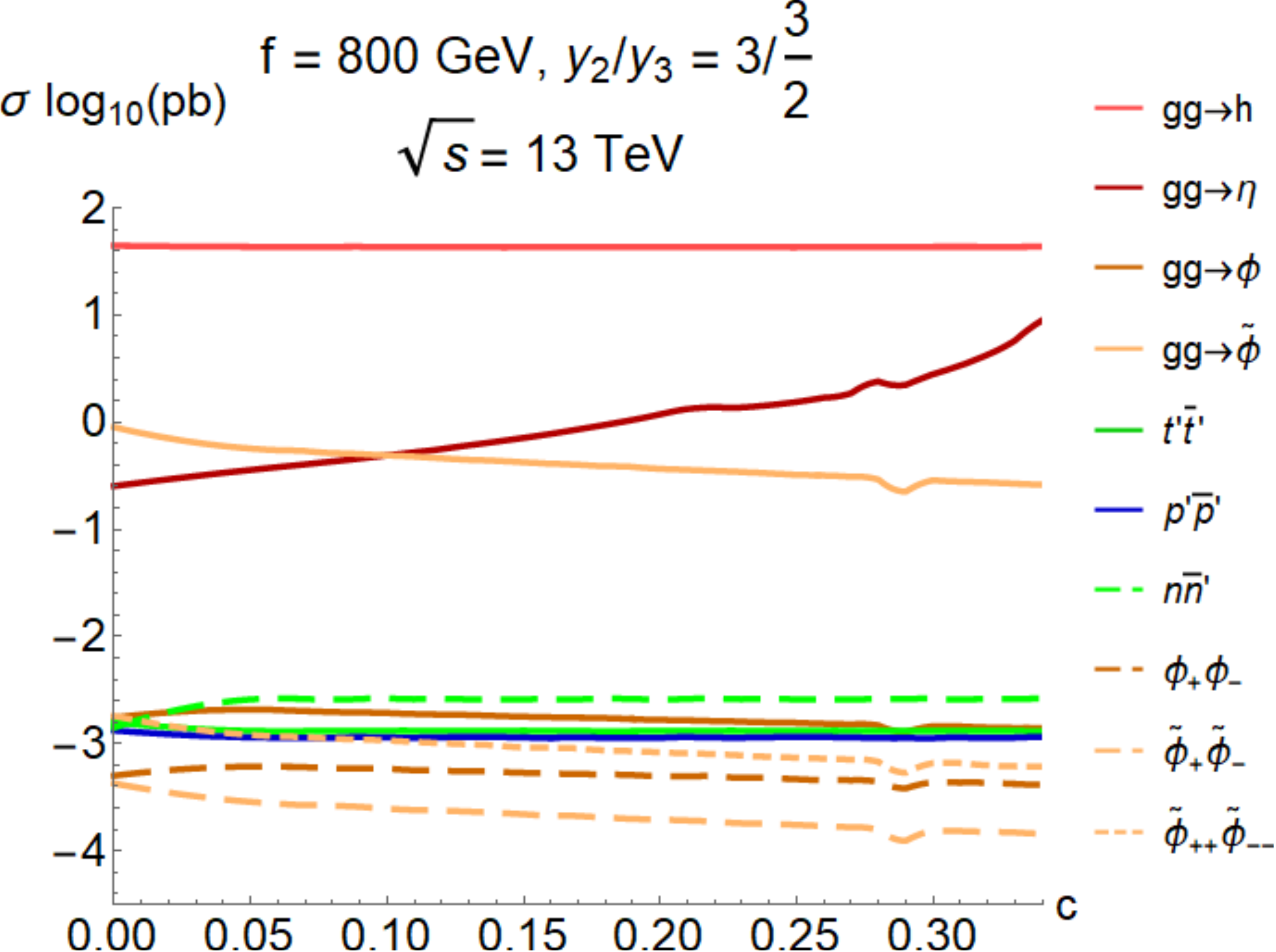}
\end{minipage}
\caption{$\sqrt{s}=13$ TeV cross sections for new particle states in their dominant production modes at $f = 800$ GeV. Cross sections are shown as a function of the gauge renormalization coefficient $c$ at two benchmark points corresponding to $y_2=y_3=2$ (left) and $y_2=3$, $y_3=3/2$ (right).}
\label{fig:sigma}
\end{figure}

In this $SU(5)/SO(5)$ model the spectrum of new states contains strongly interacting quarks with masses that scale as $M_T \sim y_i f$, hidden-sector fermions with masses that scale as $M_N \sim \tilde{y}_i f$, and weakly interacting scalars with masses that scale as $M_S \sim g f$. At $\sqrt{s}=13$ TeV the neutral spin-0 states are singly produced via gluon fusion with $\mathcal{O}( {\rm pb} )$ cross sections, and there is also zoo of new particle states that are pair produced with $\mathcal{O}( {\rm fb} )$ cross sections. These include Drell-Yan production of the charged goldstone scalars, strong production of the lightest quark partners $\tp \tpbar / \pb \pbbar$, as well as weak production of the lightest hidden-fermions $n \bar{n}$. The lightest hidden-sector states will be a spectrum of glueballs which could be long-lived, and produced via $n \bar{n}$ annihilation or through Higgs decay. The direct production cross section for these states at two benchmark points are shown in Figure \ref{fig:sigma}.

Despite the abundance of new particle states in this theory, their production cross sections and dominant decay paths would make them extremely difficult to resolve in current LHC data sets. Drell-Yan production of the doubly charged scalar $\tphi_{++}$ is appreciable in large parts of the parameter space due to the $Q_{++}^2$ enhancement to its cross section relative to the other charged states. However, existing limits on this process typically assume an $\mathcal{O}(1)$ branching fraction to pairs of same sign leptons \cite{ATLAS:2014kca,CMS:2017pet} rendering them inapplicable to the goldstone scalars in this model, which are leptophobic by assumption. Pairs of the lightest top quark partner $\tp \tpbar$ and peculiar quark $\pb \pbbar$ can also be strongly produced with cross sections as high as $\sim 5$ fb in the viable regions of parameter space. The strongest limits on vector-like quarks are set by the CMS search at $\sqrt{s}=$13 TeV in the single lepton channel~\cite{Khachatryan:2015gza,Sirunyan:2017usq}. These limits are driven by kinematically optimized searches for $t \bar{t} + X$ and $b \bar{b} + X$ where $X = W,Z,h$ is highly boosted, and are interpreted in the context of simplified models that assume $\BR(t^{\pr}\shortrightarrow Z\,t) + \BR(t^{\pr}\shortrightarrow h\,t) + \BR(t^{\pr}\shortrightarrow W\,b) = 1$. In this model the lowest lying vector-like partners are composed mostly of the hypercharge-$7/6$ doublet $( P_T, P_B )$ and their branching fractions are dominated by decays to top and bottom quarks in association with the scalar states in $\tcapphi$. These generic experimental signatures are thus qualitatively different from those of simplified models of top and bottom partners, rendering existing limits on vector-like quarks largely inapplicable. 

\subsubsection{Resonant $t \bar{t}$ Final States}

Gluon fusion of neutral pseudo-scalars $\phi_0$, $\eta$ and the complex scalar $\tphi_0$ generically leads to a resonant $t \bar{t}$ final state. The CP properties of the pNGB matrix forbids the two-body decays of the goldstone scalars into pairs of gauge and Higgs bosons and their branching fractions are thus dominated by decays to the third generation quarks whenever these channels are kinematically open. In the bulk of this parameter space we find $\BR (\tphi_0 / \phi_0 / \eta \shortrightarrow t \bar{t} ) \sim \BR (\tphi_+ / \phi_+ \shortrightarrow t \bar{b} )  \sim 1$, although the gauge singlet $\eta$ can also have a significant branching fraction to $g g$ and $b \bar{b}$ when its mass is below the top threshold as computed in Section \ref{sec:SUSPpheno}. Currently the strongest limit on the process $g g \rightarrow \tphi_0 / \phi_0 \rightarrow t \bar{t}$ comes from the ATLAS search at $\sqrt{s}=8$ TeV in the single lepton channel~\cite{Aaboud:2017hnm}. The gluon fusion process is mediated by loops of heavy vector-like quarks. Their effects may thus be parameterized by a set of dimension-5 couplings $c^{S}_{gg}$, which can be expressed as a sum over the well known gluon vertex function for the fermion triangle graph $V^S_{gg}$

\begin{figure}
\begin{minipage}{.97\textwidth}
\includegraphics[trim={0cm 0cm 0cm 0cm},clip,width=1.0\linewidth]{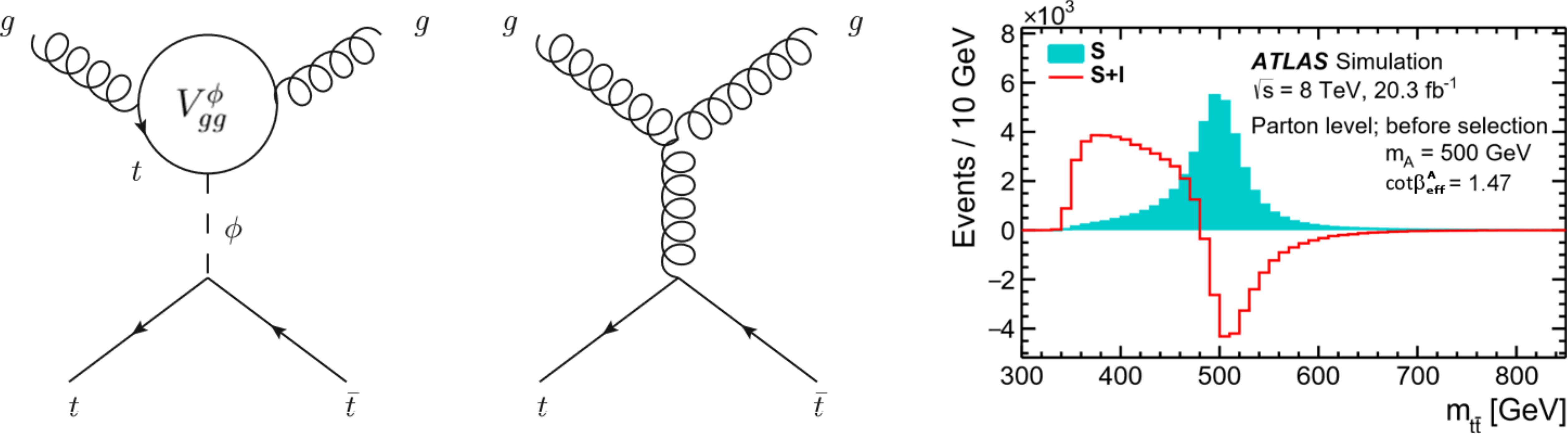}
\end{minipage}
\caption{Interfering diagrams for $t \bar{t}$ production via spin-0 and spin-1 mediators (left). Invariant mass distributions for the $t \bar{t}$ pairs (right). The contribution from the spin-0 mediator (S) is shown in blue and the combined contribution (S+I) is shown in red.}
\label{diag:ttbar}
\end{figure}

\be
\mathcal{L} \supset \displaystyle\sum_{S \in {\tphi, \phi, \eta}} - \frac{1}{4} c^{S}_{gg} \, S \, G^{\mu \nu} \widetilde{G}_{\mu \nu}
\ee
\be
c^S_{gg} = \displaystyle\sum_{\tppp \tpp t} V^S_{gg} (i) \equiv \frac{\alpha_S}{\sqrt{8}\pi} \displaystyle\sum_{\tppp \tpp t} y^S_{i i} \, \frac{\tau_i f (\tau_i)}{m_i}\\[2ex]
\ee

\noindent Searches for scalar resonances decaying to $t \bar{t}$ pairs is complicated by destructive interference between the spin-0 and spin-1 mediated s-channel diagrams in Figure \ref{diag:ttbar}. This results in a ``peak-dip'' structure for the invariant mass distribution of the $t \bar{t}$ pairs rather than the usual Breit-Wigner resonance. The analysis performed by ATLAS searches for this kinematic feature in resolved $t \bar{t}$ pairs at invariant masses in the range of $m_{t \bar{t}} = 500-800$ GeV\footnote{Invariant masses below $500$ GeV are not considered due to the importance of higher order corrections involving Higgs decays to virtual top quarks in that regime.}. The results are interpreted in the context of a Two-Higgs-Doublet model (2HDM) and the cuts are kinematically optimized for the case of a scalar, a pseudo-scalar, and the case of a degenerate scalar-pseudoscalar pair. The limits are reported as a function of the 2HDM parameter $\tan \beta$ in the alignment limit where the $W/Z$ couplings are equal to their SM values. In this limit, $\tan \beta$ can be interpreted as the ratio of the top quark Yukawa coupling to the SM Higgs and heavy Higgs bosons, $\tan \beta \rightarrow y^{H}_{tt} / y^{\widetilde{H}}_{tt}$. Here $H$ and $\widetilde{H}$ refer to the $SU(2)_L$ doublets containing the SM and heavy Higgses respectively. 

The strength of this kinematic feature is a function of ratios of the amplitudes for the two processes, which include factors involving the vertex function $V^{H}_{gg}$. For each scalar mass eigenstate $S$ we can thus define an analagous angle $\beta^S_{\rm eff}$ by replacing the vertex function in the numerator with its analogous dimension-5 coupling as shown in Equation \ref{eq:cotbeta}. If the vector-like quark masses are heavy and decoupled, the contribution to the gluon fusion vertex function is dominated by the top quark loop. In this regime the $\mathcal{O}(1)$ top-quark Yukawa coupling implies the approximate relation $\cot \beta^S_{\rm eff} \sim y_{tt}^S / y_{tt}^H$.

\begin{figure}
\begin{minipage}{.4\textwidth}
\includegraphics[trim={0cm 0cm 0cm 0cm},clip,width=1.0\linewidth]{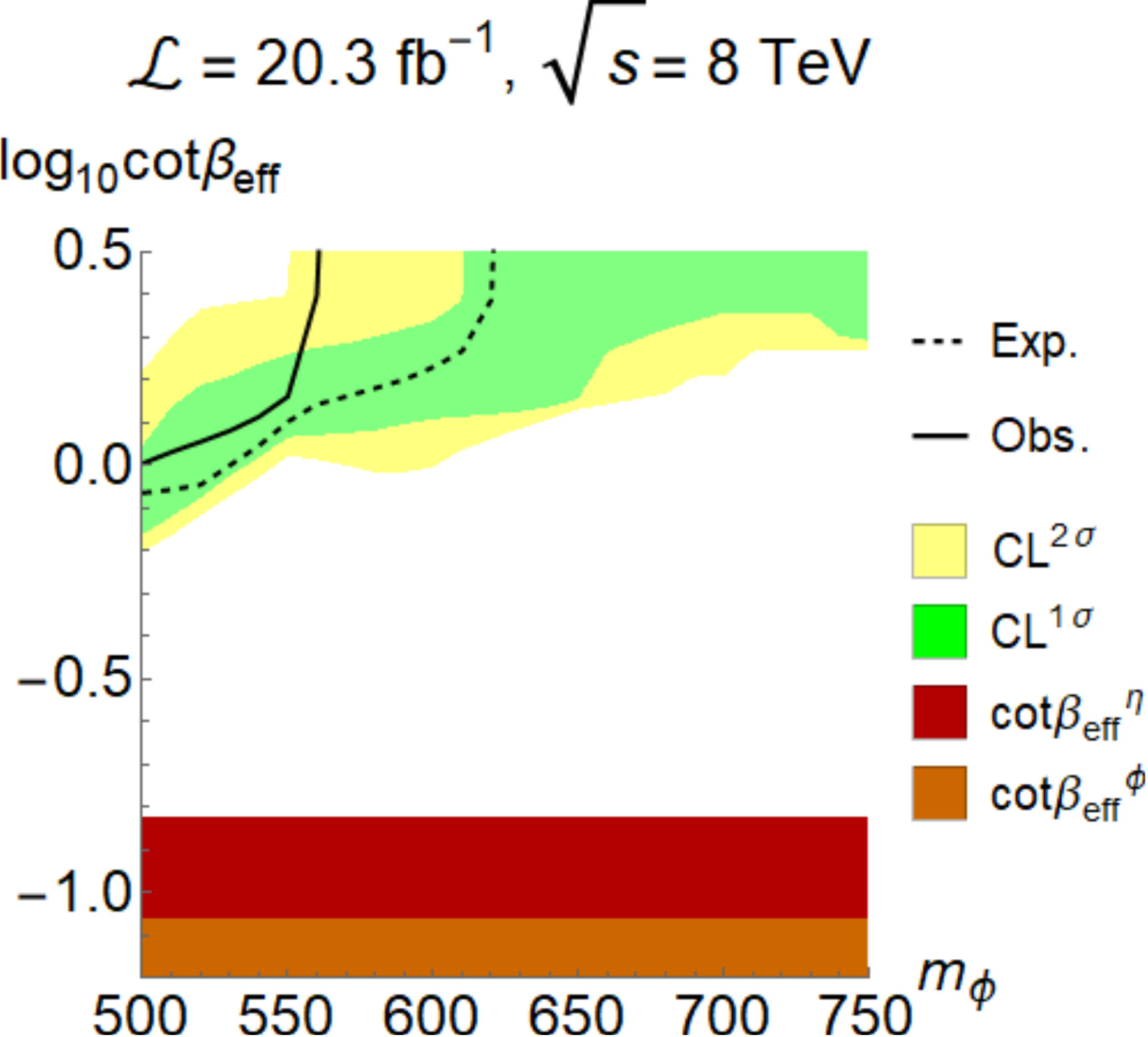}
\end{minipage}
\hspace{1cm}
\begin{minipage}{.4\textwidth}
\includegraphics[trim={0cm 0cm 0cm 0cm},clip,width=1.0\linewidth]{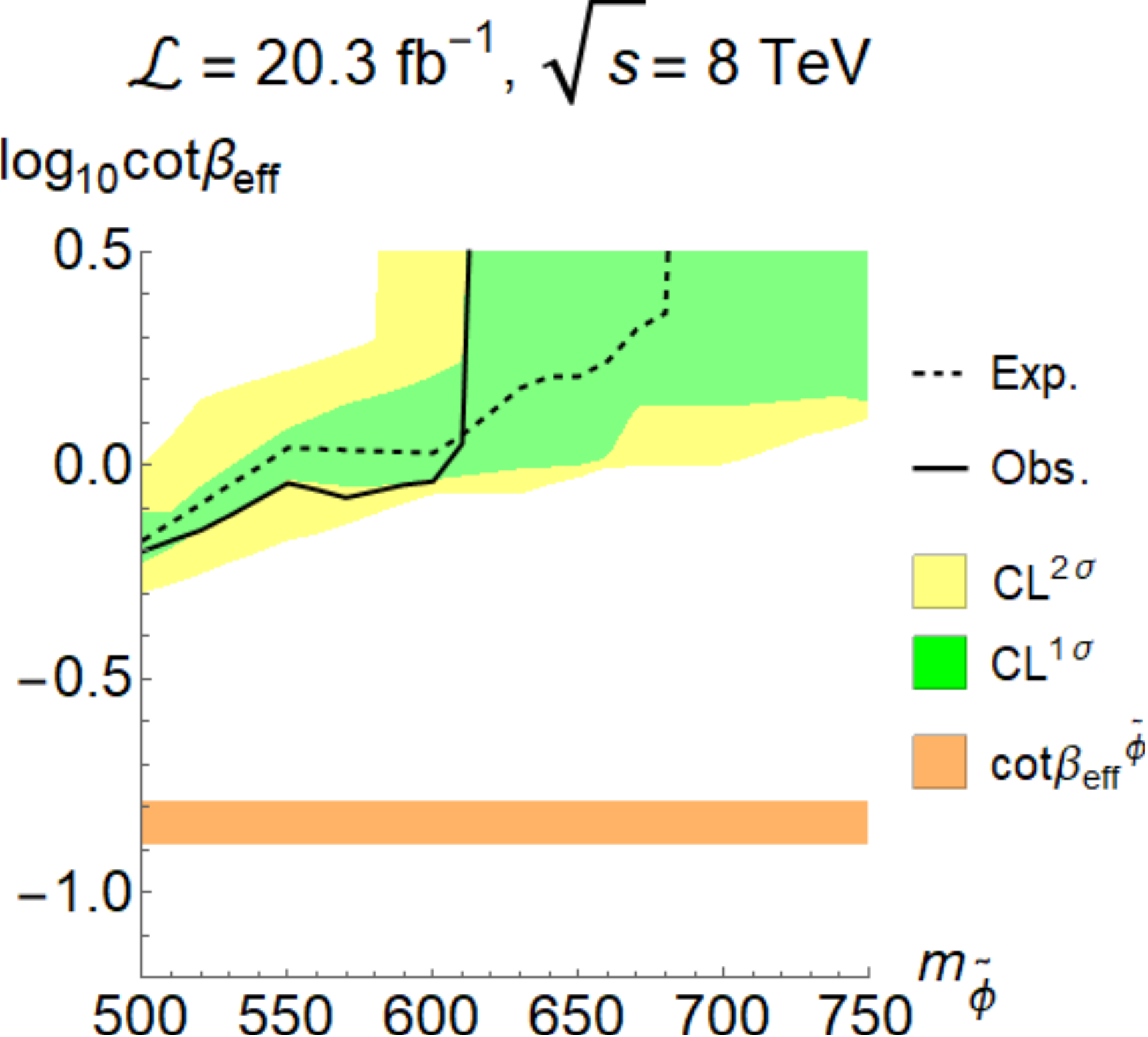}
\end{minipage}
\caption{Expected and observed limits from CMS searches for $t \bar{t}$ resonances mediated by a pseudo-scalar (left) and a degenerate scalar/pseudo-scalar (right), as a function of $\cot \beta_{\rm eff}$. Maximum values for $\cot \beta_{\rm eff}$ are also shown.}
\label{fig:yeff}
\end{figure}

\begin{align}
\cot^{2} \beta &= \bigg( \frac{y^{\widetilde{H}}_{t t}}{y^{H}_{t t}} \bigg)^2 = \frac{y^{\widetilde{H}}_{t t} \, V^{\widetilde{H}}_{gg} (t)}{y^{H}_{t t} \, V^{H}_{gg} (t)} \\[1.5ex]
\cot \beta^S_{\rm eff} &\equiv \sqrt{\frac{y^{S}_{t t}}{y^{H}_{t t}} \frac{c^{S}_{gg}}{V^{H}_{gg} (t)}} \xrightarrow{\langle \theta \rangle \rightarrow 0} \frac{y^{S}_{t t}}{y^{H}_{t t}}
\label{eq:cotbeta}
\end{align}

\noindent The ATLAS limits are strongest in the mass range of $\mathcal{O}(500 \, {\rm GeV})$, but are insensitive to new scalars with Yukawa couplings to the top quark that are below $\mathcal{O}(1)$. These limits have been reparameterized as a function of $\cot \beta^S_{\rm eff}$ in Figure \ref{fig:yeff}, along with the translated $2 \sigma$ ATLAS limits at $\sqrt{s}= 8$ TeV. The masses of the Georgi-Machacek scalars $\phi_0$ and $\tphi_0$ and the gauge-singlet $\eta$ can be varied as a function of the spurion coefficient $m_0$ throughout the mass range of interest, and are roughly independent of the magnitude of their Yukawa couplings. Throughout the parameter space of this model we find that $\cot \beta^{\phi_0}_{\rm eff} < \cot \beta^{\tphi_0}_{\rm eff} \ll 1$. The ability to resolve these neutral scalars are thus well beyond the reach of near-term LHC limits in this channel. 

\subsubsection{Multi-top Final States}

\begin{figure}
\begin{minipage}{.97\textwidth}
\includegraphics[trim={0cm 0cm 0cm 0cm},clip,width=1.0\linewidth]{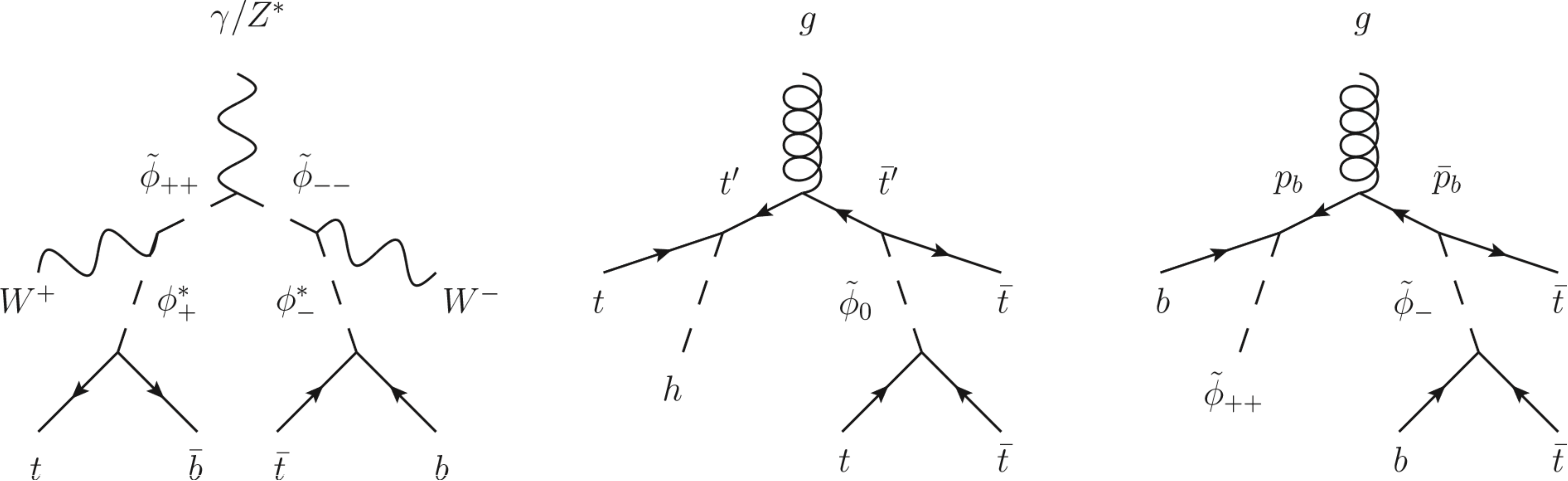}
\end{minipage}
\caption{Dominant contributions to the final state $(Wb)^n$ for $n \geq 4$. These include Drell-Yan production of charged scalars (left) as well as strong production of the lightest top partner (middle) and peculiar quark (right).}
\label{diag:multi}
\end{figure}

Pair production of doubly-charged scalars and heavy vector-like quarks democratically provide an $\mathcal{O}({\rm fb})$ contribution to $\sigma(pp \shortrightarrow t \bar{t} t \bar{t})$ via final states that include $(Wb)^n$ with $n \geq 4$. Despite the kinematic suppression of these production modes, the abundance of new particle states can lead to a measurable multi-top signature when combined over all processes. An anomalous production of final states with a high multiplicity top and bottom quarks is thus a general expectation for this class of theories. The most sensitive searches for these final states are driven by multi-lepton analyses that look for anomalously high b-jet multiplicities within samples containing three or more leptons, or two leptons of the same sign \cite{Sirunyan:2017roi}. The dominant contributions to the multi-top final state come from diagrams such as those in Figure \ref{diag:multi}.

In this model Drell-Yan production of the charged scalars are mediated by current interactions that can be extracted from the covariant derivative as shown in Equation \ref{eq:llcurrent}. The form of these couplings agree with those computed in the low energy effective theory for Georgi-Machacek scalars \cite{Hartling:2014zca} in the limit where only the Higgs gets a vacuum expectation value.  

\begin{figure}
\begin{minipage}{.45\textwidth}
\includegraphics[trim={0cm 0cm 0cm 0cm},clip,width=1.0\linewidth]{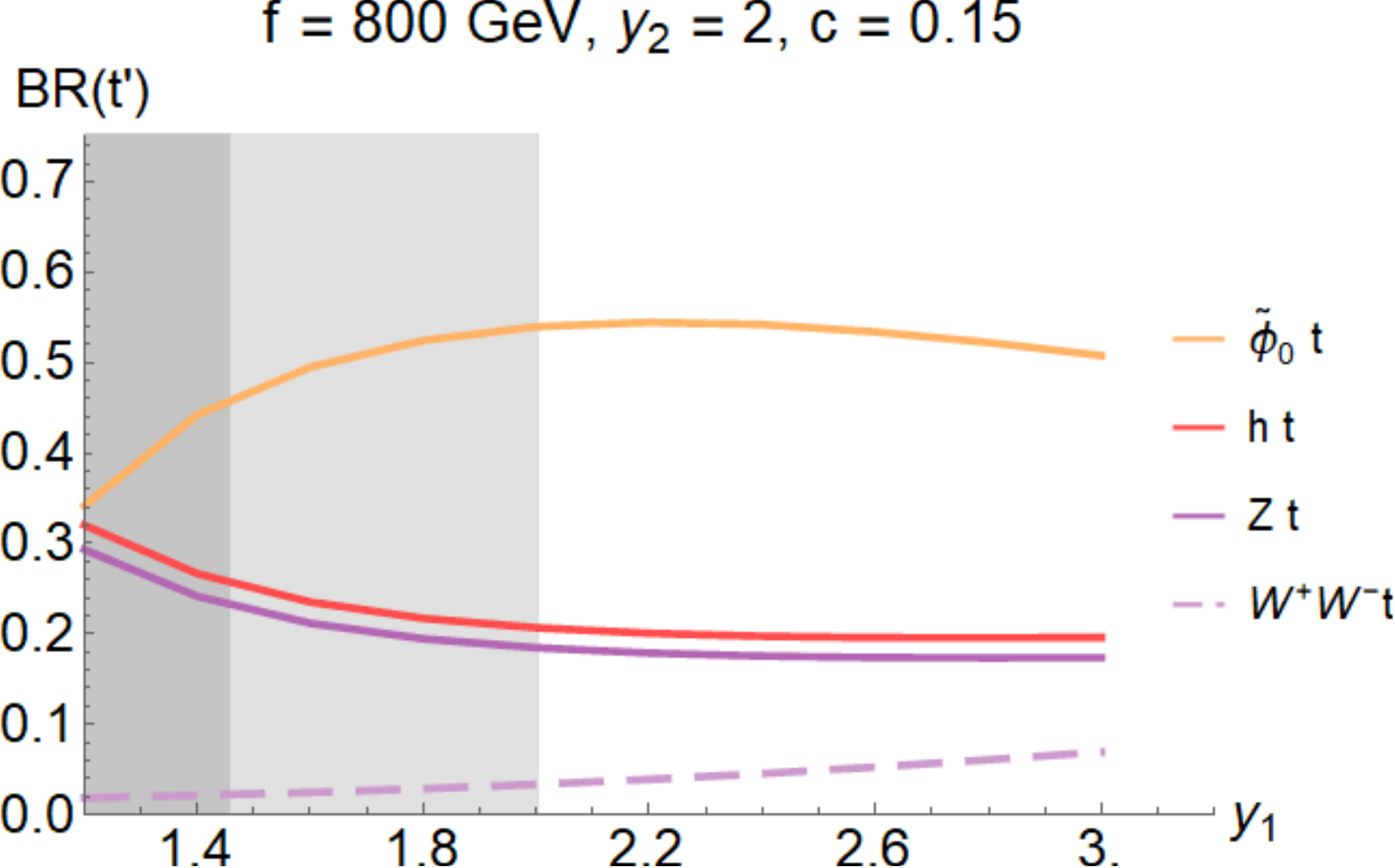}
\end{minipage}
\hspace{1cm}
\begin{minipage}{.45\textwidth}
\includegraphics[trim={0cm 0cm 0cm 0cm},clip,width=1.0\linewidth]{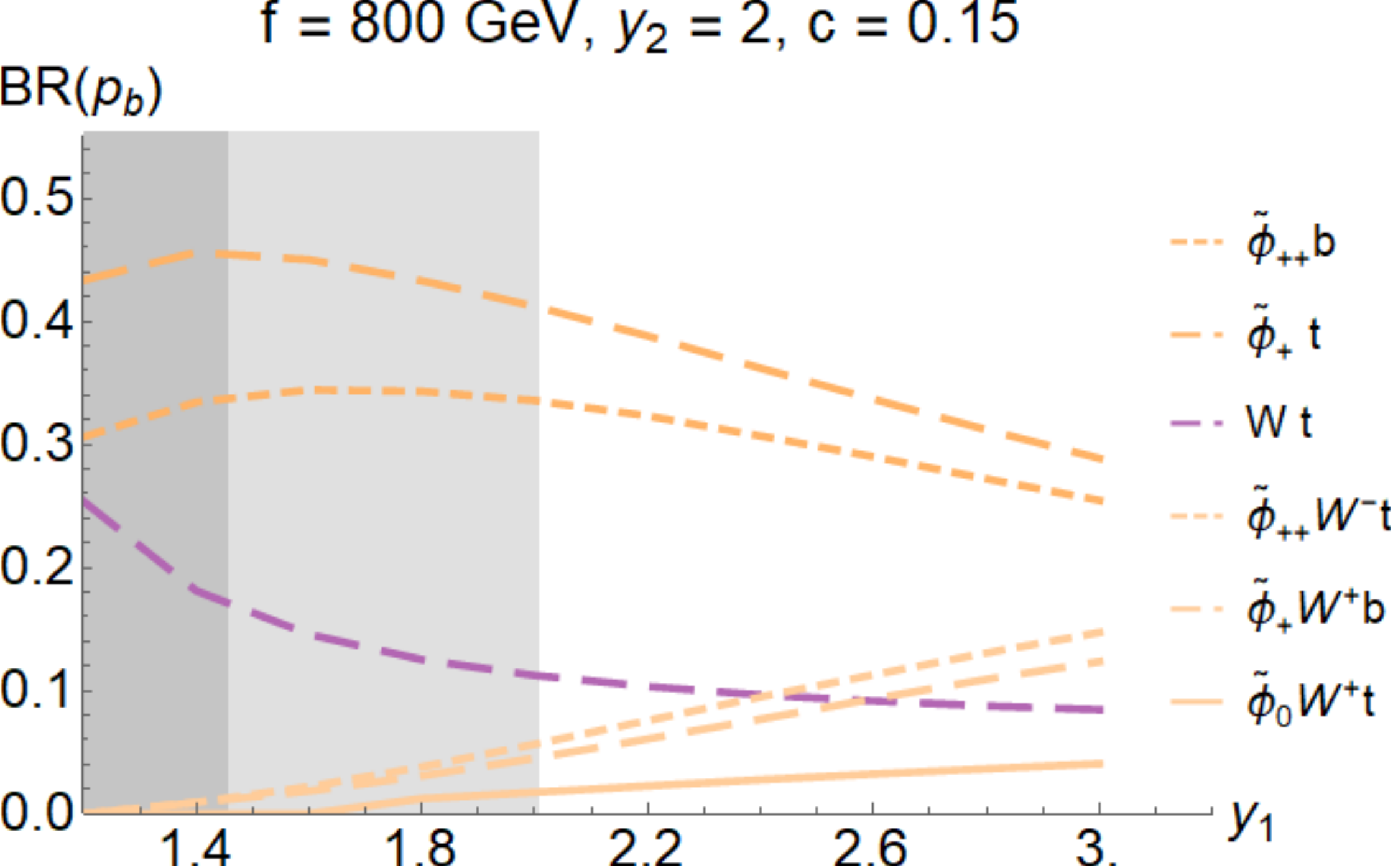}
\end{minipage}
\caption{Branching fractions of the lightest top-quark partner (left) and the peculiar bottom-quark partner (right) as a function of $y_1$ at a benchmark points corresponding to $f=800$ GeV and $y_2=2$}
\label{fig:llBRQ}
\end{figure}

\begin{align}
J_{\rm EM}^\mu &= g s_w \left( \phi_+ \fbderiv^\mu \phi_- + \tphi_+ \fbderiv^\mu \tphi_-  + 2 \, \tphi_{++} \fbderiv^\mu \tphi_{--} + \tphi_0^\dagger \fbderiv^\mu \tphi_0  \right) \\[2ex]
J_{\rm Z}^\mu &= \frac{g}{2 c_w} \bigg[ \left( 1 - 2 s_w^2 + c_{\langle \theta \rangle}^2 \right) \phi_+ \fbderiv^\mu \phi_- + \left( 1 - 2 s_w^2 - c_{\langle \theta \rangle}^2 \right) \tphi_+ \fbderiv^\mu \tphi_- \\
& \hspace{5cm} + 2 \, (1 - 2 s_w^2) \, \tphi_{++} \fbderiv^\mu \tphi_{--} + c_{\langle \theta \rangle}^2 \, \tphi_0^\dagger \fbderiv^\mu \tphi_0 \bigg] \\[2ex]
J_{\rm +}^\mu &= g \bigg[ c_{\langle \theta \rangle}^2 \, \left( \tphi_{++} \fbderiv^\mu \tphi_{-} + \tphi_+ \fbderiv^\mu \tphi_0 - \phi_+ \fbderiv^\mu \phi_0 \right)    \\
  & \hspace{4.5cm} + s_{\langle \theta \rangle}^2 \, \left( \tphi_{++} \fbderiv^\mu \phi_{-} - \tphi_{+} \fbderiv^\mu \phi_{0} + \tphi_{0} \fbderiv^\mu \phi_{+} \right) \bigg]
\label{eq:llcurrent}
\end{align}

\begin{figure}
\begin{minipage}{.45\textwidth}
\includegraphics[trim={0cm 0cm 0cm 0cm},clip,width=1.0\linewidth]{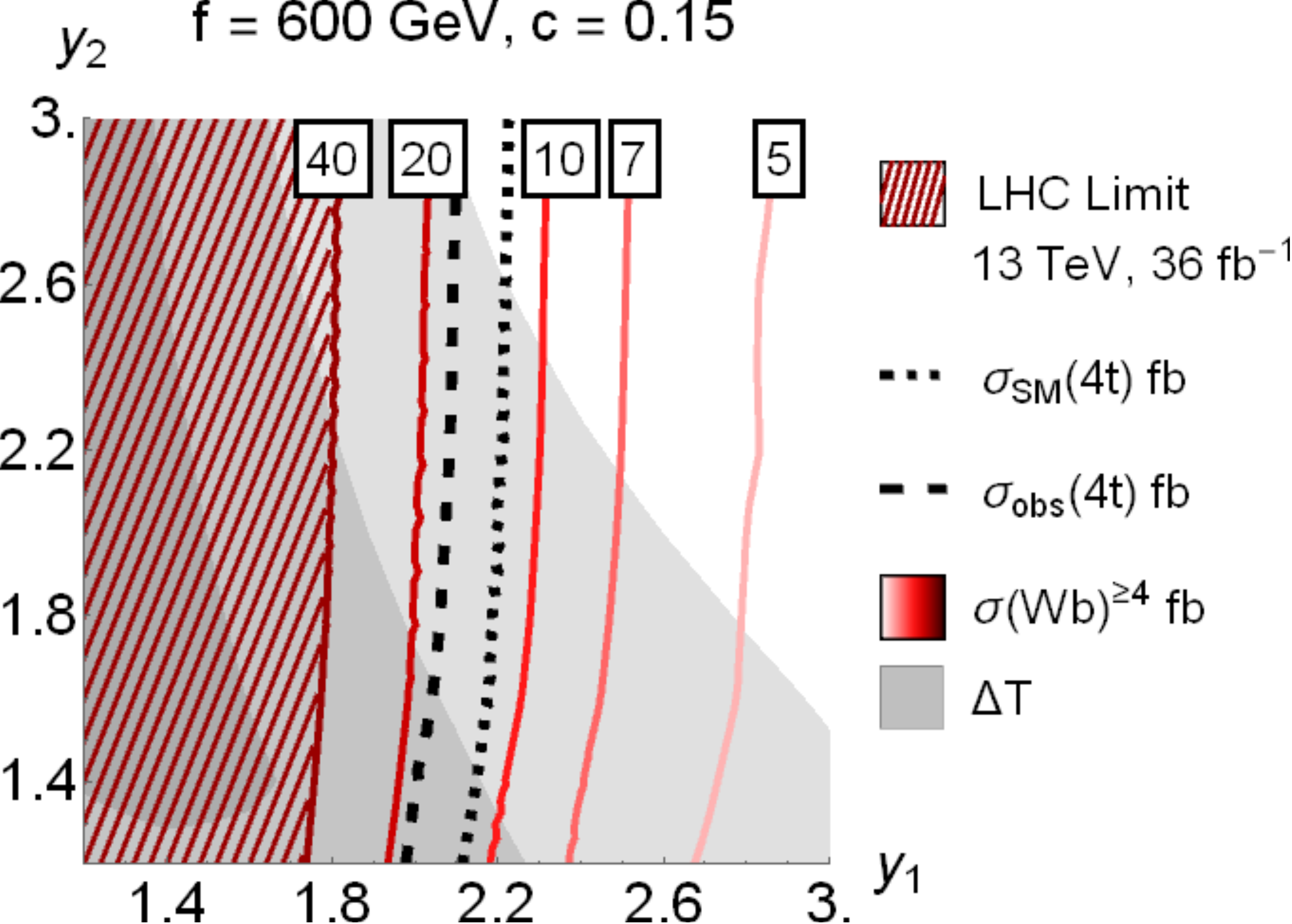}
\end{minipage}
\hspace{1cm}
\begin{minipage}{.45\textwidth}
\includegraphics[trim={0cm 0cm 0cm 0cm},clip,width=1.0\linewidth]{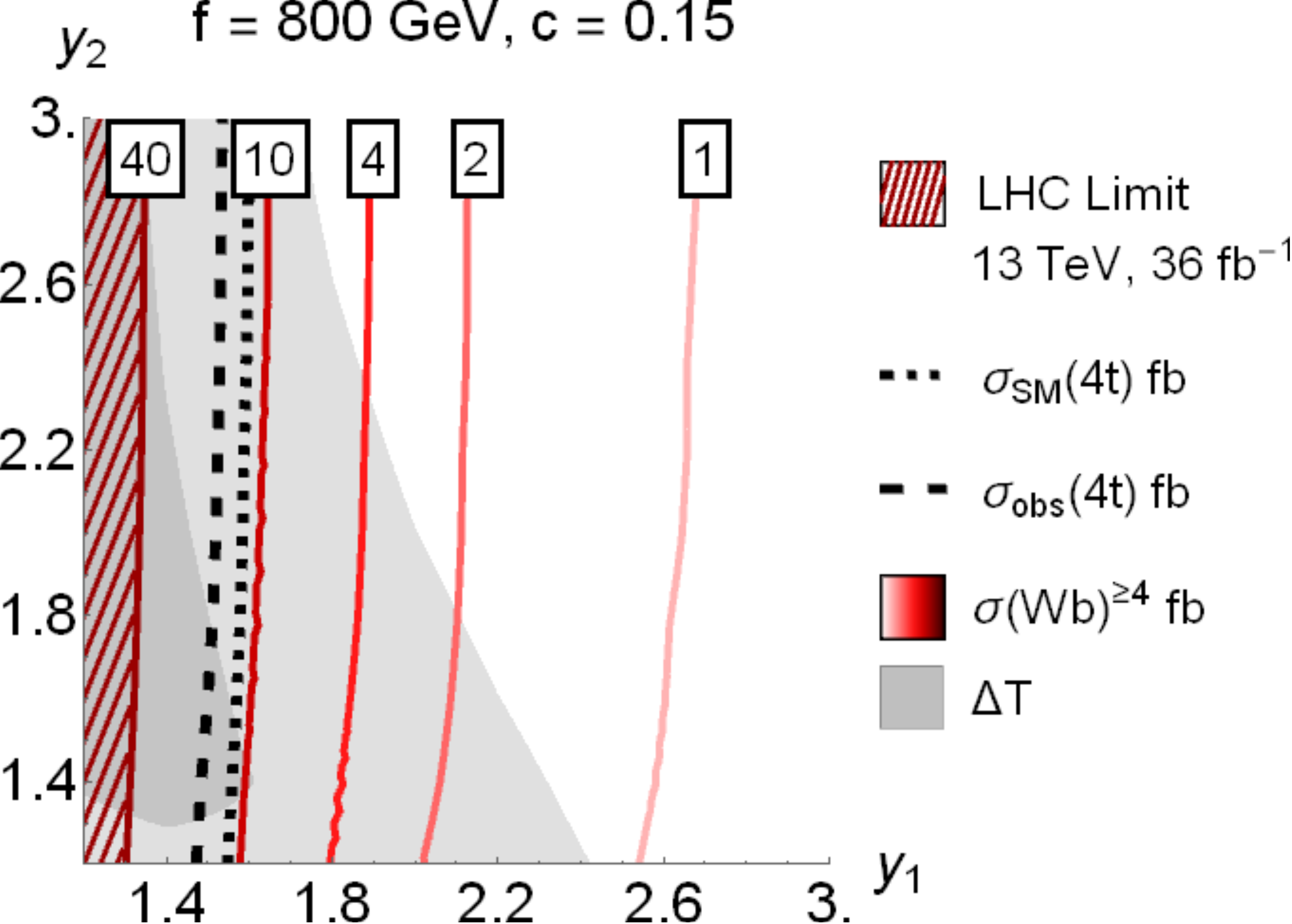}
\end{minipage}
\caption{Total cross section (fb) for production of final state $(Wb)^n$ with $n \geq 4$ at $f = 600$ GeV (left) and $f = 800$ GeV (right). Limits from multi-top searches shown in red.}
\label{fig:4top}
\end{figure}

\noindent These current interactions also mediate transitions between the different scalar mass eigenstates via ``weak-strahling" processes. However these branching fractions are sub-leading relative to their decays to pairs of fermions via $\mathcal{O}(1)$ Yukawa couplings. The mass, charge, and CP properties of the $\tphi_{++}$ scalar in this model forbid all two-body decays and its branching fraction is thus dominated by the 3-body decay ${\rm BR} (\tphi_{++} \shortrightarrow W^+ t \bar{b}) \sim 1$. Pair production of this doubly charged scalar will thus always result in the final state $\tphi_{++} \tphi_{--} \rightarrow (W b)^4$, effectively providing an $\mathcal{O}({\rm fb})$ contribution to the four top-quark cross section from the process shown in Figure \ref{diag:multi}. The other major contribution to an anomalous multi-top cross section comes from strong production of pairs of heavy vector-like quarks, which are the lightest top quark partner $\tp \tpbar$ and peculiar quark $\pb \pbbar$. The lightest top partner $\tp$ decays mostly via the neutral current $\tp \rightarrow \tphi_0 t$ but there is also a sub-leading branching fraction to the top quark via emission of $Z$ and $h$ bosons. These two sub-leading contributions are approximately equal due to the goldstone equivalence theorem, as shown in Figure \ref{fig:llBRQ}. Flavor-changing charged-current decays of $\tp$ are suppressed due to the fact that it mostly an electroweak doublet implying that $\BR (\tp\shortrightarrow W^+ b) \sim 0$. The lightest bottom-type quark partner $\pb$ has a peculiar charge of $5/3$ and therefore decays exclusively via emission of charged bosons. Pair production of these heavy color-charged fermions thus generically results in final states involving pairs of $\tcapphi$ scalars and additional third generation quarks. At large values of the $SU(5)$ symmetric Yukawa coupling $y_1$, the mass splitting between the new fermion and scalar states becomes large enough to attenuate the kinematic advantage of two-body final states. In this regime we find an increasingly significant contribution from 3-body decays, which can result in exotic signatures containing a very high multiplicity of leptons and b-jets. In this model the contribution to the multi-top cross section from any one production mode is generally below threshold for detection by current experimental searches. However we find that combining the effective contribution to this final state across multiple channels makes these searches sensitive to large parts of the parameter space. The various combinations of decay paths that contribute to the effective $t \bar{t} t \bar{t}$ cross section are enumerated in Appendix \ref{app:fourtop}. The aggregate contribution from all of these channels allows for a direct translation of the limit on anomalous $4$-top production to the parameter space of this theory, as illustrated in Figure \ref{fig:4top}.

\subsubsection{Hidden-Sector Final States}
\label{sssec:llhidden}

\begin{figure}
\begin{minipage}{.5\textwidth}
\includegraphics[trim={0cm 0cm 0cm 0cm},clip,width=1.0\linewidth]{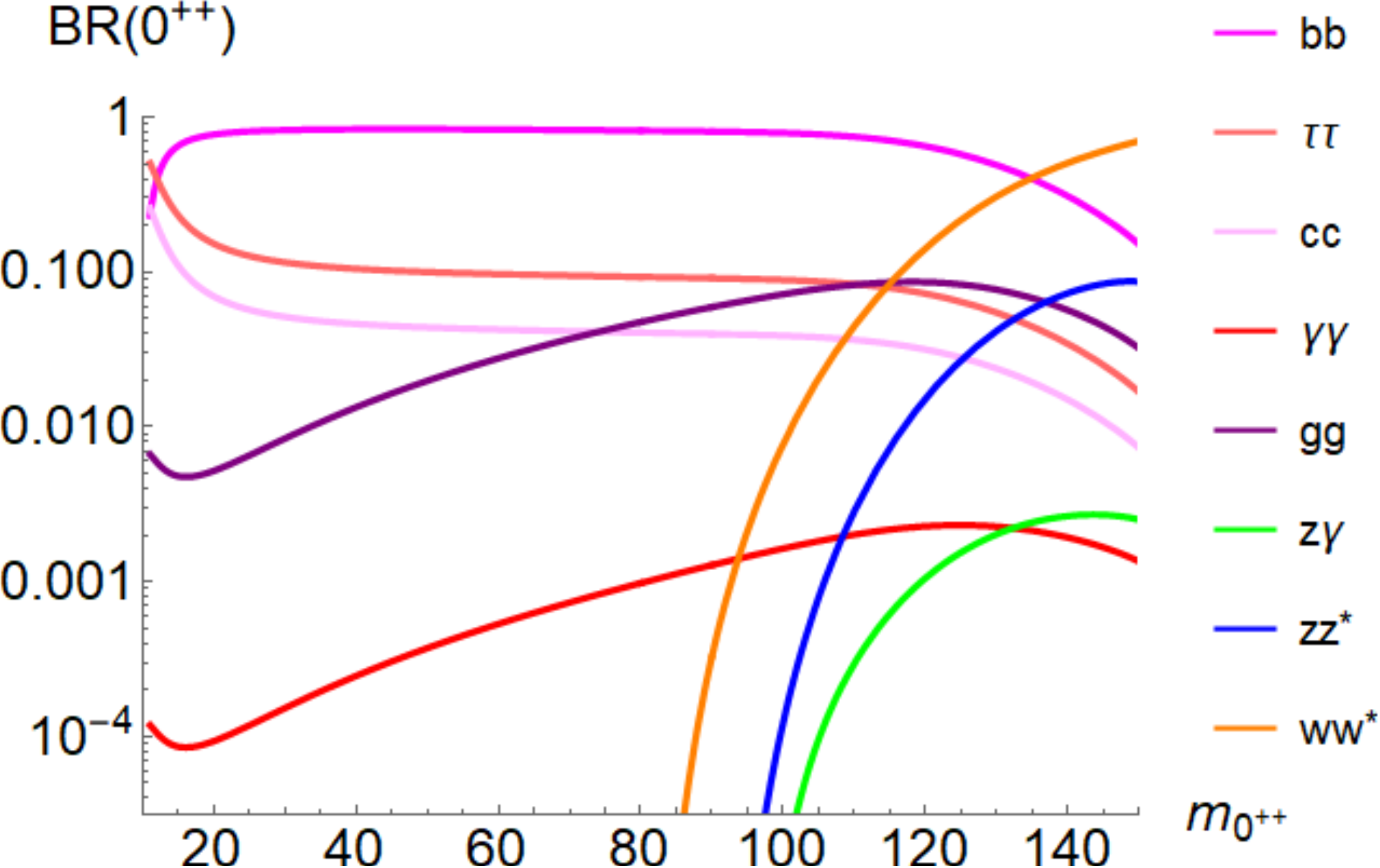}
\end{minipage}
\caption{Branching ratio of the lightest hidden-sector glueball to Standard Model particles as a function of the glueball mass $m_{0^{++}}$.}
\label{fig:llBRGB}
\end{figure}

The lightest states in the hidden-sector are a spectrum of glueballs which may be labeled by their $J^{PC}$ quantum numbers. The masses and widths of these glueballs take a large contribution from the non-perturbative dynamics of the hidden-sector gauge group $G$, thus preventing a detailed quantitative analysis of their collider phenomenology. However naive dimensional analysis implies that the glueball masses should be proportional to the hidden-sector confinement scale $\widetilde{\Lambda}$, and their widths should be proportional to some high power of $\widetilde{\Lambda}$. If the hidden-sector gauge group is $G=SU(3)$, computations on the lattice indicate the existence of at least a dozen stable glueball mass eigenstates~\cite{Morningstar:1999rf,Chen:2005mg,Gockeler:2005rv}. The lightest state has $J^{PC} = 0^{++}$ and a mass $m_{0^{++}} \sim 7 \widetilde{\Lambda}$, while the higher excited states all lie within $\mathcal{O}({\rm few}) \times m_{0^{++}}$. Despite the relative compression of this mass spectrum, if $m_{0^{++}} \gg \widetilde{\Lambda}$ then the production of higher-mass glueballs will be sub-dominant due to a Boltzmann suppression in the thermal partition function. We will thus assume going forward that an $\mathcal{O}(1)$ fraction of all glueballs produced are of the type $0^{++}$. The lowest-lying glueball $0^{++}$ can decay back to Standard Model particles through the Higgs portal and its branching ratios are thus approximately equal to those of a Higgs boson of the same mass. The widths of low-mass Higgs bosons are well known \cite{Gunion:1989we} and are illustrated in Figure \ref{fig:llBRGB}. The parametric form of the glueball width has been worked in out in \cite{Juknevich:2009gg}, and is given by the expression in Equation \ref{eq:llGBwidth}.

\be
\Gamma (0^{++} \shortrightarrow h^* \shortrightarrow X X) = \left( \frac{\kappa \, v \, m_{0^{++}}^3}{24 \pi^2 f^2 (m_h^2 - m_{0^{++}}^2)} \right)^2 \Gamma (h^* \shortrightarrow X X)
\label{eq:llGBwidth}
\ee

\noindent Here $\kappa$ is an $\mathcal{O}(1)$ constant that parameterizes the non-perturbative contribution to the width, and lattice computations for $G=SU(3)$ give $\kappa \sim 3$ \cite{Chen:2005mg,Meyer:2008tr}. For this analysis we remain agnostic about the details of the hidden-sector gauge group by treating $\kappa$ and $m_{0^{++}}$ as the input quantities that parameterize the relevant glueball properties.

If $m_{0^{++}} > m_h /2$ then the primary production mode of the $0^{++}$ glueball will be through decays of a broad $n \bar{n}$-onium resonance, which has an $\mathcal{O}({\rm fb})$ production cross section in the bulk of this parameter space, often exceeding the pair production rate of the new color-charged states $\tp \tpbar$ and $\pb \pbbar$. Weak production of $n \bar{n}$ pairs is enhanced by an abundance of new production modes via s-channel diagrams involving the new goldstone states, and also by the conditions of EWSB which generically result in a hierarchy between the color-charged and hidden-sector Yukawa couplings $m_n < m_{\tp}$. However in this regime, the hidden-sector states would be extremely difficult to resolve in any near-term LHC search due to the overwhelming QCD backgrounds to its generic final states. For example if the hidden-sector gauge group is QCD-like, then a relatively light glueball mass $m_{0^{++}} \ll m_n \sim 0.5 \, {\rm TeV}$ may result in a $n \bar{n}$-onium state that decays to a high multiplicity of glueballs, the lightest of which would decay promptly with an $\mathcal{O}(1)$ branching fraction to pairs of bottom quarks. This $\mathcal{O}({\rm fb})$ contribution to high-multiplicity bottom-quark final states would be extremely challenging to probe. However given the impressive recent advances on object identification using deep-learning algorithms, it is conceivable that future analysis techniques may become sensitive to the resonance structure of such glueball decays between correlated $b \bar{b}$ pairs. As the lightest glueball mass begins to saturate the kinematic limit $m_{0^{++}} \rightarrow m_n$, the $n \bar{n}$-onium state will decay promptly to pairs of the $0^{++}$ glueball, which has a prompt $\mathcal{O}(1)$ branching fraction to $W^+ W^-$ in this regime. The hidden sector will thus provide an $\mathcal{O}({\rm fb})$ contribution to the cross section $\sigma(pp \shortrightarrow n \bar{n} \shortrightarrow W^+ W^- W^+ W^-)$ that would be similarly challenging to resolve. The non-resonant same-sign dilepton and multi-lepton final states are suppressed by $\mathcal{O}(10^{-2})$ branching fractions and the rates are thus orders of magnitude too low to be constrained by current non-resonant multi-lepton searches~\cite{Khachatryan:2016kod,Aaboud:2017dmy}. In the regime $m_{0^{++}} > 2 m_W$ where all $W$-bosons are on-shell, then the fully hadronic decay modes of the $W^+ W^- W^+ W^-$ final state would contain interesting hierarchical resonance structures that could also conceivably be exploited by future advances in analysis techniques.

\begin{figure}
\begin{minipage}{.31\textwidth}
\includegraphics[trim={0cm 0cm 0cm 0cm},clip,width=1.0\linewidth]{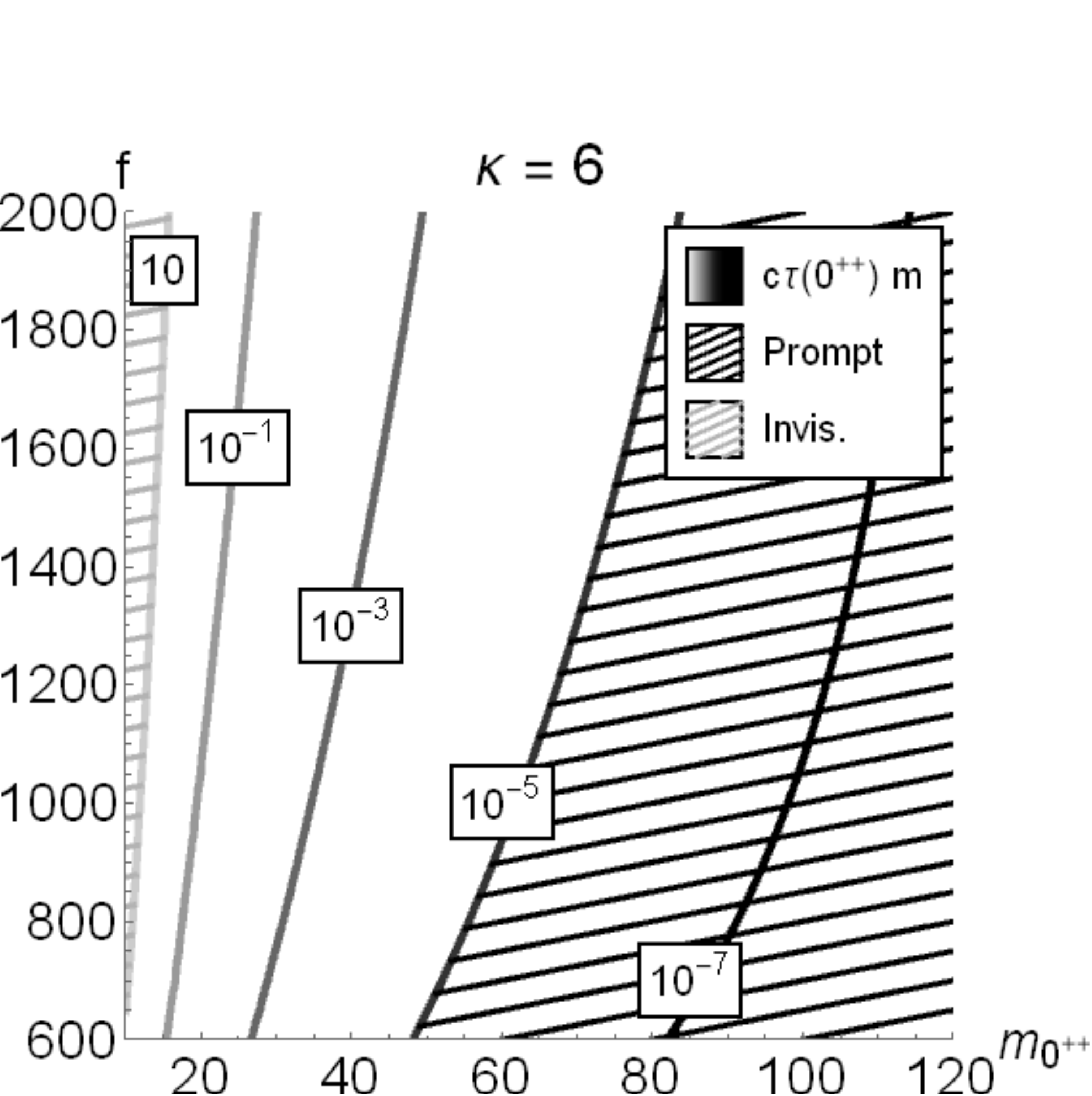}
\end{minipage}
\hspace{0.25cm}
\begin{minipage}{.31\textwidth}
\includegraphics[trim={0cm 0cm 0cm 0cm},clip,width=1.0\linewidth]{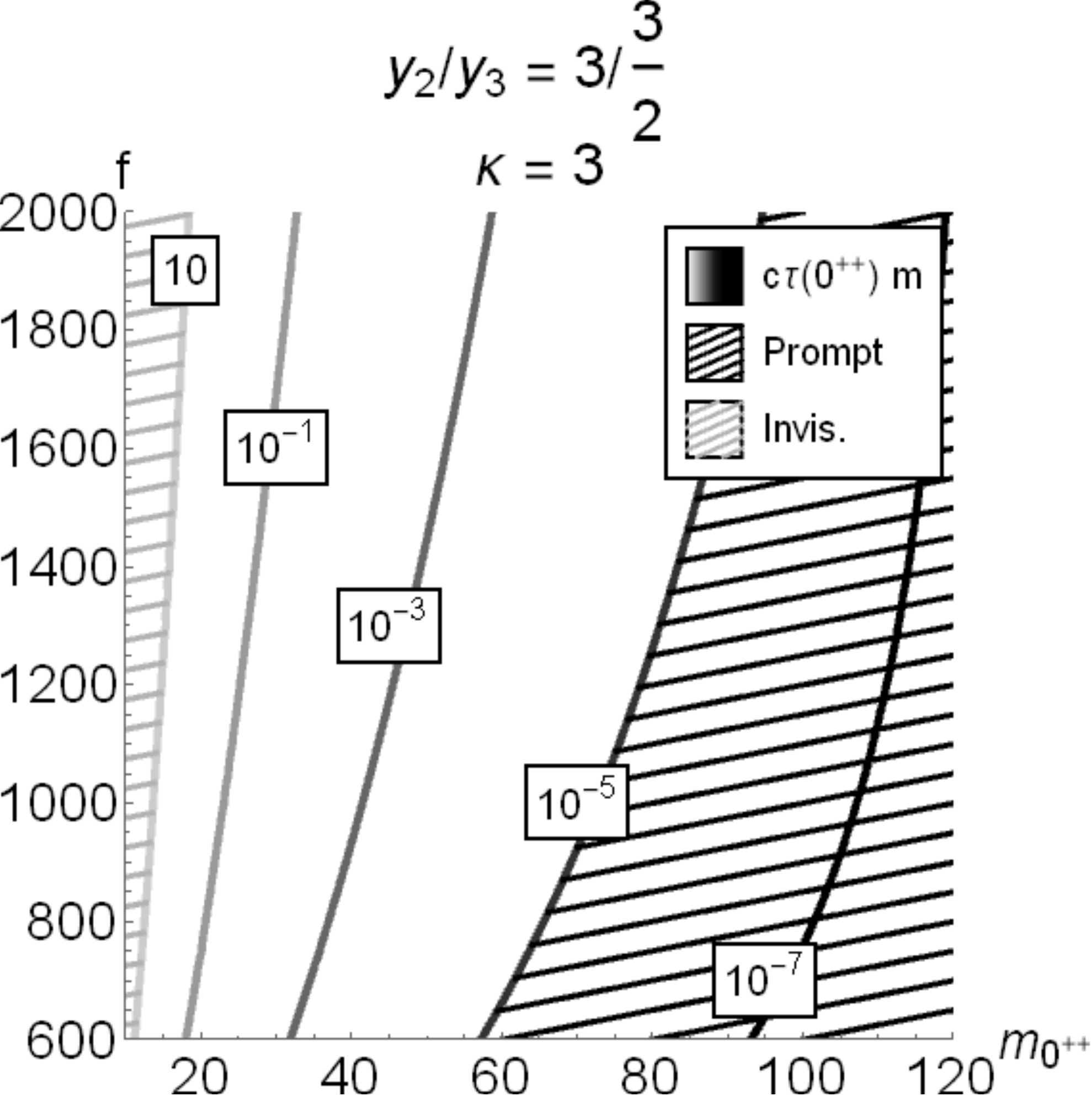}
\end{minipage}
\hspace{0.25cm}
\begin{minipage}{.31\textwidth}
\includegraphics[trim={0cm 0cm 0cm 0cm},clip,width=1.0\linewidth]{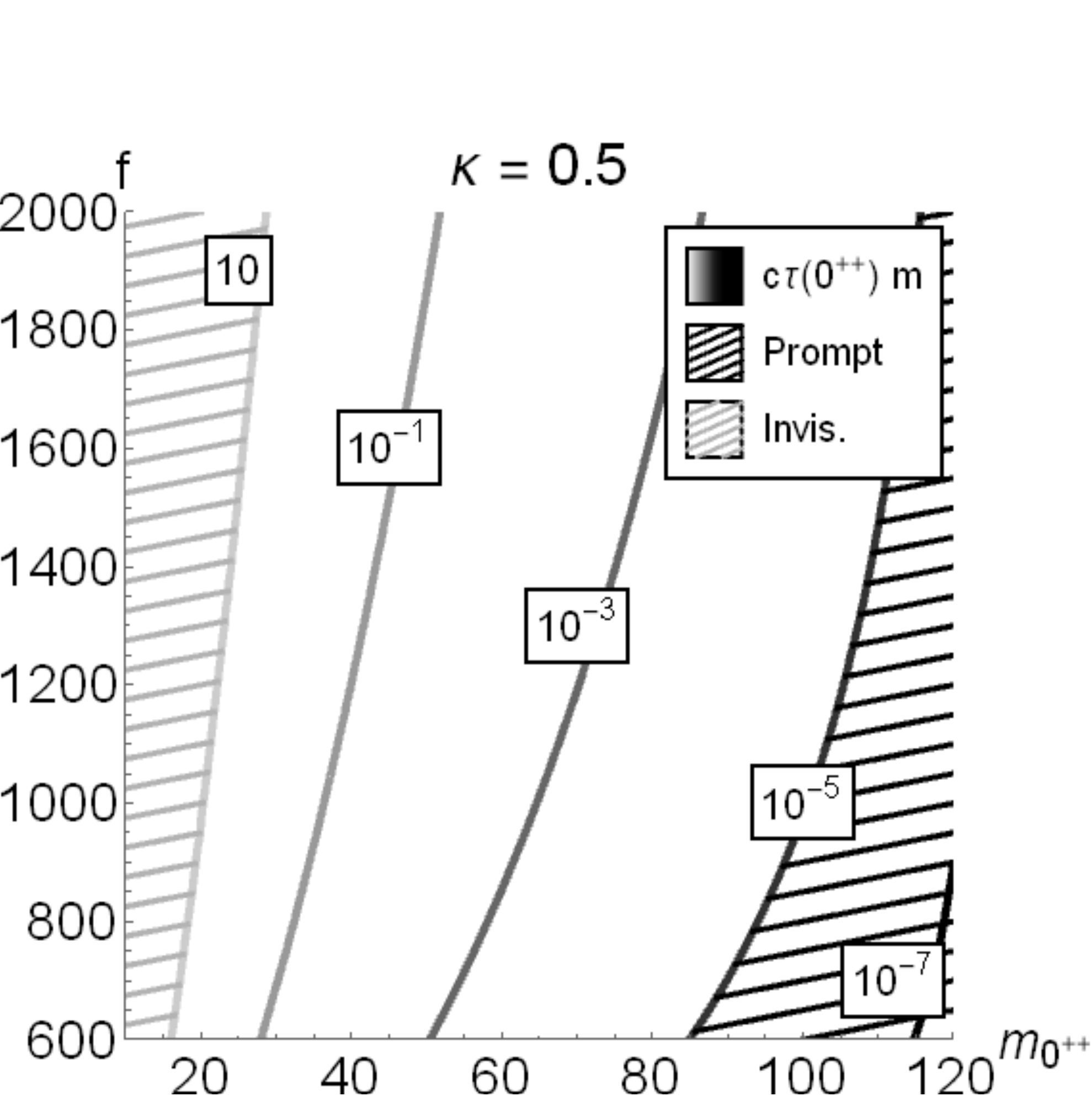}
\end{minipage}
\caption{The glueball decay length $c \tau$ (m) as a function of $m_{0^{++}}$ and $f$. Results are shown for various values of the constant $\kappa$ pameterizing non-perturbative effects with $\kappa = 6$ (left), $\kappa = 3$ (middle), and $1/2$ (right). Here we define ``prompt" decays as those corresponding to $c \tau < 10 \, \mu \rm{m}$ and ``invisible" as those corresponding to $c \tau > 10 \, \rm{m}$. }
\label{fig:llctaGB}
\end{figure}

As a consequence of Equation \ref{eq:llGBwidth}, the lifetime of the $0^{++}$ glueball is a strong function of the hidden-sector confinement scale $\widetilde{\Lambda}$ and its lifetime can either be prompt or extremely long-lived. The decay lengths at a generic value of the Yukawa couplings are illustrated in Figure \ref{fig:llctaGB}, with the middle plot corresponding to the QCD-like scenario with $\kappa =3$.  Here we define ``prompt" decays as those corresponding to $c \tau < 10 \, \mu \rm{m}$ and ``invisible" as those corresponding to $c \tau > 10 \, \rm{m}$. In the regime where $m_{0^{++}} < m_h /2$ then the $0^{++}$ glueball can be very long-lived~\cite{Burdman:2008ek,Craig:2015pha,Curtin:2015fna,Lichtenstein:2018kno}, and Big-Bang Nucleosynthesis (BBN) places a limit on long-lived particles for which $c \tau > 10^7 \, {\rm m}$. For a QCD-like hidden-sector this would place a lower limit on the lightest glueball mass $m_{0^{++}}$ of about $\mathcal{O}({\rm few}) \, {\rm GeV}$. A simplified parameterization of the decay length relative to the BBN limit is given by Equation \ref{eq:llGBcta}.

\be
c \tau (0^{++}) \sim 1.5 \times 10^7 \, {\rm m} \, \left( \frac{1.5 \, {\rm GeV}}{m_{0^{++}}} \right)^7 \left( \frac{f}{800 \, {\rm GeV}} \right)^4 \\[2ex]
\label{eq:llGBcta}
\ee

\noindent In this parameter space $c \tau (0^{++}) \sim 10 \, {\rm m}$ corresponds approximately to $m_{0^{++}} \sim 10 \, {\rm GeV}$, thus placing the glueball decay vertex outside of the LHC detectors at lower masses. For simplicity we thus restrict our analysis to glueball masses $m_{0^{++}} > 2 m_b$ where we have approximately ${\rm BR}(0^{++} \shortrightarrow b \bar{b}) \sim \mathcal{O} (1)$, although similar limits hold at lower masses when ${\rm BR}(0^{++} \shortrightarrow \tau^+ \tau^-) \sim \mathcal{O} (1)$. The primary production mode of the $0^{++}$ glueball in this regime occurs via decays of the Higgs mediated by loops of hidden-sector fermions. The branching ratio of the Higgs boson to hidden-sector glueballs in this framework can be estimated by the Higgs partial width to pairs of hidden-sector gluons, and is parametrically given by the relation in Equation \ref{eq:llBRHGB}.

\be
\frac{{\rm BR} (h \rightarrow 0^{++} 0^{++})}{{\rm BR} (h \rightarrow g g)} \sim \left( \frac{\tilde{\alpha} (m_h)}{\alpha_S} \frac{v^2}{f^2} \right)^2 \sqrt{1 - \frac{4 m_{0^{++}}^2}{m_h^2}} \\[2ex]
\label{eq:llBRHGB}
\ee

\begin{figure}
\begin{minipage}{.45\textwidth}
\includegraphics[trim={0cm 0cm 0cm 0cm},clip,width=1.0\linewidth]{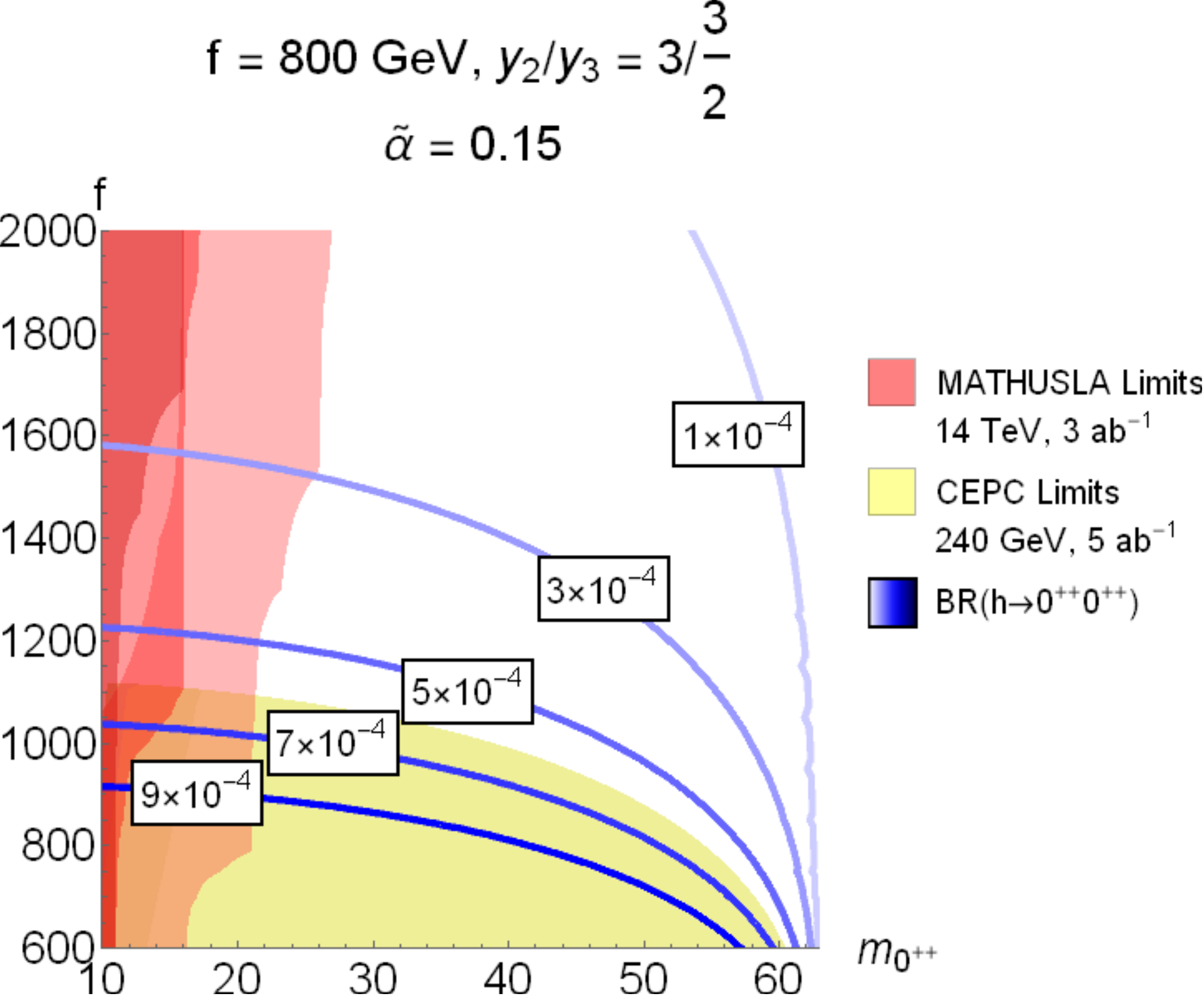}
\end{minipage}
\hspace{1cm}
\begin{minipage}{.45\textwidth}
\includegraphics[trim={0cm 0cm 0cm 0cm},clip,width=1.0\linewidth]{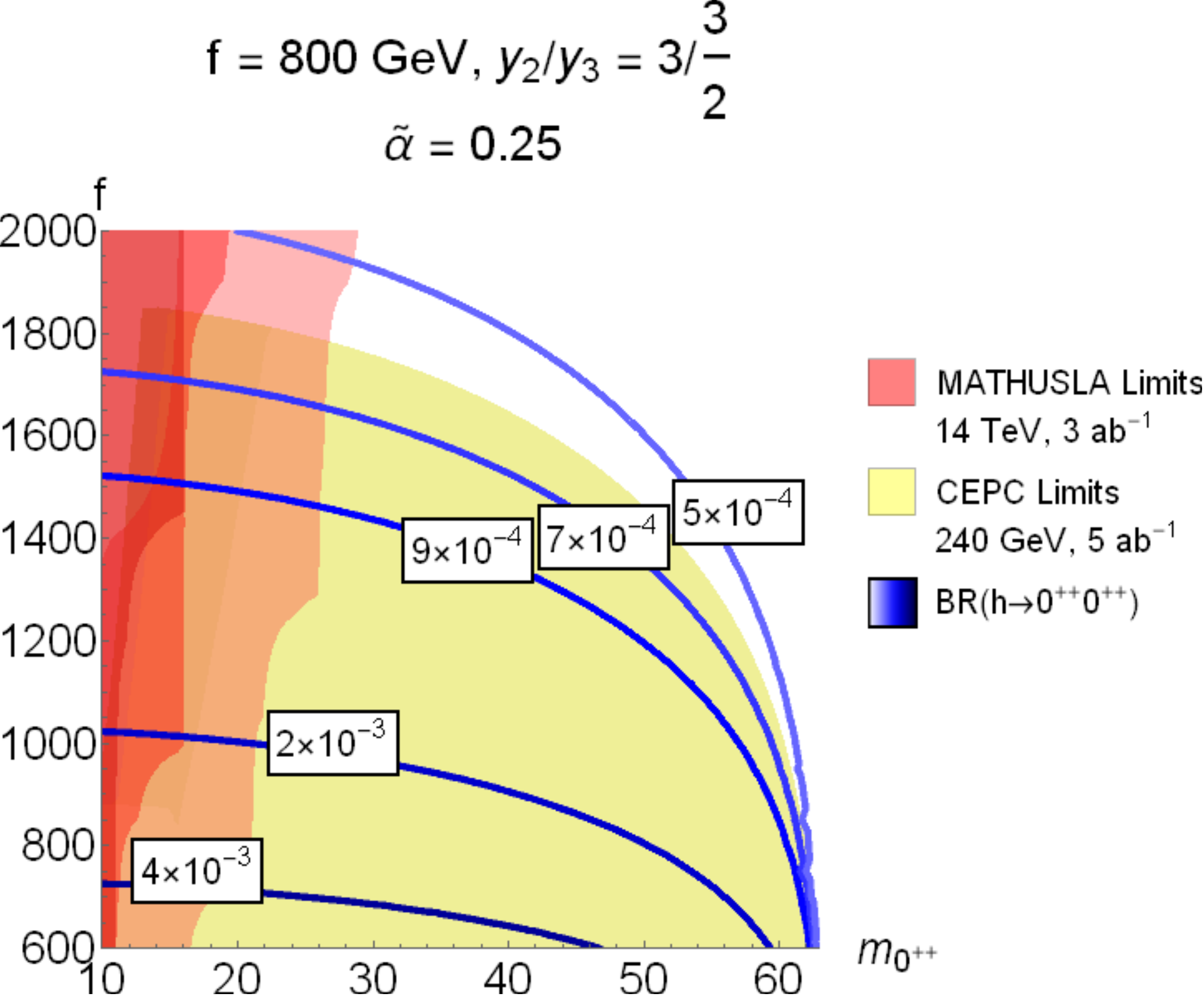}
\end{minipage}
\caption{Higgs branching ratio to hidden-glueballs as a function of $m_{0^{++}}$ and $f$ for a fast and slow running hidden-sector gauge couplings $\tilde{\alpha} = 0.2$ (left) and $0.4$ GeV (right). Projected $95 \%$ confidence limits are shown for the proposed MATHUSLA surface detector (red) as well as the proposed CEPC Higgs factory (yellow). The light/medium/dark shades correspond to different values of the glueball width parameter $\kappa = 6/3/0.5$.}
\label{fig:llBRHGB}
\end{figure}

\noindent This branching ratio ${\rm BR} (h \rightarrow 0^{++} 0^{++})$ depends on the running of the hidden-sector coupling constant $\tilde{\alpha}$ from the hidden confinement scale $\widetilde{\Lambda}$ to the Higgs mass $m_h$. In this model we have assumed that the lightest states charged under the hidden-sector gauge group $G$ have masses on the order of $f > \widetilde{\Lambda}$ and the $\beta$-function for $\tilde{\alpha}$ thus depends purely on the quadratic Casimir of $G$. Weak-scale perturbativity of the coupling constant $\tilde{\alpha} (m_h) < 0.5$ generically implies a small branching fraction for $h \rightarrow 0^{++} 0^{++}$ that is below sensitivity projections for the HL-LHC~\cite{Dawson:2013bba,CMS:2013xfa,CMS-DP-2016-064,ATL-PHYS-PUB-2013-014,ATL-PHYS-PUB-2014-016}. However future Higgs factories such as the proposed Circular Electron-Positron Collider (CEPC) or the International Linear Collider (ILC)~\cite{CEPC-SPPCStudyGroup:2015csa,Fujii:2015jha,Gomez-Ceballos:2013zzn} will be able to exclude a wide range of values for $\tilde{\alpha} (m_h)$. The expected sensitivities to a variety of rare Higgs decays have been projected for a long-term run at the CEPC~\cite{Liu:2016zki}, which will be sensitive to the decay topology ${\rm BR} (h \shortrightarrow 0^{++} 0^{++} \shortrightarrow b \bar{b} b \bar{b}) \sim 6 \times 10^{-4}$, as illustrated in Figure \ref{fig:llBRHGB}. Here we have restricted these limits to regions in which the $0^{++}$ glueballs decay within the tracker radius of the proposed CEPC detectors $\sim 1810 \, {\rm mm}$, and assume that object identification would be inefficient from calorimeter data alone. The large region of parameter space with macroscopic glueball decay lengths also provides additional motivation for a more careful exploration of the lifetime frontier via surface detectors such as the proposed MATHUSLA experiment~\cite{Curtin:2018mvb}. Preliminary studies indicate that such a surface detector could exclude rare branching fractions of the Higgs boson as low as ${\rm BR} (h \shortrightarrow XX) \sim 10^{-5}$ for detector-optimized values of $c \tau (0^{++})$. The long-term projections for these rare Higgs decays assuming $3 \, {\rm ab}^{-1}$ of MATHUSLA data are also illustrated in Figure \ref{fig:llBRHGB}. 

\newpage

\section{$SU(4)/Sp(4)$ Intermediate Model}
\label{sec:SUSP}

The symmetry breaking pattern $SU(4)/Sp(4)$ produces $15 - 10  = 5$ pseudo Nambu-Goldstone Bosons (pNGBs). The nonlinear sigma model describing the low energy effective theory may be expressed in terms of an antisymmetric unitary matrix $\Sigma$, which transforms under $SU(4)$ as $V \Sigma V^T$ where $V$ is an $SU(4)$ matrix.  It is convenient to specify a background field $\Sigma_0$, which is invariant under the $Sp(4)$ subgroup containing $SU(2)_L \otimes SU(2)_R$. The generators for the $SU(2)_L\otimes U(1)_Y$ gauge symmetry $L^a$ and $Y$ are embedded into an approximate global custodial symmetry via $Y=R^3$, which remains unbroken in the reference vacuum $\Sigma=\Sigma_0$.

\begin{equation}
  \Sigma_0 = 
  \begin{pmatrix} 
  i\sigma_2 & \quad \\ \quad  & i\sigma_2 \\ 
\end{pmatrix}
\qquad \qquad 
L^a =
  \frac{1}{2}\begin{pmatrix}
    \sigma^a &\quad \\ \quad& ~~0
  \end{pmatrix} \qquad R^a =
  \frac{1}{2}\begin{pmatrix}
    0~~ &\quad \\ \quad& \sigma^{a}
  \end{pmatrix}
  \label{eq: SUSPgauge}
\end{equation}

\noindent The pNGB's thus transform under the electroweak gauge group as $1_0 \oplus 2_{\pm1/2}$ giving one real electroweak singlet and a SM Higgs doublet. The Nambu-Goldstone bosons are fluctuations about this background in the direction of the broken generators, $\Pi \equiv \pi^a X^a$, and may be parameterized as

\beq
\Pi  &=& \frac{1}{2} \begin{pmatrix} 
  \frac{1}{\sqrt{2}} \eta ~\mathbbm{1} & ~~(H^c ~H) \\
  (H^c ~H)^\dagger  & ~~-\frac{1}{\sqrt{2}} \eta ~\mathbbm{1} \\ 
\end{pmatrix}
\eeq

\noindent Here $H = (H_+ ~H_0)$ and $H^c = i \sigma_2 H^*$. Vacuum misalignment with respect to the weak gauge interactions can be parameterized by an angle $\theta$. This vacuum, which is invariant under the custodial $SU(2)_c$ generated by $R^a+L^a$, can be written as 

\be
\theta^2 \equiv \frac{1}{2} \frac{H^\dagger H}{f^2} \qquad \qquad \langle \Sigma \rangle =
\begin{pmatrix} 0 & \cos\theta & 0 & i \sin\theta \\
  -\cos\theta & 0 & -i \sin\theta & 0 \\
  0 & i \sin\theta & 0 & \cos\theta \\
  -i \sin\theta & 0 & -\cos\theta & 0
\end{pmatrix},
\ee

\noindent The gauge couplings explicitly break the $SU(4)$ global symmetry. The covariant derivative can be expanded to second order in the vacuum giving tree level masses to the weak gauge bosons in terms of the $SU(2)_L \times U(1)_Y$ couplings $g, g^\prime$. The tree level relation $m_W/m_Z=\cos\theta_w$ is guaranteed by the preservation of the remnant custodial symmetry $SU(2)_c \subset Sp(4)$

\begin{align}
m_W^2 = \frac{g^2}{2}\,f^2 \sin^2 \theta && m_Z^2 = \frac{g^2 + g'^2}{2}\, f^2 \sin^2 \theta.
\end{align}

\subsection{Fermion Sector}

In this model the Yukawa sector of the Standard Model is extended to include interactions between the $\Sigma$ field and composite fermions in complete $SU(4)$ multiplets. Additional gauge invariant interactions softly break the global symmetry, generating a significant negative contribution to the Higgs potential, thus driving electroweak symmetry breaking. The mass and quartic couplings of the Higgs can be tuned to their experimentally measured values by extending the fermion sector of this theory to include two vector-like multiplets of fermions in the fundamental representation of $SU(4)$. The first multiplet $(\psi, \,\bar{\psi})$ is charged under $SU(3)_c$ and mixes with the third generation of Standard Model quarks resulting in a partially composite top and bottom. The second multiplet $(\chi, \, \bar{\chi})$ is charged under a hidden-sector gauge group $G$ that confines at some scale $\widetilde{\Lambda} < f$.

\subsubsection{Color-charged Fermions}

The new color-charged vector-like states $(\psi , \, \bar{\psi})$ mix with the chiral third-generation of SM quarks, which are singlets under the global $SU(4)$ symmetry. The gauge eigenstate components of these multiplets are fixed by the embedding of $SU(2)_L \times U(1)_Y \subset SU(4)$, and are shown in Equation \ref{eq:vlq}. The gauge quantum numbers of the new color-charged fermions are given in Table \ref{tab:qnumbers}.

\be
\psi = \left( Q  ~~T ~~B \right) \qquad \qquad \bar{\psi} = \left( \overline{Q} ~~\overline{T} ~~\overline{B} \right)^{T}
\label{eq:vlq}
\ee
\begin{center}
  \begin{table}[ht]
    \begin{tabular}{| c | c c c | c c c |}
      \hline
      & ~$\left( Q ~\overline{Q} \right)$~ & ~$\left( T ~\overline{T} \right)$~ & ~$\left( B ~\overline{B} \right)$~ & ~$q$~ & ~$\bar{t}$~ & ~$\bar{b}$~ \\ [0.5ex] 
      \hline
      $SU(3)_c$ & $(3 ~\overline{3})$ & $(3 ~\overline{3})$ & $(3 ~\overline{3})$ & $3$ & $\overline{3}$ & $\overline{3}$ \\
      $SU(2)_L$ & $(2 ~2)$ & $(\mathbbm{1} ~\mathbbm{1})$ & $(\mathbbm{1} ~\mathbbm{1})$ & $2$ & $\mathbbm{1}$ & $\mathbbm{1}$ \\
      $U(1)_Y$ & $\left( \frac{1}{6} ~\text{-}\frac{1}{6} \right)$ & $\left( \frac{2}{3} ~\text{-}\frac{2}{3} \right)$ & $\left( \text{-}\frac{1}{3} ~\frac{1}{3} \right)$ & $\frac{1}{6}$ & $\text{-}\frac{2}{3}$ & $\frac{1}{3}$ \\
      \hline
      $G$ & $(\mathbbm{1} ~\mathbbm{1})$ & $(\mathbbm{1} ~\mathbbm{1})$ & $(\mathbbm{1} ~\mathbbm{1})$ & $\mathbbm{1}$ & $\mathbbm{1}$ & $\mathbbm{1}$ \\
      \hline
    \end{tabular}
    \caption{Gauge quantum numbers for the new vector-like quarks. These include a new $SU(2)_L$ doublet $Q = (Q_T, \, Q_B)$ in addition to two vector-like singlets $T$ and $B$. The new vector-like states mix with the SM chiral doublet $q = (q_T, \, q_B)$ and singlets $(\bar{t}, \, \bar{b})$.}
    \label{tab:qnumbers}
  \end{table}
\end{center}

\noindent The most general gauge invariant Yukawa couplings for the color-charged fermions are given by Equation \ref{topYukawa}, and collective symmetry breaking properties of these interactions guarantee the absence of quadratic divergences to the scalar potential from fermion-loops. The resulting top-like mass matrix $M_T$ has the property $\det M_T^\dagger M_T \propto \sin^2 \theta$, thus producing one massless state when $\theta = n \pi$ for $n \in \mathbb{Z}$. 

\be
\CL_{\psi} = y_1\,f\,\psi \,\Sigma\,\bar{\psi} + y_2 f \, q \,\overline{Q} + y_3\,f\,T\,\bar{t}\, + y_4\,f\,B\,\bar{b} + {\rm h.c.}
\label{topYukawa}
\ee

\begin{center}
  \begin{table}[ht]
    \begin{tabular}{| c | c c c |}
      \hline
      & $~~\overline{Q}_T~~$ & $~~\overline{T}~~$ & $~~\bar{t}~~$ \\ [0.5ex] 
      \hline
      $Q_T$  & $y_1 f \cos \theta$ & $-i y_1 f \sin \theta$ & $0$ \\
      $T$  & $-i y_1 f \sin \theta$ & $y_1 f \cos \theta$ & $y_3 f$ \\
      $q_T$  & $y_2 f$ & $0$ & $0$ \\
      \hline
    \end{tabular} \qquad \qquad \qquad \begin{tabular}{| c | c c c |}
      \hline
      & $~~\overline{Q}_B~~$ & $~~\overline{B}~~$ & $~~\bar{b}~~$ \\ [0.5ex] 
      \hline
      $Q_B$  & $y_1 f \cos \theta$ & $-i y_1 f \sin \theta$ & $0$ \\
      $B$  & $-i y_1 f \sin \theta$ & $y_1 f \cos \theta$ & $y_3 f$ \\
      $q_B$  & $0$ & $y_2 f$ & $0$ \\
      \hline
    \end{tabular}
    \caption{Mass matrices for the top-like quarks $M_T$ (left) and bottom-like quarks $M_B$ (right) in the gauge eigenbasis.}
    \label{tab:qmass}
  \end{table}
\end{center}
\vspace{-1cm}

\noindent Below the scale of electroweak symmetry breaking the fermion interactions can be expressed in terms of their low energy gauge eigenstates. The top-like and bottom-like mass matrices $M_T$ and $M_B$ are a function of Higgs' vacuum expectation value $\theta = \langle h\rangle/ (\sqrt{2} f)$, and produce six mass eigenstates upon diagonalization. The up-type sector contains three charge $\pm 2/3$ states which we label in order of descending mass as $\tpp$, $\tp$, $t$, and the lightest of these states corresponds to the SM top quark. The down-type sector contains three charge $\mp 1/3$ mass eigenstates which we label in order of descending mass as $\bpp$, $\bp$, $b$, and the lightest of these states corresponds to the SM bottom quark. The effective Yukawa couplings for the SM top and bottom quarks may be expressed as
\be
y_t = \frac{y_1 y_2 y_3}{\sqrt{y_1^2 + y_2^2} \sqrt{y_1^2 + y_3^2}} \qquad \qquad \qquad y_b = \frac{y_1 y_2 y_4}{\sqrt{y_1^2 + y_2^2} \sqrt{y_1^2 + y_3^2}}
\ee
Note that a non-zero top quark mass requires $y_1, y_2, y_3 > 0$ simultaneously, revealing the collective symmetry breaking properties of the top sector mass matrix. The masses of the color-charged fermions are computed in Appendix \ref{app:scalar} to second order in $\sin^2 \theta$, and the numerically diagonalized mass spectrum for the heavy fermions are shown in Figure (\ref{fig:fmasses}). To leading order in $\sin^2 \theta$ these masses scale approximately as
\be
\begin{aligned}[c]
m_{t}^2 &\sim f^2 y_t^2 \sin^2 \theta \\
m_{\tp}^2 &\sim f^2 \left(y_1^2 + y_2^2 \right) \\
m_{\tpp}^2 &\sim f^2 \left( y_1^2 + y_3^2 \right) 
\end{aligned}
\qquad\qquad\qquad
\begin{aligned}[c]
m_{b}^2 &\sim f^2 y_b^2 \sin^2 \theta \\
m_{\bp}^2 &\sim f^2 \left(y_1^2 + y_4^2 \right) \\
m_{\bpp}^2 &\sim f^2 \left(y_1^2 + y_2^2 \right)
\end{aligned}
\ee

\subsubsection{Hidden-Sector Fermions}

The new hidden-sector vector-like states $(\chi, \, \bar{\chi})$ transform in the $(\Box,4) \oplus (\ovB{\Box},4)$ representation of $G \times SU(4)$. The hidden-sector gauge group $G$ confines at some scale $\widetilde{\Lambda} < f$ resulting in a low energy spectrum comprised of unstable mesons, stable baryons, and potentially long-lived glueballs. The lightest baryons in this sector are a natural candidate for dark matter if they are electrically neutral, however a detailed analysis of hidden-baryon spectrum is beyond the scope of this analysis. The gauge quantum numbers of the hidden-fermions are given in Table \ref{tab:qnumbersdark}.

\be
\chi = \left( X  ~~N ~~C \right) \qquad \qquad \bar{\chi} = \left( \overline{X} ~~\overline{N} ~~\overline{C} \right)^{T}
\label{eq:vlqdark}
\ee
\begin{center}
  \begin{table}[ht]
    \begin{tabular}{| c | c c c | c |}
      \hline
      & ~$\left( X ~\overline{X} \right)$~ & ~$\left( N ~\overline{N} \right)$~ & ~$\left( C ~\overline{C} \right)$~ & ~$\left( n ~\bar{n} \right)$~  \\ [0.5ex] 
      \hline
      $SU(3)_c$ & $(\mathbbm{1} ~\mathbbm{1})$ & $(\mathbbm{1} ~\mathbbm{1})$ & $(\mathbbm{1} ~\mathbbm{1})$ & $(\mathbbm{1} ~\mathbbm{1})$  \\
      $SU(2)_L$ & $(2 ~2)$ & $(\mathbbm{1} ~\mathbbm{1})$ & $(\mathbbm{1} ~\mathbbm{1})$ & $(\mathbbm{1} ~\mathbbm{1})$ \\
      $U(1)_Y$ & $\left( \frac{1}{2} ~\text{-}\frac{1}{2} \right)$ & $\left( 0 ~0 \right)$ & $\left( \text{-}1 ~1 \right)$ & $\left( 0 ~0 \right)$  \\
      \hline
      $G$ & $(\Box ~\ovA{\Box})$ & $(\Box ~\ovA{\Box})$ & $(\Box ~\ovA{\Box})$ & $(\Box ~\ovA{\Box})$  \\
      \hline
    \end{tabular}
    \caption{Gauge quantum numbers for the new vector-like quarks. These include one $SU(2)_L$ doublet $X = (X_+ ~~ X_0)$, one charged singlet $C$, and two neutral singlets $N$ and $n$.}
    \label{tab:qnumbersdark}
  \end{table}
\end{center}

\noindent A Standard Model-like Higgs boson can be obtained from fermionic loop-corrections by introducing symmetry breaking interactions between the multiplet $(\chi, \, \bar{\chi})$ and an additional massless vector-like pair of hidden-sector fermions $(n, \, \bar{n})$. The Yukawa interactions thus include an $SU(4)$ symmetric coupling to the $\Sigma$-field in addition to terms that softly break the global symmetry.

\be
\CL_{\chi} = \tilde{y}_1\,f\,\chi \,\Sigma\,\bar{\chi} + \tilde{y}_2 f \, n \,\overline{N} + \tilde{y}_3\,f\,N\,\bar{n}  + {\rm h.c.}
\label{topYukawadark}
\ee

\begin{center}
  \begin{table}[ht]
   \begin{tabular}{| c | c c c |}
      \hline
      & $~~\overline{X}_0~~$ & $~~\overline{N}~~$ & $~~\bar{n}~~$ \\ [0.5ex] 
      \hline
      $X_0$  & $\tilde{y}_1 f \cos \theta$ & $-i \tilde{y}_1 f \sin \theta$ & $0$ \\
      $N$  & $-i \tilde{y}_1 f \sin \theta$ & $\tilde{y}_1 f \cos \theta$ & $\tilde{y}_3 f$ \\
      $n$  & $0$ & $\tilde{y}_2 f$ & $0$ \\
      \hline
    \end{tabular}  \qquad \qquad \qquad
    \begin{tabular}{| c | c c |}
      \hline
      & $~~\overline{X}_+~~$ & $~~\overline{C}~~$ \\ [0.5ex] 
      \hline
      $X_+$  & $\tilde{y}_1 f \cos \theta$ & $-i \tilde{y}_1 f \sin \theta$ \\
      $C$  & $-i \tilde{y}_1 f \sin \theta$ & $\tilde{y}_1 f \cos \theta$ \\
      \hline
    \end{tabular}
    \caption{Mass matrices for the neutral hidden sector fermions $M_N$ (left) and charged hidden-sector fermions $M_C$ (right) in the gauge eigenbasis.}
    \label{tab:qmassdark}
  \end{table}
\end{center}
\vspace{-1cm}

The hidden-sector interactions are constructed in such a way that the neutral mass matrix $M_N$ obeys the condition $\det M_N^\dagger M_N \propto \cos^2 \theta$, thus producing one massless state when $\theta = (n+1/2) \pi$. Diagonalizing $M_N$ produces three electrically neutral mass eigenstates which we label in in order of descending mass as $\npp$, $\np$, $n$. Diagonalizing the charged hidden-sector mass matrix $M_C$ results in a charge $\pm 1$ vector-like pair $(\xp, \, \xpbar)$ and a charge $\mp 1$ vector-like pair $(\cm, \, \cmbar)$. The precision electroweak constraints described in the following section push the fermion masses of this theory to very high scales, and the details of their mass spectrum are thus irrelevant for hadron-collider phenemenology. We thus consider a simplified parameter space in which $\tilde{y}_3 = \tilde{y}_2 = \tilde{y}_1$, though the masses are computed for general Yukawa couplings in Appendix \ref{app:scalar} to second order in $\sin^2 \theta$. In this simplified parameter space the eigenvalues of the mass matrix scale approximately as
\be
\begin{aligned}[c]
m_n^2 &\sim \frac{f^2}{4} \tilde{y}_1^2 \cos^2 \theta \\[10pt]
m_{\np}^2 &\sim m_{\npp}^2 \sim 2 f^2 \tilde{y}_1^2
\end{aligned} \qquad \qquad
\begin{aligned}[c]
& \\[10pt]
m_{x_+}^2 &\sim m_{c_-}^2 \sim f^2 \tilde{y}_1^2
\end{aligned}
\ee

\subsubsection{Theory Space}

\begin{figure}
\begin{minipage}{.45\textwidth}
\includegraphics[trim={0cm 0cm 0cm 0cm},clip,width=1.0\linewidth]{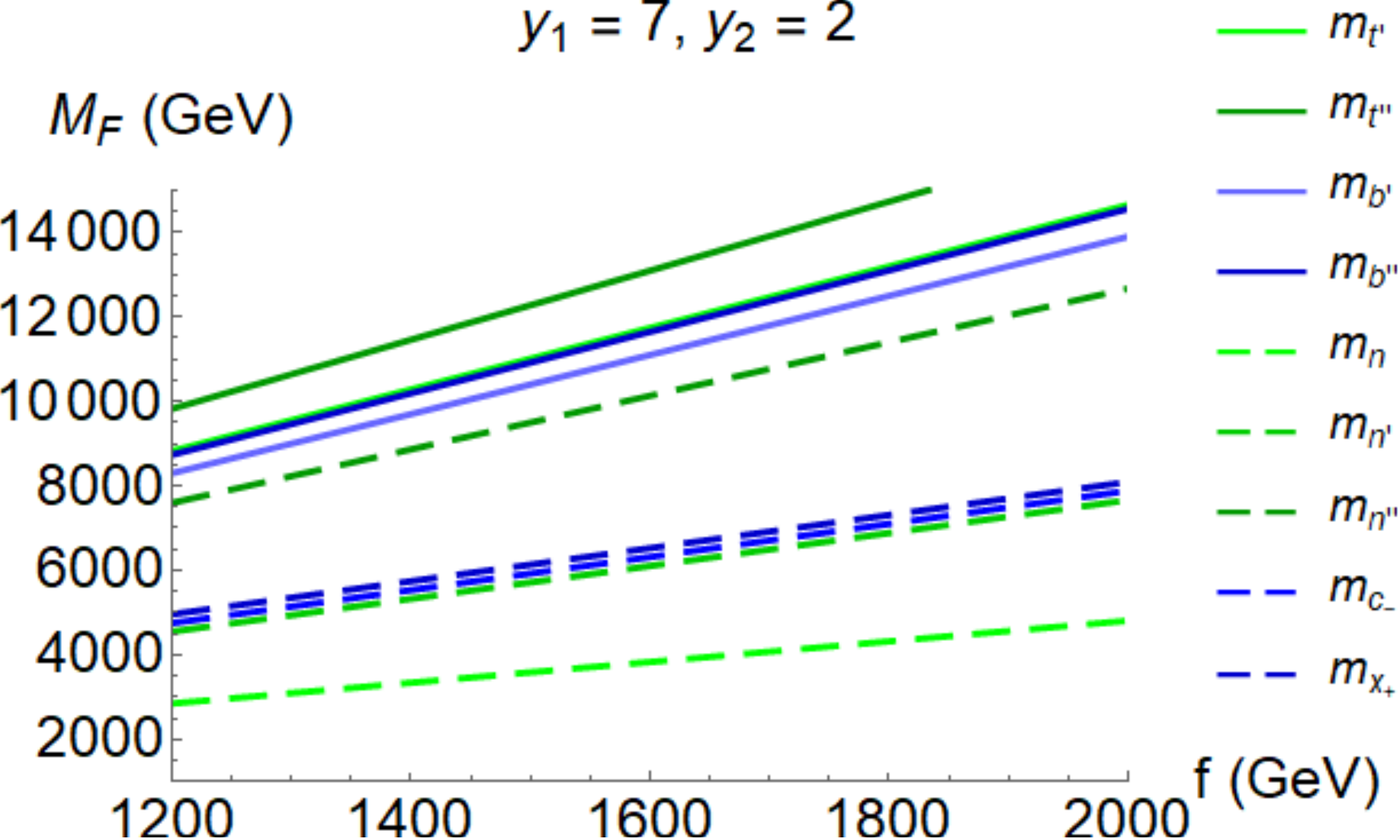}
\end{minipage}
\hspace{1cm}
\begin{minipage}{.45\textwidth}
\includegraphics[trim={0cm 0cm 0cm 0cm},clip,width=1.0\linewidth]{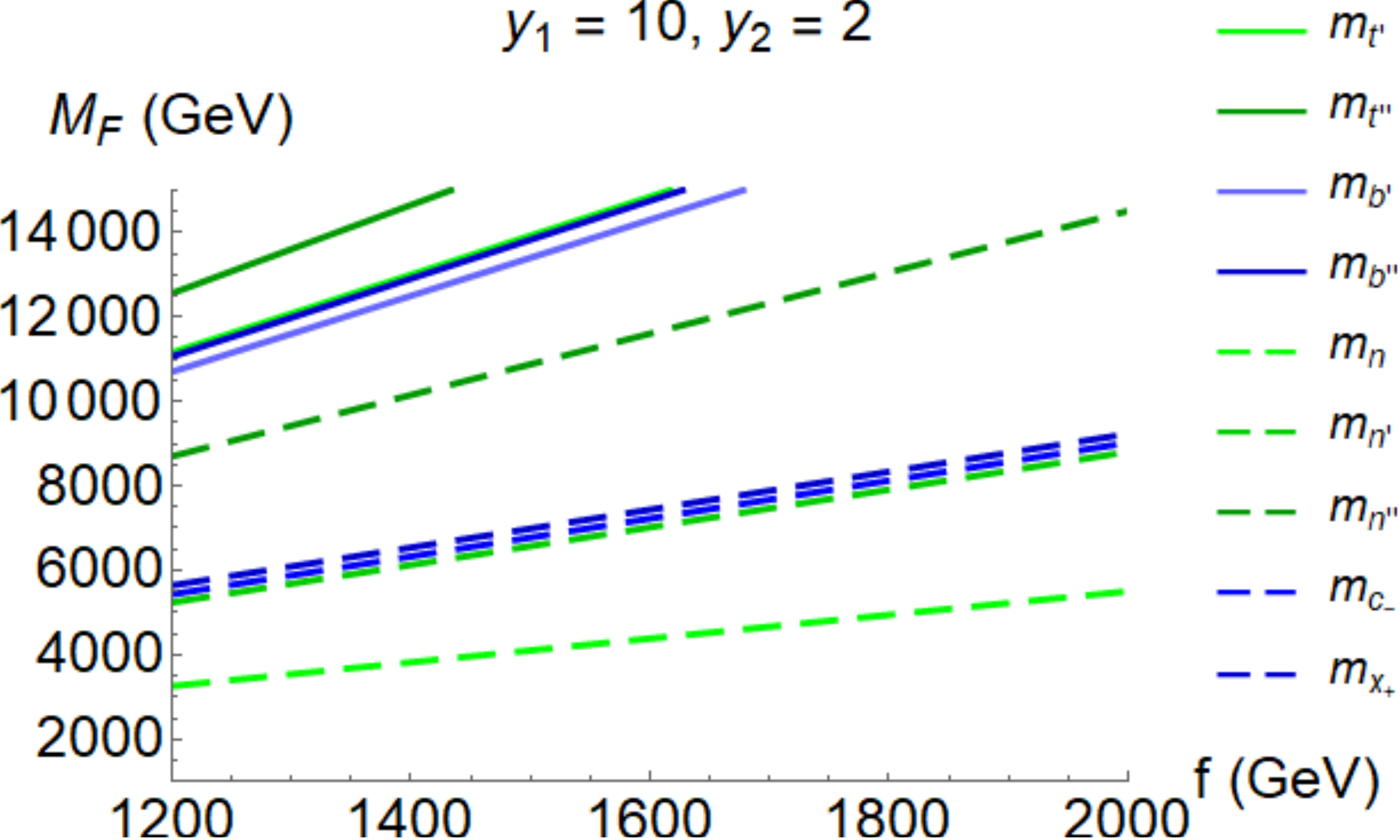}
\end{minipage}
\caption{Mass spectrum for the new quark partners at two benchmark points. One corresponding to $y_1 / y_2= 7/2$ (left) and $y_1 / y_2 = 10/2$ GeV (right). The hidden-fermion masses are fixed relative to the color-charged states by the Higgs mass.}
\label{fig:fmasses}
\end{figure}

The new color-charged fermion interactions introduce four Yukawa couplings $y_i$, two of which are fixed by the Standard Model top and bottom quark masses as implied by Equation \ref{topYukawa}. This leaves two free parameters $y_1$ and $y_2$ which are bound from below by the top-quark Yukawa coupling $y_t < y_1 y_2 / \sqrt{y_1^2 + y_2^2}$. The new hidden-sector interactions introduce three new Yukawa couplings, which we reduce to a single parameter by assuming $\tilde{y}_1 = \tilde{y}_2 = \tilde{y}_3$. The conditions of EWSB described in the following section thus fixes a unique value for $\tilde{y}_1$ at each point in the two-dimensional parameter space of $(y_1, \, y_2)$. In this framework, the experimentally measured Higgs mass implies a mass scale for the new hidden-sector fermions that is $\mathcal{O}({\rm few})$ times lower than their color-charged counterparts. The masses of the new fermionic states as a function of the chiral symmetry breaking scale $f$ are illustrated for two benchmark points in Figure \ref{fig:fmasses}

\begin{figure}
\begin{minipage}{.31\textwidth}
\includegraphics[trim={0cm 0cm 0cm 0cm},clip,width=1.0\linewidth]{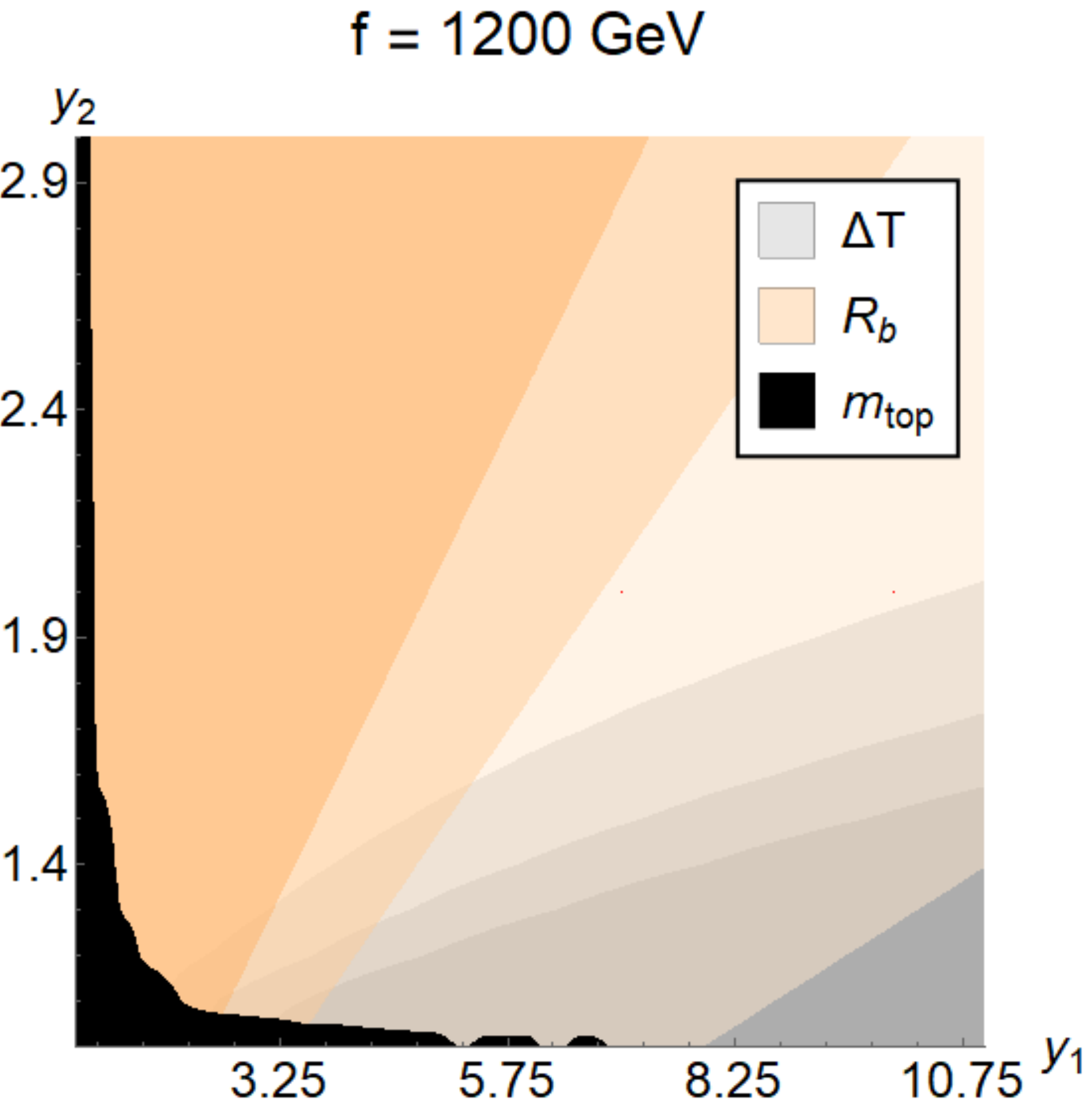}
\end{minipage}
\hspace{0.25cm}
\begin{minipage}{.31\textwidth}
\includegraphics[trim={0cm 0cm 0cm 0cm},clip,width=1.0\linewidth]{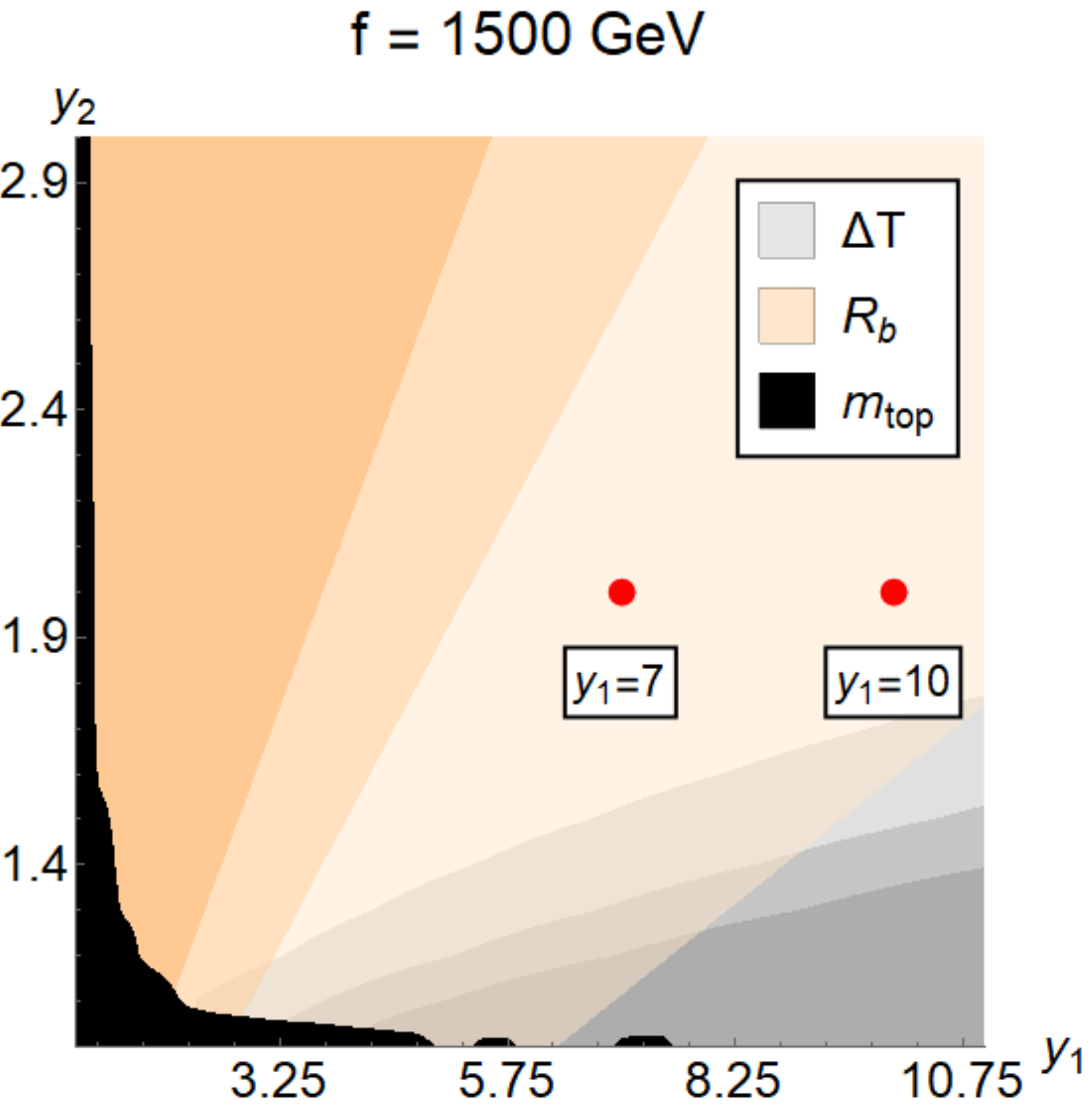}
\end{minipage}
\hspace{0.25cm}
\begin{minipage}{.31\textwidth}
\includegraphics[trim={0cm 0cm 0cm 0cm},clip,width=1.0\linewidth]{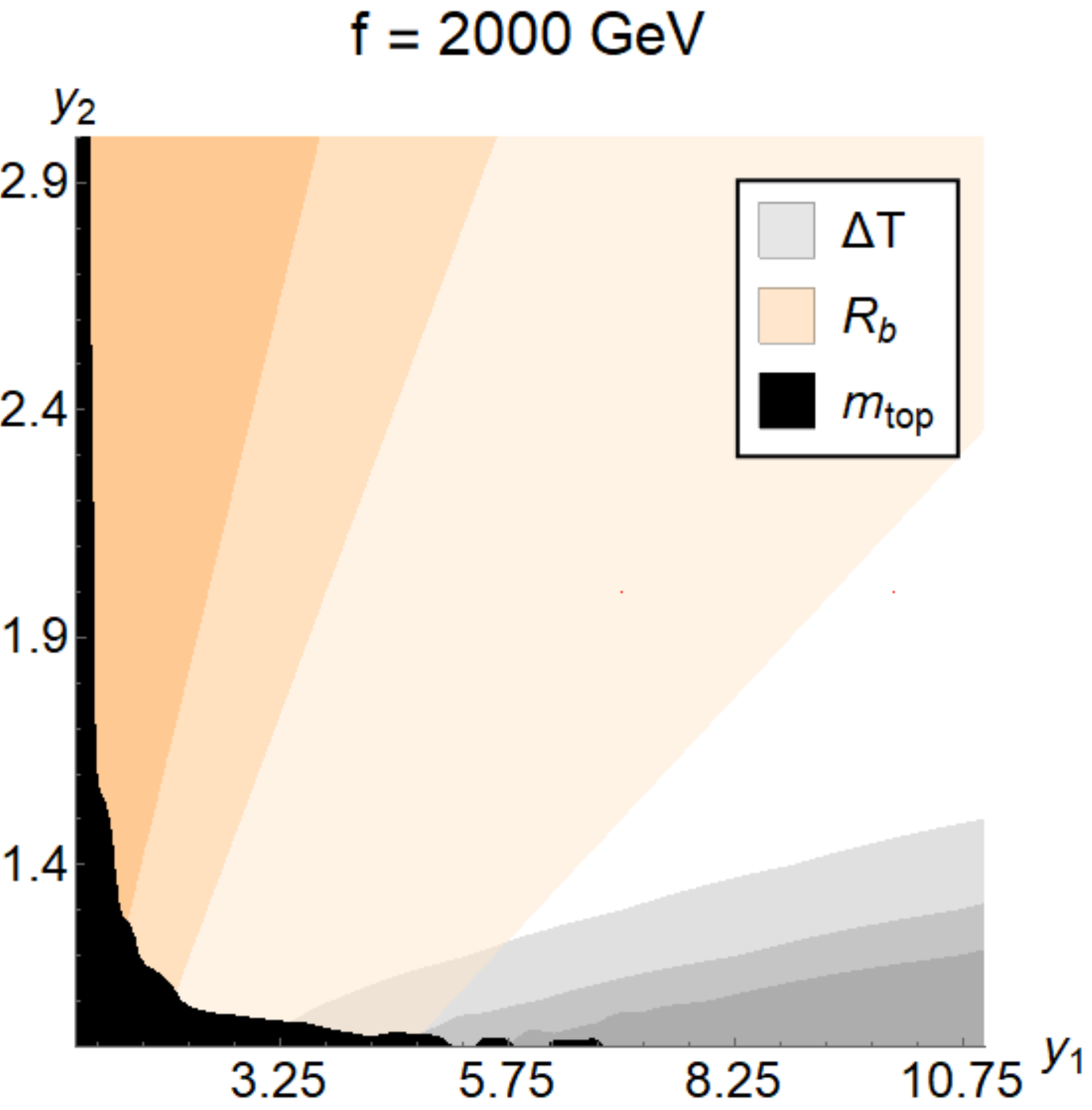}
\end{minipage}
\caption{Constraints from precision electroweak measurements at $f = 1200$ GeV (left), $f = 1500$ GeV (middle), and $f = 2000$ GeV (right). The light/medium/dark shades correspond to the $1 \sigma / 2 \sigma / 3 \sigma$ deviations from experimentally measured values. The black corresponds to regions inconsistent with the top quark mass.}
\label{fig:PEW}
\end{figure}

The most constraining precision electroweak observable for the $SU(4)/Sp(4)$ model comes from the branching fraction of $Z \rightarrow b \bar{b}$. The doublet-singlet mixing between the SM bottom quark and its vector-like partners will generically induce large corrections to the bottom quark neutral currents if $y_2$ is significantly larger than $y_1$. The extreme precision of experimental measurements on the parameter $R_b = \Gamma (Z \rightarrow b \bar{b}) / \Gamma (Z \rightarrow {\rm had}) = 0.21629 \pm 0.00066$ thus excludes all regions except those in which $y_1 \gg y_2$, as shown in Figure \ref{fig:PEW}. The precision electroweak constraints from $R_b$ thus push the masses of the vector-like fermions to very high scales that are likely beyond the reach of a $14 \, {\rm TeV}$ machine via direct production. These constraints may be relaxed by introducing additional states with bottom-like gauge quantum numbers that reduce the doublet-singlet mixing. However, for the remainder of this paper we will assume the minimal matter content for this model in which the fermions are heavy and decoupled. The contributions to the $T$-parameter from fermion loops are less constraining in this model and are also shown in Figure \ref{fig:PEW}.

\subsection{Electroweak Symmetry Breaking}

The one-loop corrections to the scalar effective potential can  be extracted from the quadratically and logarithmically sensitive terms in the Coleman Weinberg potential. The Higgs mass  receives a large positive contribution from gauge loops that is quadratically sensitive to the compositeness scale $\Lambda$, and proportional to the gauge boson mass matrix squared. Taking $\Lambda \sim 4 \pi f$ we find 
\begin{align}
  \delta V_G (c, \theta) &= f^2 \tr M_V^2 (\Sigma) \\
  &= -f^2 f_G^2 \cos 2 \theta \\
  f_G^2 &= \frac{f^2}{4} c \, (3 g^2 + g'^2)
  \label{eq:gaugecw}
\end{align}
Collective symmetry breaking ensures the absence of quadratically divergent contributions from the Yukawa sector, and a logarithmic contribution that is well approximated by the leading order terms in $\det M_T^\dagger M_T$ and $\det M_N^\dagger M_N$. As argued in Appendix \ref{app:scalar} the higher order contributions are suppressed by corresponding powers of $(\tr M_T^\dagger M_T)^3$, and are thus proportional to a function of the Yukawa couplings that has a numerically negligible global maximum. The leading contribution to the Higgs effective potential from the color-charged and hidden-sector fermions are thus numerically well approximated by a simple $\delta V_F \propto \pm \cos 2 \theta$ dependence, and contribute with opposite sign. Analytical expressions for the coefficients of this potential are computed for general values of the Yukawa couplings $y_i$ and $\tilde{y}_i$ in Appendix \ref{app:scalar}, as well as higher order corrections. In the simplified parameter space, the fermionic contribution to the scalar effective potential may be expressed to leading order as

\begin{align}
  \delta V_F (\theta) &\approx f^2 \left( f_T^2 - f_N^2 \right) \cos 2 \theta
\end{align}
\be
f_T^2 = \frac{3 f^2}{16 \pi^2} \frac{y_1^2 y_2^2 y_3^2}{y_2^2  - y_3^2} \log \frac{y_1^2 + y_2^2}{y_1^2 + y_3^2} \qquad \qquad f_N^2 = \frac{3 f^2}{16 \pi^2} \tilde{y}_1^4
\ee

\noindent A realistic pattern of EWSB in this model requires the introduction of a $U(1)_a$ violating spurion analogous to the operator of Equation \ref{eq:llspurion}. This spurion descends naturally from fermion mass terms in the UV complete theory, and gives a contribution to the Higgs potential that is periodic in $\theta \mod 2 \pi$

\begin{align}
  \delta V_0 (m_0, \theta) &= - f^3 \tr[ M_0 \Sigma ] + {\rm h.c.} \\
  &= -f^2 f_0^2 \cos \theta \\
  f_0^2 &= 4 f m_0
\label{eq:spurion}
\end{align}

\noindent In the absence of additional matter content at the scale $f$, the leading contributions to the scalar effective potential are fixed by the Yukawa coupings $y_i$ and $\tilde{y}_i$, the 1-loop gauge coefficient $c$, and spurion coefficient $m_0$. This potential has a generic EWSB vacuum when $f_T^2 - f_N^2 - f_G^2 \gg f_0^2 > 0$ and the vacuum expectation value of the order parameter $\theta$ can be expressed as a function of these scales.

\be
  V_{\rm H} (\theta) = - f^2 f_0^2 \cos \theta + f^2 \left( f_T^2 - f_N^2 - f_G^2 \right) \cos 2 \theta
\label{eq:scalarV}
\ee
\be
\langle \theta^2 \rangle = \frac{\langle H^\dagger H \rangle}{2 f^2} = 6 \left[ \frac{4 \left( f_T^2 - f_N^2 - f_G^2 \right) - f_0^2}{16 \left( f_T^2 - f_N^2 - f_G^2 \right) - f_0^2} \right]
\ee

\noindent Fixing $v^2 = 2 f^2 \langle \theta^2 \rangle = (246 \text{ GeV})^2$ and expanding about this symmetry breaking vacuum, we find that the coefficients of the one-loop corrections are fixed by the measured value of the physical Higgs boson $m_h = 126$ GeV.

\begin{align}
f_T^2 - f_N^2 - f_G^2&= \frac{m_h^2}{12} \left( \frac{1}{2} \frac{1}{\langle \theta^2 \rangle} - 1 \right) \\[5pt]
f_0^2 &= \frac{4 m_h^2}{3} \left( \frac{3}{2} \frac{1}{\langle \theta^2 \rangle} - 1 \right) 
\label{eq:higgs}
\end{align}

In this framework the mass of the gauge-singlet pseudoscalar $\eta$ takes a large contribution from the spurion coefficient $m_0$ via Equation \ref{eq:higgs}. At leading order in $\theta$, the size of this coefficient is fixed by the Higgs mass to be $m_0 \sim f m_h^2 / v^2$ resulting in a heavy pseudoscalar $m_\eta \sim \mathcal{O}({\rm TeV})$. The pseudoscalar mass can be made into a free parameter by introducing an additional symmetry breaking spurion that is proportional to an alternative $Sp(4)$ preserving background field $\widetilde{M}_0 \equiv \tilde{m}_0 \widetilde{\Sigma}_0$ 
\begin{align}
  \delta \widetilde{V}_0 (\tilde{m}_0, \alpha) &= - f^2 \tr[ \widetilde{M}_0 \Sigma ]^2 + {\rm h.c.} \\
  &= 8 f^2 \tilde{m}_0^2 \, \eta^2 \, \frac{\sin^2 \alpha}{\alpha^2}
 \label{eq:spurion2}
\end{align}
\be
\widetilde{\Sigma}_0 = 
 \begin{pmatrix} 
    i\sigma_2 & \quad \\ \quad  & -i\sigma_2 \\ 
 \end{pmatrix} \qquad \alpha =\sqrt{\frac{\eta^2 + H^2}{2 f^2}}
\ee 
The spurion given by Equation \ref{eq:spurion2} would descend naturally from a four-fermion interaction in the UV completion at the compositeness scale $\Lambda$. This term provides a contribution to the potential for $\eta$ only. In the absence of additional sources of symmetry breaking terms, the pseudoscalar mass is given as a function of the spurion coefficients $m_0$ and $\tilde{m}_0$
\be
m_\eta^2 \sim 4 f^2 \left( \frac{m_0}{f} - \frac{8 \tilde{m}_0^2}{f^2} \right)
\ee

\subsection{Phenomenology}
\label{sec:SUSPpheno}

In the $SU(4)/Sp(4)$ model, precision electroweak constraints push the masses of the BSM fermions to very high values, putting them firmly beyond the reach of near-term LHC limits. These constraints can be relaxed by introducing additional vector-like partners in a way that suppresses mixing between the SM bottom quark and the heavy singlet. However the branching fractions of the lightest BSM states to the singlet $\eta$ are generically small. The lightest color-charged states in such models would thus be virtually indistinguishable from the top and bottom partners of simplified models ~\cite{Khachatryan:2015gza,Sirunyan:2017usq} and their phenomenology will not be discussed further here. The phenomenological signatures of of the hidden-sector in this model are similar to those described in Section \ref{sssec:llhidden} and will also not be reiterated here. In the absence of kinematically accessible BSM fermions, the production of new particles is limited to weak production of the goldstone mode $\eta$, which can be singly produced via gluon fusion, or pair produced via Higgs decay $h \rightarrow \eta \eta$ if it is sufficiently light. The large tree-level Yukawa couplings in this model result in ${\rm BR}(\eta \shortrightarrow t \bar{t}) \sim 1$ when $m_\eta > 2 m_t$ and ${\rm BR}(\eta \shortrightarrow b \bar{b}) \sim 1$ when $m_\eta < 2 m_t$.

\begin{figure}
\begin{minipage}{.4\textwidth}
\includegraphics[trim={0cm 0cm 0cm 0cm},clip,width=1.0\linewidth]{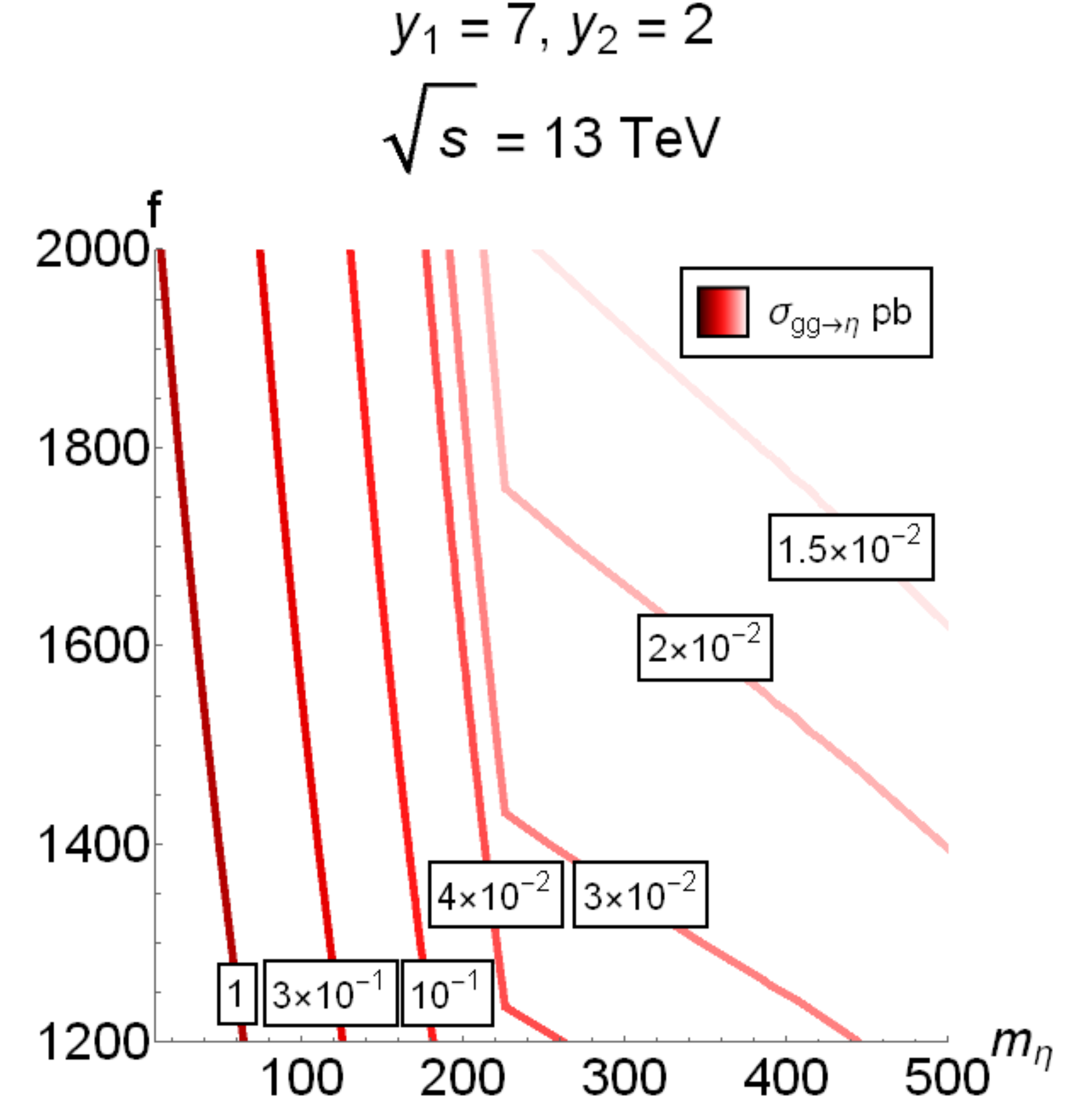}
\end{minipage}
\hspace{1.5cm}
\begin{minipage}{.4\textwidth}
\includegraphics[trim={0cm 0cm 0cm 0cm},clip,width=1.0\linewidth]{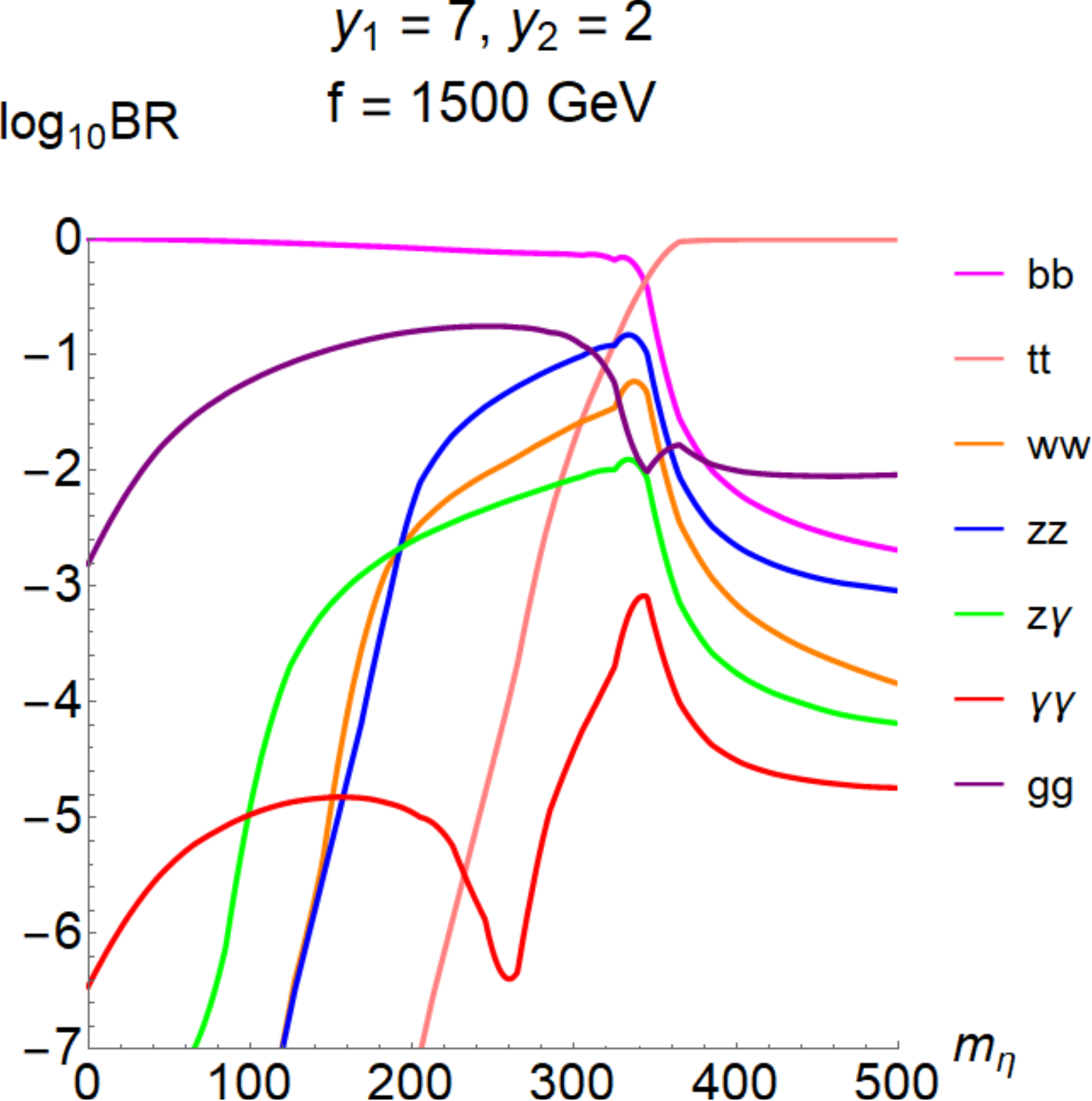}
\end{minipage}
\caption{Cross sections for single pseudo-scalar production at 13 TeV (left) and branching fractions as a function of the pseudo-scalar mass $m_\eta$ (right)}
\label{fig:etabranch}
\end{figure}

In the high-mass regime the discovery of such a state would be extremely challenging due to the low production cross section and the interference effects of its two-body decays, as described in Section \ref{sec:LLexp}. The loop-induced interactions of the $\eta$ boson are roughly independent of the Yukawa-couplings due to the dominant invariant contribution from the top-quark loop. The cross sections and branching fractions of the $\eta$ boson are shown in Figure \ref{fig:etabranch} for a generic value of the Yukawa couplings. The pseudo-scalar $\eta$ only has tree level couplings to the third generation quarks and their BSM partners. Its branching fractions are thus dominated by decays to top and bottom quark pairs whenever these channels are kinematically open. However there will also be effective couplings to pairs of gauge bosons from dimension-5 operators, which arise via 1-loop diagrams analogous to gluon fusion. The relevant dimension-5 interactions are given in Equation \ref{eq:dim5op}. The couplings $\tilde{c}^\eta_{V_1 V_2}$ have been computed in FeynCalc~\cite{Shtabovenko:2016sxi}, and the partial widths to various pairs of gauge bosons are given by the expression in Equation \ref{eq:widths}.

\begin{eqnarray}
\mathcal{L} &\supset& \frac{1}{4} \eta \left( \tilde{c}^\eta_{w w} \,W_{\mu\nu}\tilde{W}^{\mu\nu} + \tilde{c}^\eta_{z z} \,Z_{\mu\nu}\tilde{Z}^{\mu\nu} + \tilde{c}^\eta_{z \gamma} \,Z_{\mu\nu}\tilde{F}^{\mu\nu} + \tilde{c}^\eta_{\gamma \gamma} \,F_{\mu\nu}\tilde{F}^{\mu\nu} +  \tilde{c}^\eta_{g g} \,G_{\mu\nu}\tilde{G}^{\mu\nu} \right)
\label{eq:dim5op}
\end{eqnarray}
\be
\Gamma (\eta \rightarrow V_1 V_2) = \frac{1}{16 \pi m_\eta} \left| \tilde{c}^{\eta}_{V_1 V_2} \right|^2 \left[ \lambda (m_\eta^2, m_{V_1}^2, m_{V_2}^2) + 6 m_{V_1}^2 m_{V_2}^2 \right]
\label{eq:widths}
\ee

\subsubsection{Rare Higgs Decays}

\begin{figure}
\begin{minipage}{.35\textwidth}
\includegraphics[trim={0cm 0cm 0cm 0cm},clip,width=1.0\linewidth]{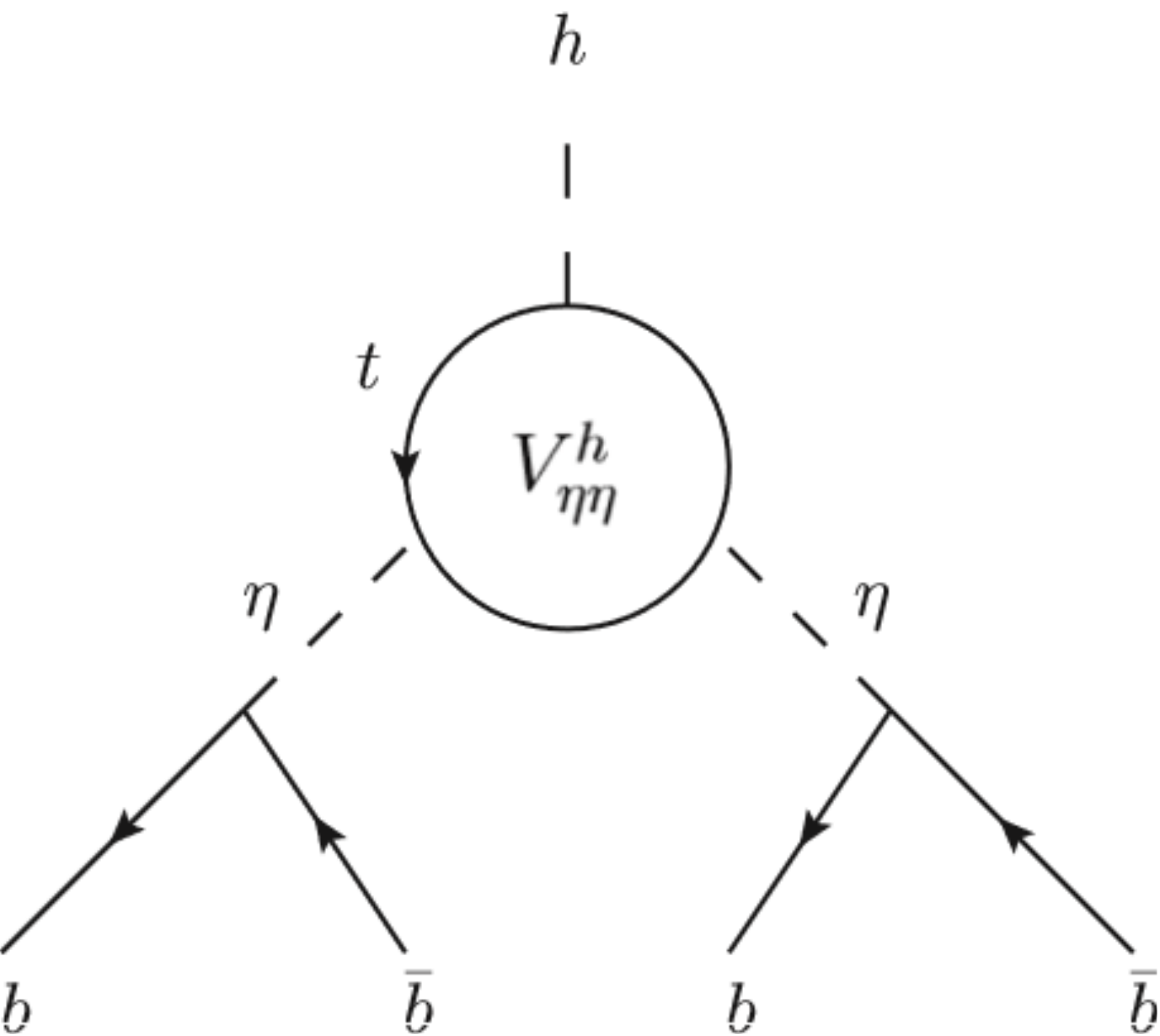}
\end{minipage}
\caption{Feynman diagram for $h \rightarrow (b \bar{b}) (b \bar{b})$ via the rare Higgs decay mode $h \rightarrow \eta \eta$}
\label{diag:Haa}
\end{figure}

If the pseudo-scalar mass is comparable to that of the Higgs boson then its dominant branching fraction to bottom quarks would make it vulnerable to sophisticated searches for boosted $b \bar{b}$ pairs such as those recently performed by CMS \cite{Sirunyan:2017dgc,Vernieri:2017jqy}. However these techniques are unable to resolve boosted objects with low invariant masses $m_\eta < 50$ GeV due to overwhelming multijet backgrounds \cite{CMS:2018qbg}. In this regime, the most viable kinematic pathway for resolving the pseudoscalar $\eta$ would be via its interactions with the Higgs boson, which can decay to pairs of $\eta$ bosons via loop induced couplings as shown in Figure \ref{diag:Haa}. The branching ratio ${\rm BR}(h \rightarrow \eta \eta)$ is thus loop-suppressed and such decays would likely be impossible to resolve at a future hadron collider such as the HL-LHC, which will constrain the decay modes of the Higgs at a relative precision not exceeding $\mathcal{O}(10 \%)$ \cite{Dawson:2013bba,CMS:2013xfa,CMS-DP-2016-064,ATL-PHYS-PUB-2013-014,ATL-PHYS-PUB-2014-016}.

\begin{figure}
\begin{minipage}{.45\textwidth}
\includegraphics[trim={0cm 0cm 0cm 0cm},clip,width=1.0\linewidth]{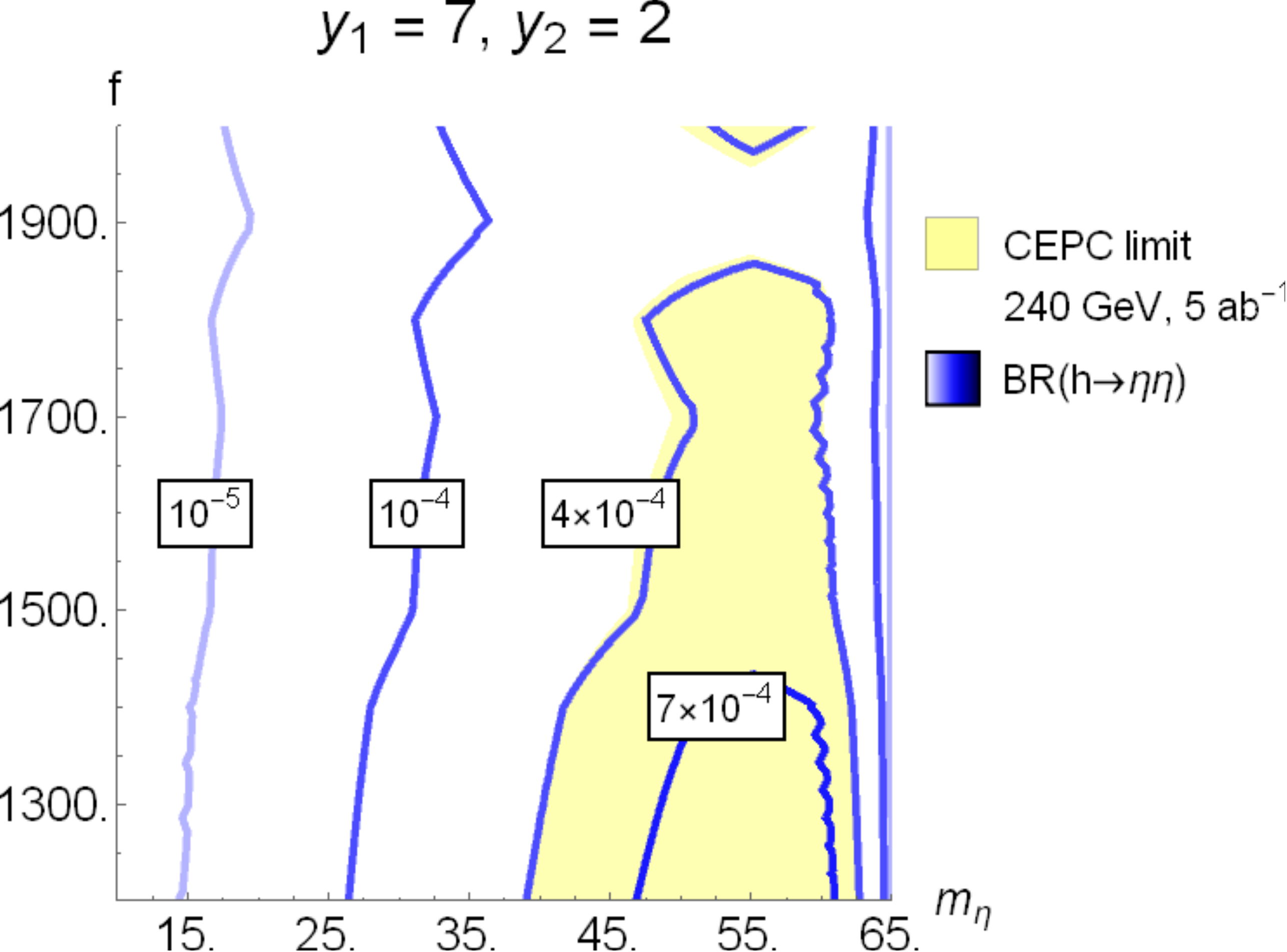}
\end{minipage}
\hspace{1cm}
\begin{minipage}{.45\textwidth}
\includegraphics[trim={0cm 0cm 0cm 0cm},clip,width=1.0\linewidth]{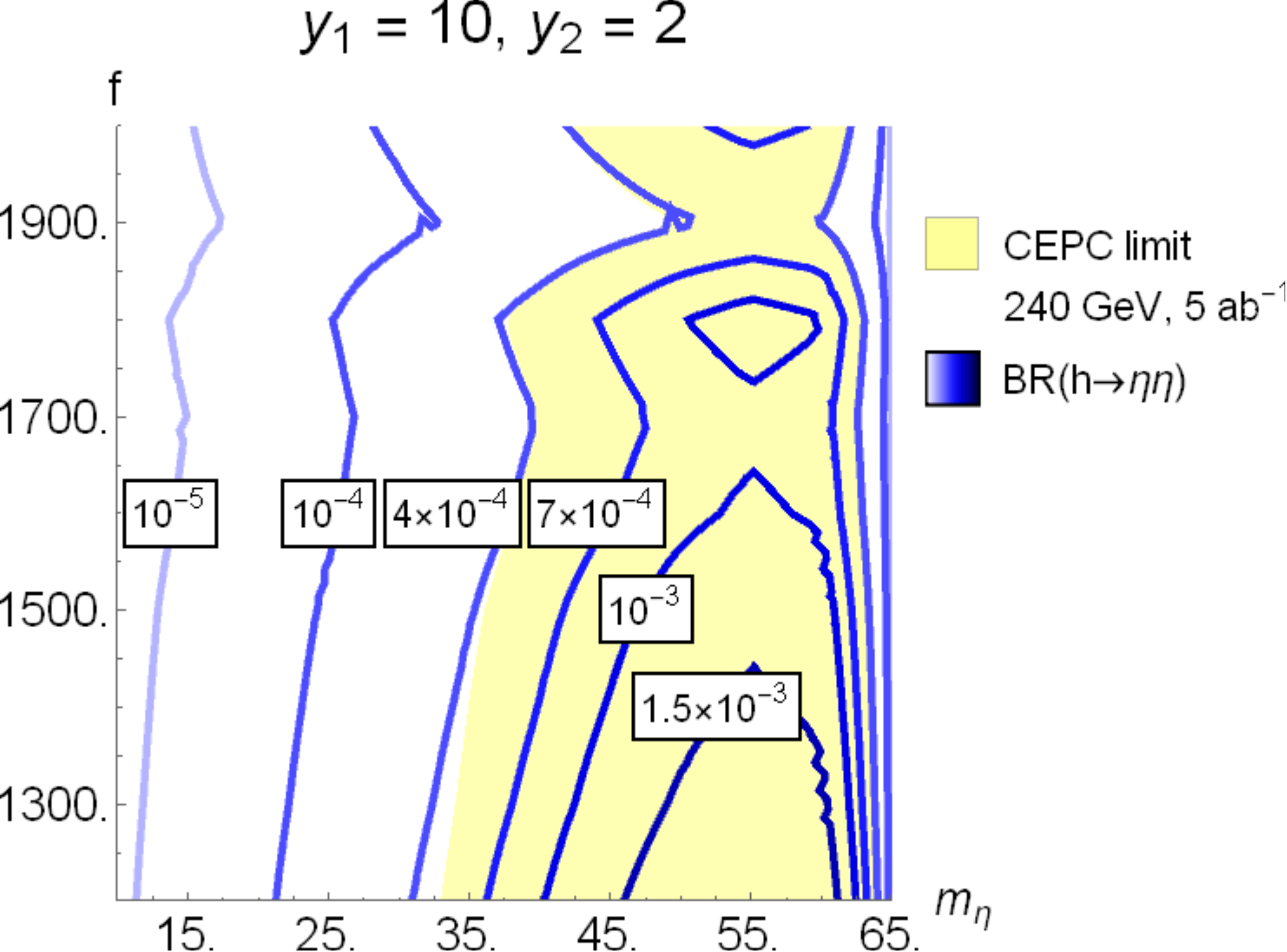}
\end{minipage}
\caption{Branching ratios for rare decays of the Higgs boson to pairs of pseudoscalars. The expected reach at CEPC from rare Higgs decays $h \rightarrow (b \bar{b}) (b \bar{b})$ are also shown at $5 \text{ ab}^{-1}$.}
\label{fig:cepc}
\end{figure}

The presence of such a low-mass goldstone mode provides additional motivation for precision measurements at a future Higgs factory. One existing proposal for such a machine is the Circular Electron-Positron Collider (CEPC), which is expected to measure various couplings of the Higgs boson at a relative precision of $\mathcal{O}(0.1 - 1 \%)$ \cite{CEPC-SPPCStudyGroup:2015csa,Fujii:2015jha,Gomez-Ceballos:2013zzn}. The partial width for the rare Higgs decay $h \rightarrow \eta \eta$ is an increasing function of the Yukawa couplings $y_i$ and is saturated at the kinematic limit, as shown in Figure \ref{fig:cepc}. The branching fraction for this process lies in the range of ${\rm BR}(h \shortrightarrow \eta \eta) \sim 10^{-5} - 10^{-3}$ throughout the parameter space of this model. The dominant branching fraction of the pseudoscalar to $b \bar{b}$ pairs would thus lead to a rare $h \rightarrow (b \bar{b}) (b \bar{b})$ signal that could be effectively probed at an $e^+ e^-$ collider. An estimate of the expected reach for this process at the CEPC has been computed at $5 \text{ ab}^{-1}$ \cite{Liu:2016zki}, and we find that translating these limits puts a large part of this parameter space within reach of a long-term run at the CEPC, as shown in Figure \ref{fig:cepc}. 

\newpage


 \section{Conclusions and Outlook}
\label{sec:con}

A  pseudo Nambu-Goldstone Higgs with a top coupling arising from mixing with top partners offers the theoretically compelling possibility of a natural and calculable model for electroweak symmetry breaking. We have considered two of the most economical scenarios, expanded to include a Higgs coupling to a hidden fermion sector. The addition of a hidden sector coupled to the pseudo Nambu-Goldstone bosons allows for a relatively simple way to obtain a Higgs potential that is consistent with the experimentally measured masses of Higgs and weak gauge bosons, as well as a possible dark matter candidate. We have performed analyses on the most relevant collider signatures of the   new  fermions and new spinless particles, and find that for most of the viable parameter space a new particle discovery will be challenging at the LHC. Given low energy precision electroweak constraints, the lack of an LHC discovery of any non-standard model particles to date would thus be a general expectation for both of these composite Higgs models, though for different reasons. Nonetheless our analysis suggests several options for probing these nonstandard phenomena at the LHC and future colliders.

In the $SU(5) / SO(5)$ model these challenges arise due to the serendipitous subtlety of the final state signatures which arise from electroweak interactions that are preferential to the third generation of quarks. The backgrounds from QCD processes generically overwhelm or interfere with the most obvious processes that could be used to resolve new particle states beyond the Standard Model. However the high multiplicity of new BSM states will generally aggregate to produce a measurable deviation in the production of final states with a high multiplicity of third generation quarks in the high-luminosity limit. The presence of a confining hidden-sector gauge group could also lead to a potentially measurable branching fraction of the Higgs boson to displaced pairs of bottom quarks, thus providing additional motivation for experiments focused on the long-lived particle frontier, such as the proposed MATHUSLA experiment. In the $SU(4) / Sp(4)$ model, the experimental challenges arise primarily due to the stark absence of new particle states involved with the EWSB dynamics. The goldstone sector of this theory consists a single spin-0 gauge-singlet state, and precision electroweak constraints imply a mass range for new strongly interacting particles that is likely well beyond the reach of a $14 \, {\rm TeV}$ hadron collider. However the Higgs boson generically has a decay rate to the gauge-singlet goldstone mode that could be resolved at a future Higgs factory such as the CEPC. 

\section*{Acknowledgments}
The work of A.E.N. is partially supported by the DOE under grant  DE-SC0011637 and by the Kenneth K. Young Memorial Endowed Chair. The work of D.G.E.W. was partially supported by Burke faculty fellowship.  We would like to thank Calibourne D.~Smith and Scott Thomas for helpful conversations. A.E.N. acknowledges the hospitality of the Aspen Center for Physics, which is supported by National Science Foundation grant PHY-1607611.  We thank the Galileo Galilei Institute for Theoretical Physics for the hospitality and the INFN for partial support during the completion of this work. This work was partially supported by a grant from the Simons Foundation (341344, LA).

\newpage
 
\appendix

\section{Fermion Contribution to $SU(4) / Sp(4)$ Higgs Potential}
\label{app:scalar}

In the $SU(4) / Sp(4)$ model, the top-sector mass matrix for the color-charged quarks $M_T$ takes the form given in Table \ref{tab:appqmass}. The mass matrices can be expressed in terms of a single angle $\theta \equiv v / \sqrt{2} f$, representing the magnitude of the Higgs VEV relative to the chiral symmery breaking scale.
\begin{center}
  \begin{table}[ht]
    \begin{tabular}{| c | c c c |}
      \hline
      & $~~\overline{X}_0~~$ & $~~\overline{N}~~$ & $~~\bar{n}~~$ \\ [0.5ex] 
      \hline
      $X_0$  & $\tilde{y}_1 f \cos \theta$ & $-i \tilde{y}_1 f \sin \theta$ & $0$ \\
      $N$  & $-i \tilde{y}_1 f \sin \theta$ & $\tilde{y}_1 f \cos \theta$ & $\tilde{y}_3 f$ \\
      $n$  & $0$ & $\tilde{y}_2 f$ & $0$ \\
      \hline
    \end{tabular}
    \caption{Mass matrices for the color-charged (left) and hidden-sector (right) top-like fermions in the gauge eigenbasis.}
    \label{tab:appqmass}
  \end{table}
\end{center}
\vspace{-1cm}
The UV insensitivity of the Yukawa sector provides powerful constraints on the form of the scalar effective potential, and is guaranteed by the following properties of the mass matrix

\be
\partial_{\theta} \tr M_T^\dagger M_T = \partial_{\theta} \tr (M_T^\dagger M_T)^2 = 0 
\label{eq:nodiv2}
\ee

\noindent The fermion masses are periodic with respect to $\theta \mod 2 \pi$. The color-charged mass matrix $M_T$ produces one massless state when $\theta = n \pi$ for $n \in \mathbb{Z}$. For notational simplicity we define the following combinations of Yukawa couplings
\be
y_{\pm}^2=(y_1^2 + y_2^2) \pm (y_1^2 + y_3^2) \qquad \qquad \qquad y_{123}^2 \equiv y_1^2 y_2^2 y_3^2
\label{eq:yukshort}
\ee
In terms of these Yukawa parameters, the traces and determinants of the fermion mass matrices take a simple form
\be
\tr M_T^\dagger M_T = f^2 y_{+}^2 \qquad \qquad \tr (M_T^\dagger M_T)^2 = \frac{f^2}{2} ( y_{+}^4 + y_{-}^4 )
\ee
\be 
\det M_T^\dagger M_T = f^6 y_{123}^2 \sin^2 \theta
\label{eq:detM}
\ee
These expressions allow for a direct computation of the one-loop fermion corrections order-by-order in the Higgs VEV. We thus expand the fermion masses as a power series in $\sin^2 \theta$

\begin{align}
m_t^2 &= \displaystyle\sum_{n=0}^{\infty} m_{n}^2 \sin^{2 n} \theta \qquad \qquad m_0^2 = 0 \\
m_{\tp}^2 &= \displaystyle\sum_{n=0}^{\infty} m^{\prime\,2}_{n} \sin^{2 n} \theta \qquad \qquad m^{\prime 2}_0 = \frac{f^2}{2} ( y_{+}^2 + y_{-}^2 ) \\
m_{\tpp}^2 &= \displaystyle\sum_{n=0}^{\infty} m^{\prime \prime\,2}_{n} \sin^{2 n} \theta \qquad \qquad m^{\prime \prime 2}_0 = \frac{f^2}{2} ( y_{+}^2 - y_{-}^2 )
\end{align}

\begin{figure}
\begin{minipage}{.45\textwidth}
\includegraphics[trim={0cm 0cm 0cm 0cm},clip,width=1.0\linewidth]{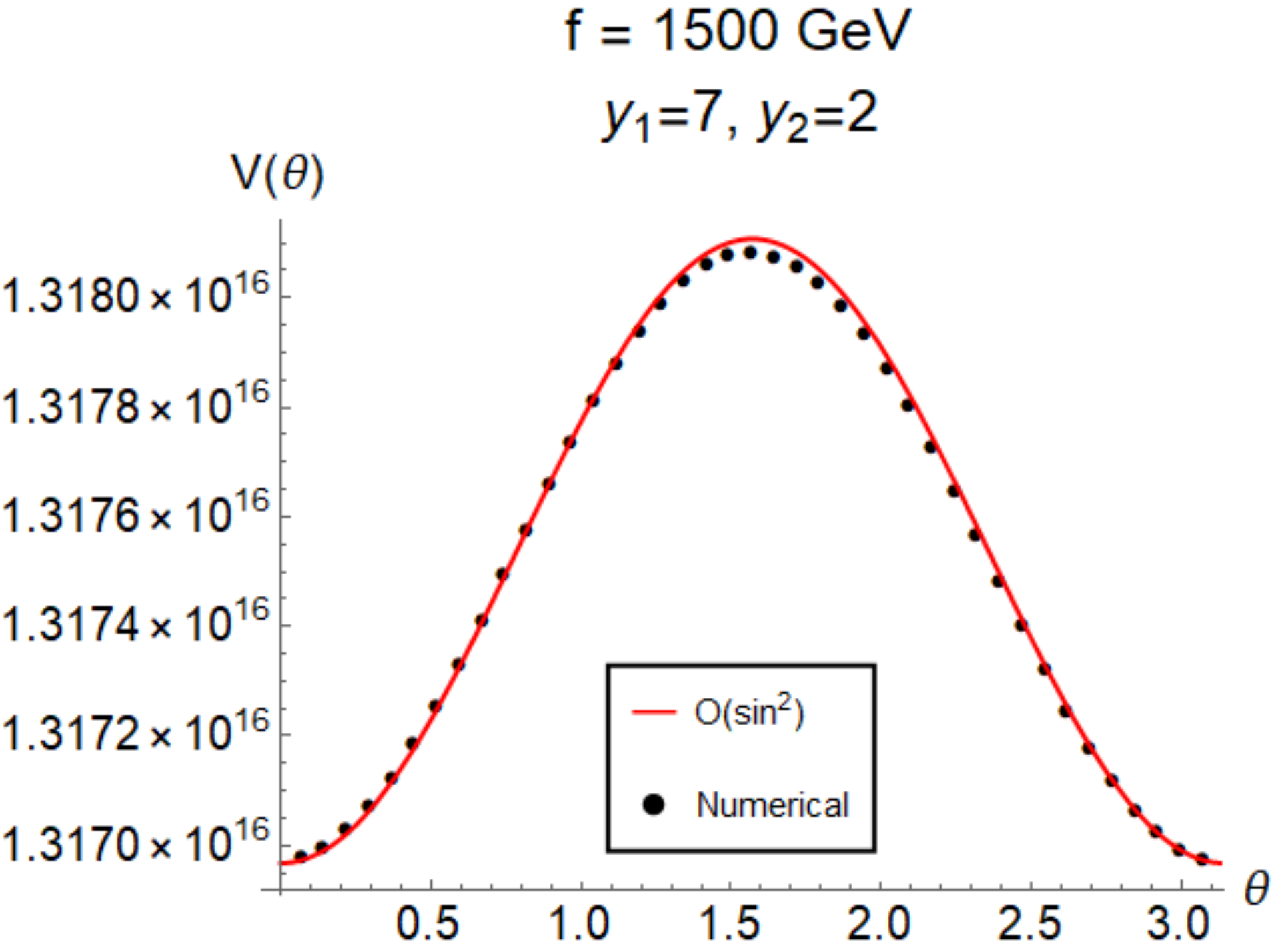}
\end{minipage}
\hspace{1cm}
\begin{minipage}{.45\textwidth}
\includegraphics[trim={0cm 0cm 0cm 0cm},clip,width=1.0\linewidth]{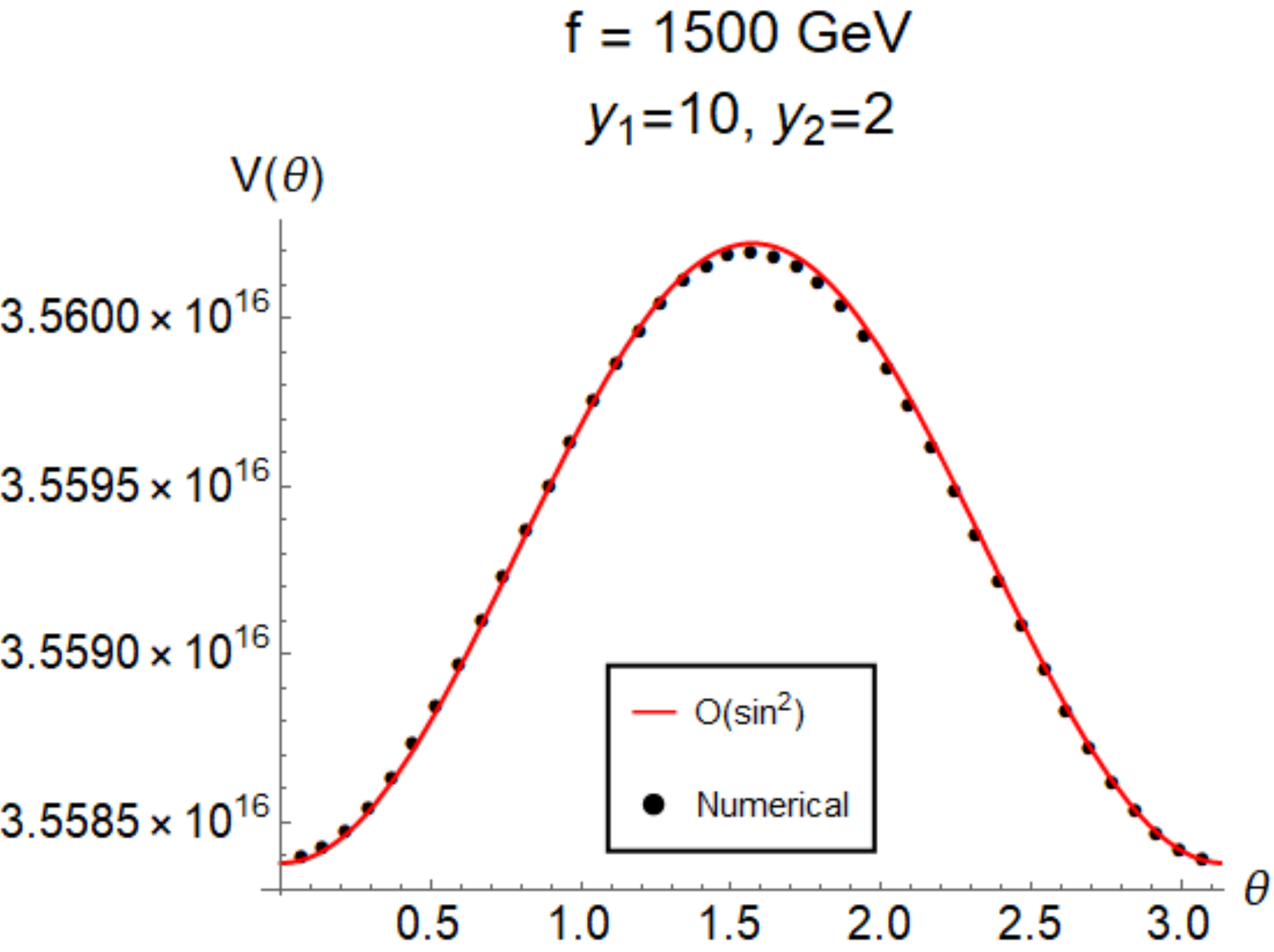}
\end{minipage}
\caption{Fermionic contribution to the Higgs potential from the color-charge top-like sector. The numerical result is shown in black dots while the $\mathcal{O} (\sin^2 \theta)$ approximation is shown in red.}
\label{fig:su4sp4numerical}
\end{figure}

Computing to fourth order in $\sin \theta$, we find that the color-charged fermion mass eigenstates take the following form
\begin{align}
\frac{m_t^2}{f^2} &= \frac{4 y_{123}^2}{y_{+}^4 - y_{-}^4} \sin^2 \theta - \frac{64 y_{123}^4 y_{+}^2}{(y_{+}^4 - y_{-}^4)^3} \sin^4 \theta \\[2ex]
\frac{m_{\tp}^2}{f^2} &= \frac{y_{+}^2 - y_{-}^2}{2} - \frac{2 y_{123}^2}{y_{-}^2 (y_{+}^2 - y_{-}^2)} \sin^2 \theta + \frac{4 y_{123}^4 (y_{+}^2 - 3 y_{-}^2)}{y_{-}^6 (y_{+}^2 - y_{-}^2)^3} \sin^4 \theta \\[2ex]
\frac{m_{\tpp}^2}{f^2} &= \frac{y_{+}^2 + y_{-}^2}{2} + \frac{2 y_{123}^2}{y_{-}^2 (y_{+}^2 + y_{-}^2)} \sin^2 \theta - \frac{4 y_{123}^4 (y_{+}^2 + 3 y_{-}^2)}{y_{-}^6 (y_{+}^2 + y_{-}^2)^3} \sin^4 \theta
\end{align}
The UV insensitivity of the Yukawa sector additionally guarantees that their contributions to the scalar effective potential can be expressed in terms of ratios of mass eigenstates. Computing to fourth order in $\sin \theta$, we find that the top-sector contribution to the Higgs effective potential takes the following form
\begin{align}
 \delta V_T(y_i, \theta) &= - \frac{3 }{16 \pi^2} \tr(M_T^\dagger\!M_T)^2 \log \frac{M_T^\dagger\!M_T}{\Lambda^2} \\[1.0ex]
  &= -\frac{3}{16 \pi^2} \displaystyle\sum_{\tpp \tp t} |m_q^2 (y_i, \theta)|^2 \log |m_q^2(y_i, \theta)| \\[1.0ex]
  &\approx C_0 + C_2 \sin^2 \theta + C_4 \sin^4 \theta 
\end{align}	
The coefficients of the $\theta$ dependent terms may be expressed as
\begin{align}
  C_2 &= -\frac{3 f^4}{16 \pi^2} \frac{2 y_{123}^2}{y_{-}^2} \log \frac{y_{+}^2 - y_{-}^2}{y_{+}^2 + y_{-}^2} \\[1.0ex]
  C_4 &= -\frac{3 f^4}{16 \pi^2} \frac{4 y_{123}^4}{(y_{+}^2 - y_{-}^2)^2} \Bigg( \frac{2 y_{+}^2}{y_{-}^2} - \frac{y_{+}^6 - 3 y_{-}^4 y_{+}^2 -  6 y_{-}^6}{3 y_{-}^6} \log \frac{8 y_{123}^2}{(y_{+}^2 - y_{-}^2)^3} \\
  &\hspace{5cm} + \frac{y_{+}^6 - 3 y_{-}^4 y_{+}^2 +  6 y_{-}^6}{3 y_{-}^6} \log \frac{8 y_{123}^2}{(y_{+}^2 + y_{-}^2)^3}  \Bigg)
\end{align}	
The collective symmetry breaking properties of $M_T$ guarantees that the potential is proportional to $\det M_T^\dagger M_T$. Higher order contributions are thus suppressed by powers of $y_{123}^2 / y_{+}^6$, which is a function of the Yukawa couplings that has a global maximum at $1/54$
\be
\frac{C_4}{C_2} \sim \frac{y_{123}^2}{y_{+}^6} \left( 1 + \mathcal{O} \left( \frac{y_{-}^4}{y_{+}^4} \right) \right) \qquad \qquad \qquad \left| \frac{y_{123}^2}{y_{+}^6} \right| \leq \frac{1}{54}
\ee
For the color-charged contribution we thus find that the potential is numerically well approximated by the second order approximation in $\sin \theta$, which has been plotted against the numerical result in Figure \ref{fig:su4sp4numerical}. The fermionic contribution to the Higgs potential is thus extremely well approximated by a $\cos 2 \theta$ dependence.

\begin{align}
 \delta V_T(y_i, \theta) &= - 2 f^2 f_T^2 \sin^2 \theta \\[1ex]
 &\sim f^2 f_T^2 \cos 2 \theta \\[2ex]
  f_T^2 &= \frac{3 f^2}{16 \pi^2} \frac{y_{123}^2}{y_{-}^2} \log \frac{y_{+}^2 - y_{-}^2}{y_{+}^2 + y_{-}^2}
\end{align}

The contribution to the Higgs effective potential from hidden-sector fermions follows a similar story to the color-charged case. The neutral hidden-sector fermion mass matrix $M_N$ is given in Table \ref{tab:appqmassdark}, and the insensitivity of the Higgs potential to loops of hidden-sector fermions is guaranteed by Equation \ref{eq:nodiv2dark}

\begin{center}
  \begin{table}[ht]
    \begin{tabular}{| c | c c c |}
      \hline
      & $~~\overline{X}_0~~$ & $~~\overline{N}~~$ & $~~\bar{n}~~$ \\ [0.5ex] 
      \hline
      $X_0$  & $\tilde{y}_1 f \cos \theta$ & $-i \tilde{y}_1 f \sin \theta$ & $0$ \\
      $N$  & $-i \tilde{y}_1 f \sin \theta$ & $\tilde{y}_1 f \cos \theta$ & $\tilde{y}_3 f$ \\
      $n$  & $0$ & $\tilde{y}_2 f$ & $0$ \\
      \hline
    \end{tabular}
    \caption{Mass matrices for the color-charged (left) and hidden-sector (right) top-like fermions in the gauge eigenbasis.}
    \label{tab:appqmassdark}
  \end{table}
\end{center}

\be
\partial_{\theta} \tr M^\dagger_{N} M_{N} = \partial_{\theta} \tr (M^\dagger_{N} M_{N})^2 = 0
\label{eq:nodiv2dark}
\ee

\noindent The mass matrix for the neutral states $M_N$ produces one massless state when $\theta = (n+1/2) \pi$, for $n \in \mathbb{Z}$ which can be seen from the form of the mass determinant.
\be
\tilde{y}_{\pm}^2=(\tilde{y}_1^2 + \tilde{y}_2^2) \pm (\tilde{y}_1^2 + \tilde{y}_3^2) \qquad \qquad \qquad \tilde{y}_{123}^2 \equiv \tilde{y}_1^2 \tilde{y}_2^2 \tilde{y}_3^2
\ee
\be
\det ( M_N^\dagger M_N ) = f^6 \tilde{y}_{123}^2 \cos^2 \theta
\label{eq:darkdetM}
\ee
The hidden-sector fermion mass eigenstates thus take a simple form when expressed as an expansion in $\cos^2 \theta$	
\begin{align}
\frac{m_n^2}{f^2} &= \frac{4 \tilde{y}_{123}^2}{\tilde{y}_{+}^4 - \tilde{y}_{-}^4} \cos^2 \theta - \frac{64 \tilde{y}_{123}^4 \tilde{y}_{+}^2}{(\tilde{y}_{+}^4 - \tilde{y}_{-}^4)^3} \cos^4 \theta \\[2ex]
\frac{m_{\np}^2}{f^2} &= \frac{\tilde{y}_{+}^2 - \tilde{y}_{-}^2}{2} - \frac{2 \tilde{y}_{123}^2}{\tilde{y}_{-}^2 (\tilde{y}_{+}^2 - \tilde{y}_{-}^2)} \cos^2 \theta + \frac{4 \tilde{y}_{123}^4 (\tilde{y}_{+}^2 - 3 \tilde{y}_{-}^2)}{\tilde{y}_{-}^6 (\tilde{y}_{+}^2 - \tilde{y}_{-}^2)^3} \cos^4 \theta \\[2ex]
\frac{m_{\npp}^2}{f^2} &= \frac{\tilde{y}_{+}^2 + \tilde{y}_{-}^2}{2} + \frac{2 \tilde{y}_{123}^2}{\tilde{y}_{-}^2 (\tilde{y}_{+}^2 + \tilde{y}_{-}^2)} \cos^2 \theta - \frac{4 \tilde{y}_{123}^4 (\tilde{y}_{+}^2 + 3 \tilde{y}_{-}^2)}{\tilde{y}_{-}^6 (\tilde{y}_{+}^2 + \tilde{y}_{-}^2)^3} \cos^4 \theta
\end{align}
Similarly for the hidden-sector fermions, we find that the contribution is well approximated by a $\cos 2 \theta$ dependence, but with an opposite sign that descends from the form of the determinants in Equation \ref{eq:darkdetM}
\begin{align}
  \delta V_N(\tilde{y}_i, \theta) &= - 2 f^2 f_N^2 \cos^2 \theta \\[1ex]
  &\sim -f^2 f_N^2 \cos 2 \theta \\[2ex]
  f^2_N &= \frac{3 f^2}{16 \pi^2} \frac{\tilde{y}_{123}^2}{\tilde{y}_{-}^2} \log \frac{\tilde{y}_{+}^2 - \tilde{y}_{-}^2}{\tilde{y}_{+}^2 + \tilde{y}_{-}^2}
\end{align}

\newpage

\section{Fermion Contribution to $SU(5) / SO(5)$ Higgs Potential}
\label{app:llscalar}

\begin{figure}
\begin{minipage}{.45\textwidth}
\includegraphics[trim={0cm 0cm 0cm 0cm},clip,width=1.0\linewidth]{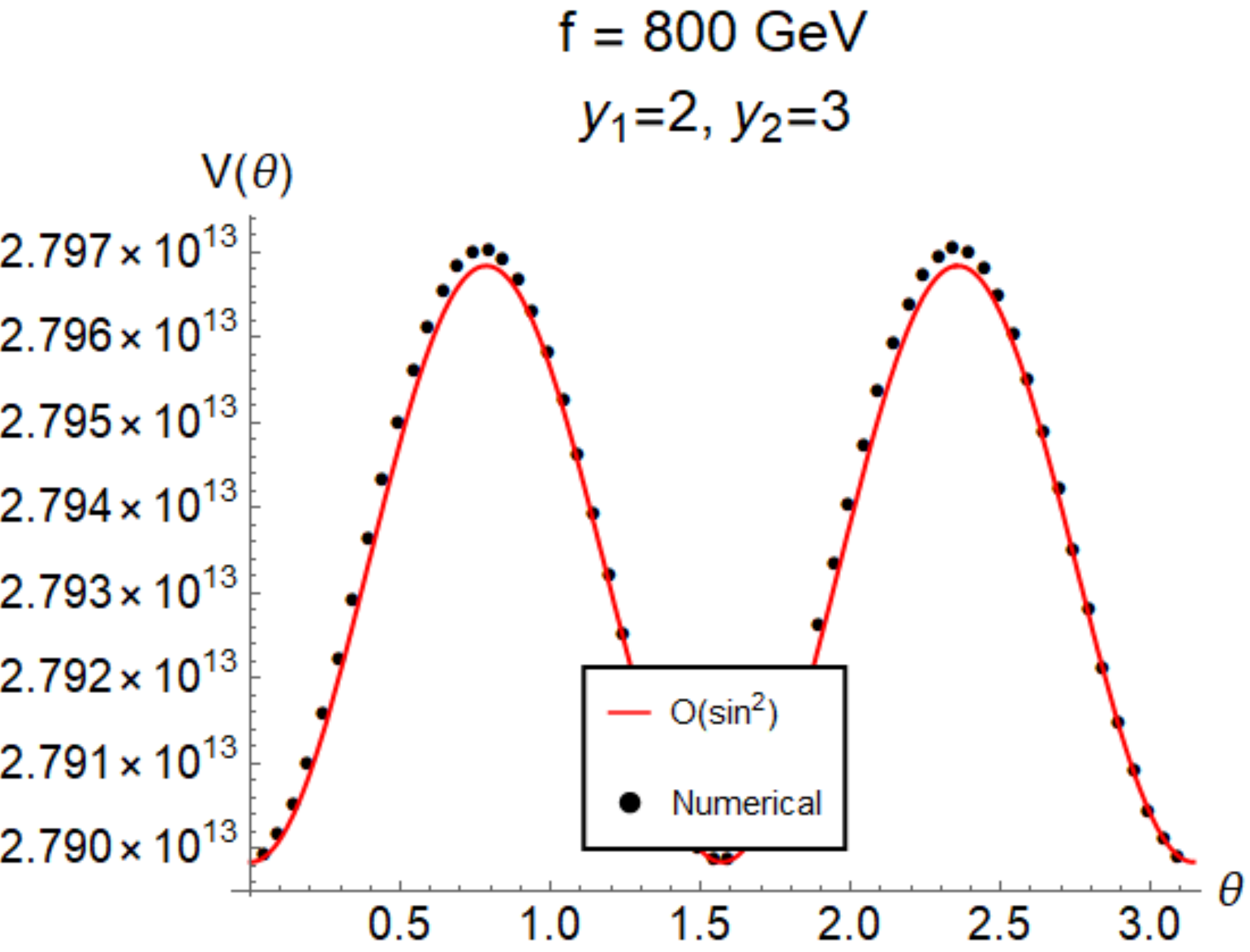}
\end{minipage}
\hspace{1cm}
\begin{minipage}{.45\textwidth}
\includegraphics[trim={0cm 0cm 0cm 0cm},clip,width=1.0\linewidth]{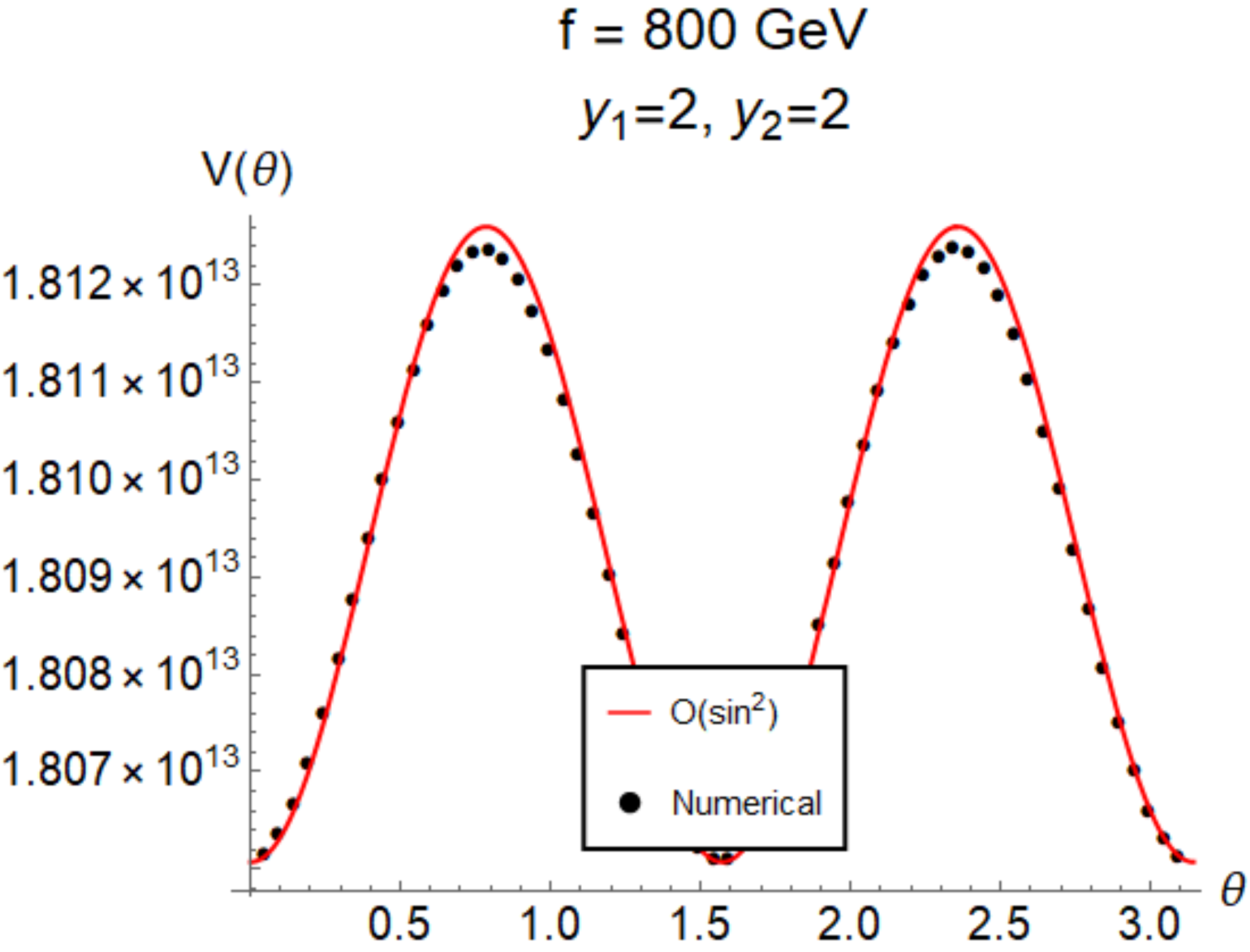}
\end{minipage}
\caption{Fermionic contribution to the Higgs potential from the color-charge top-like sector. The numerical result is shown in black dots while the $\mathcal{O} (\sin^2 2 \theta)$ approximation is shown in red.}
\label{fig:su5so5numerical}
\end{figure}

In the $SU(5) / SO(5)$ model, the top-sector mass matrix for the color-charged quarks $M_T$ and the hidden sector fermions $\widetilde{M}_T$ takes the form given in Table \ref{tab:llappqmass}. The mass matrices can be expressed in terms of a single angle $\theta \equiv v / 2 f$, representing the magnitude of the Higgs VEV relative to the chiral symmery breaking scale.
\begin{center}
  \begin{table}[ht]
    \begin{tabular}{| c | c c c c |}
      \hline
      & $~~\overline{T}~~$ & $~~\overline{Q}_T~~$ & $~~\overline{P}_T~~$ & $~~\bar{t}~~$ \\ [0.5ex] 
      \hline
      $T$  & $y_1 f c_{2 \theta}$ & $\frac{i}{\sqrt{2}} y_1 f s_{2 \theta}$ & $\frac{i}{\sqrt{2}} y_1 f s_{2 \theta}$ & $y_3 f$ \\
      $Q_T$  & $\frac{i}{\sqrt{2}} y_1 f s_{2 \theta}$ & $y_1 f c^2_{\theta}$ & $-y_1 f s^2_{\theta}$ & $0$ \\
      $P_T$  & $\frac{i}{\sqrt{2}} y_1 f s_{2 \theta}$ & $-y_1 f s^2_{\theta}$ & $y_1 f c^2_{\theta}$ & $0$ \\
      $q_T$  & $0$ & $y_2 f$ & $0$ & $0$ \\
      \hline
    \end{tabular} \qquad \qquad   \begin{tabular}{| c | c c c c |}
      \hline
      & $~~\overline{N}~~$ & $~~\overline{X}_0~~$ & $~~\overline{Y}_0~~$ & $~~\bar{n}~~$ \\ [0.5ex] 
      \hline
      $N$  & $\tilde{y}_1 f c_{2 \theta}$ & $\frac{i}{\sqrt{2}} \tilde{y}_1 f s_{2 \theta}$ & $\frac{i}{\sqrt{2}} \tilde{y}_1 f s_{2 \theta}$ & $\tilde{y}_3 f$ \\
      $X_0$  & $\frac{i}{\sqrt{2}} \tilde{y}_1 f s_{2 \theta}$ & $\tilde{y}_1 f c^2_{\theta}$ & $-\tilde{y}_1 f s^2_{\theta}$ & 0 \\
      $Y_0$  & $\frac{i}{\sqrt{2}} \tilde{y}_1 f s_{2 \theta}$ & $-\tilde{y}_1 f s^2_{\theta}$ & $\tilde{y}_1 f c^2_{\theta}$ & $0$ \\
      $n$ & $\tilde{y}_2 f$ & $0$ & $0$ & $0$ \\
      \hline
    \end{tabular}
    \caption{Mass matrix for the top-sector fermions $M_T$ in the gauge eigenbasis. Mass matrices for the color-charged (left) and hidden-sector (right) top-like fermions.}
    \label{tab:llappqmass}
  \end{table}
\end{center}
\vspace{-1cm}
The fermion masses are periodic with respect to $\theta \mod \pi$. The determinant of the color-charged mass matrix guarantees the existence of a massless fermion whenever $\theta = n \pi / 2$ for $n \in \mathbb{Z}$, which we may associated with the Standard Model top quark. Similarly the determinant of the hidden-sector mass matrix squared produces one massless state when $\theta = (n+1/2) \pi / 2$
\be
\det M_T^\dagger M_T = f^8 y_1^2 y_{123}^2 \sin^2 2 \theta \qquad \qquad
\det M^\dagger_{N} M_{N} = f^8 \tilde{y}_1^2 \tilde{y}_{123}^2 \cos^2 2 \theta
\ee
Diagonalizing these mass matrices produces four charge $\pm 2/3$ mass eigenstates and four neutral mass eigenstates, which can be computed order by order in $\sin 2 \theta$ using the constraints described in Appendix \ref{app:scalar}. In terms of the Yukawa couplings that were defined in Equation \ref{eq:yukshort}, the eigenvalues of the top-sector mass matrices take the following form to $\mathcal{O} (\sin^2 2 \theta)$.
\be
\begin{aligned}[c]
\frac{m_t^2}{f^2} &= \frac{2 y_{123}^2}{y_{+}^4 - y_{-}^4} \sin^2 2 \theta \\
\frac{m_{\tp}^2}{f^2} &= y_1^2 \\
\frac{m_{\tpp}^2}{f^2} &= \frac{y_{+}^2 - y_{-}^2}{2} - \frac{y_{123}^2}{y_{-}^2 (y_{+}^2 - y_{-}^2)}\sin^2 2 \theta \\
\frac{m_{\tppp}^2}{f^2} &= \frac{y_{+}^2 + y_{-}^2}{2} + \frac{y_{123}^2}{y_{-}^2 (y_{+}^2 + y_{-}^2)} \sin^2 2 \theta
\end{aligned}
\qquad
\begin{aligned}[c]
\frac{m_n^2}{f^2} &= \frac{2 \tilde{y}_{123}^2}{\tilde{y}_{+}^4 - \tilde{y}_{-}^4}  \cos^2 2 \theta \\
\frac{m_{\np}^2}{f^2} &= \tilde{y}_1^2 \\
\frac{m_{\npp}^2}{f^2} &= \frac{\tilde{y}_{+}^2 - \tilde{y}_{-}^2}{2} - \frac{\tilde{y}_{123}^2}{\tilde{y}_{-}^2 (\tilde{y}_{+}^2 - \tilde{y}_{-}^2)}  \cos^2 2 \theta \\
\frac{m_{\nppp}^2}{f^2} &= \frac{\tilde{y}_{+}^2 + \tilde{y}_{-}^2}{2} + \frac{\tilde{y}_{123}^2}{\tilde{y}_{-}^2 (\tilde{y}_{+}^2 + \tilde{y}_{-}^2)} \cos^2 2 \theta
\end{aligned}
\ee
The insensitivity of these mass matrices to UV scales is again guaranteed by the relations in Equation \ref{eq:nodiv2}. The contributions from these states to the scalar potential may thus be expressed in terms of ratios of mass eigenstates, and are extremely well approximated by their leading order terms in $\sin 2 \theta$, as shown in Figure \ref{fig:su5so5numerical}. This contribution to the Higgs potential is thus approximately given by the $\cos 4 \theta$ dependence in Equation \ref{eq:llfermhiggs}. 
\begin{align}
 \delta V_T(y_i, \theta) +  \delta V_N(\tilde{y}_i, \theta) &= - 2 f^2 \left( f_T^2 \sin^2 2 \theta + f_N^2 \cos^2 2 \theta \right) \\[1.5ex]
  &\sim f^2 (f_T^2 - f_N^2) \cos 4 \theta
  \label{eq:llfermhiggs}
\end{align}
\be
f_T^2 = \frac{3 f^2}{32 \pi^2} \frac{y_{123}^2}{y_{-}^2} \log \frac{y_{+}^2 - y_{-}^2}{y_{+}^2 + y_{-}^2} \qquad \qquad
f^2_N = \frac{3 f^2}{32 \pi^2} \frac{\tilde{y}_{123}^2}{\tilde{y}_{-}^2} \log \frac{\tilde{y}_{+}^2 - \tilde{y}_{-}^2}{\tilde{y}_{+}^2 + \tilde{y}_{-}^2}
\ee
The fermion contribution to the Higgs potential is thus a function of two independent scales $f_T^2$ and $f^2_N$ for the color-charged and hidden Yukawa sectors respectively

\newpage

\section{Multi-top Final States in $SU(5) / SO(5)$}
\label{app:fourtop}

The $SU(5) / SO(5)$ theory generically predicts an enhancement of the total $t \bar{t} t \bar{t}$ cross section due to the Drell-Yan production of doubley charged scalars and the strong production of pairs of vector-like quarks. Here we enumerate all of the possible production and decay modes that can contribute to the final state $(Wb)^n$ with $n \geq 4$. 
These decay channels can result in complex correlations between the final state particles that could be resolved with targeted kinematic techniques in the high-luminosity limit.
\begin{center}
  \begin{table}[ht]
    \begin{tabular}{| l | l | l |}
      \hline
      Production Mode & Decay $\hspace{9cm}$ & Final State \\ [0.7ex] 
      \hline
      $g g \rightarrow \pb \pbbar$ & $\pb \pbbar \rightarrow (\tphi_{++} b)(\tphi_{--} W^+ \bar{t}) \rightarrow (W^+ t \bar{b} b)(W^- \bar{t} b W^+ \bar{t})$ & $(Wb)^6$ \\ [0.1ex] 
      & $\pb \pbbar \rightarrow (\tphi_{++} b)(\tphi_{0} W^- \bar{t}) \rightarrow (W^+ t \bar{b} b)(t \bar{t} W^- \bar{t})$ & \\ [0.1ex] 
      & $\pb \pbbar \rightarrow (\tphi_{+} t)(\tphi_{--} W^+ \bar{t}) \rightarrow (t \bar{b} t)(W^- \bar{t} b W^+ \bar{t})$ & \\ [0.1ex] 
      & $\pb \pbbar \rightarrow (\tphi_{+} t)(\tphi_{0} W^- \bar{t}) \rightarrow (t \bar{b} t)(t \bar{t} W^- \bar{t})$ & \\ [0.1ex] 
      & $\pb \pbbar \rightarrow (\tphi_{++} W^- t)(\tphi_{-} W^- \bar{b}) \rightarrow (W^+ t \bar{b} W^- t)(\bar{t} b W^- \bar{b})$ & \\ [0.1ex] 
      & $\pb \pbbar \rightarrow (\tphi_{+} W^+ b)(\tphi_{0} W^- \bar{t}) \rightarrow (t \bar{b} W^+ b)(t \bar{t} W^- \bar{t})$ & \\ [0.1ex] 
      
      & $\pb \pbbar \rightarrow (\tphi_{++} b)(W^- \bar{t}) \rightarrow (W^+ t \bar{b} b)(W^- \bar{t})$ & $(Wb)^4$ \\ [0.1ex] 
      & $\pb \pbbar \rightarrow (\tphi_{+} t)(W^{-} \bar{t}) \rightarrow (t \bar{b} t)(W^{-} \bar{t})$ & \\ [0.1ex] 
      & $\pb \pbbar \rightarrow (W^+ t)(\tphi_{-} W^- \bar{b}) \rightarrow (W^+ t)(\bar{t} b W^- \bar{b})$ & \\ [0.1ex] 
      
      & $\pb \pbbar \rightarrow (\tphi_{++} b)(\tphi_{--} \bar{b}) \rightarrow (W^+ t \bar{b} b)(W^- \bar{t} b \bar{b})$ & $(Wb)^4 b b$ \\ [0.1ex] 
      & $\pb \pbbar \rightarrow (\tphi_{+} t)(\tphi_{-} \bar{t}) \rightarrow (t \bar{b} t)(\bar{t} b\bar{t})$ & \\ [0.1ex] 
      & $\pb \pbbar \rightarrow (\tphi_{+} W^+ b)(\tphi_{-} W^- \bar{b}) \rightarrow (t \bar{b} W^+ b)(\bar{t} b W^- \bar{b})$ & \\ [0.1ex] 
      & $\pb \pbbar \rightarrow (\tphi_{++} b)(\tphi_{-} \bar{t}) \rightarrow (W^+ t \bar{b} b)(\bar{t} b \bar{t})$ & \\ [0.1ex] 
      & $\pb \pbbar \rightarrow (\tphi_{++} b)(\tphi_{-} W^- \bar{b}) \rightarrow (W^+ t \bar{b} b)(\bar{t} b W^- \bar{b})$ & \\ [0.1ex] 
      & $\pb \pbbar \rightarrow (\tphi_{+} t)(\tphi_{-} W^- \bar{b}) \rightarrow (t \bar{b} t)(\bar{t} b W^- \bar{b})$ & \\ [0.1ex] 
      
      & $\pb \pbbar \rightarrow (W^+ t)(\tphi_{--} W^+ \bar{t}) \rightarrow (W^+ t)(W^- \bar{t} b W^+ \bar{t})$ & $(Wb)^4 W W$ \\ [0.1ex] 
      & $\pb \pbbar \rightarrow (W^+ t)(\tphi_{0} W^- \bar{t}) \rightarrow (W^+ t)(t \bar{t} W^- \bar{t})$ & \\ [0.1ex] 
      
      & $\pb \pbbar \rightarrow (\tphi_{0} W^+ t)(\tphi_{0} W^- \bar{t}) \rightarrow (t \bar{t} W^+ t)(t \bar{t} W^- \bar{t})$ & $(Wb)^6 W W$ \\ [0.1ex] 
      & $\pb \pbbar \rightarrow (\tphi_{++} W^- t)(\tphi_{--} W^+ \bar{t}) \rightarrow (W^+ t \bar{b} W^- t)(W^- \bar{t} b W^+ \bar{t})$ & \\ [0.1ex] 
      & $\pb \pbbar \rightarrow (\tphi_{++} W^- t)(\tphi_{0} W^- \bar{t}) \rightarrow (W^+ t \bar{b} W^- t)(t \bar{t} W^- \bar{t})$ & \\ [0.1ex] 
      \hline
    \end{tabular}
  \end{table}
\end{center}

\begin{center}
  \begin{table}[ht]
    \begin{tabular}{| l | l | l |}
      \hline
      Production Mode & Decay $\hspace{9cm}$ & Final State \\ [0.7ex] 
      \hline
      $g g \rightarrow \tp \tpbar$ & $\tp \tpbar \rightarrow (\tphi_0 t)(\tphi_0 \bar{t}) \rightarrow (t \bar{t} t)(t \bar{t} \bar{t})$ & $(Wb)^6$ \\ [0.1ex] 
      & $\tp \tpbar \rightarrow (W^+ W^- t)(h \bar{t}) \rightarrow (W^+ W^- t)(b \bar{b} \bar{t})$ & $(Wb)^4$ \\ [0.1ex] 
      & $\tp \tpbar \rightarrow (\tphi_0 t)(h \bar{t}) \rightarrow (t \bar{t} t)(b \bar{b} \bar{t})$ & $(Wb)^4 b b$ \\ [0.1ex] 
      & $\tp \tpbar \rightarrow (\tphi_0 t)(Z \bar{t}) \rightarrow (t \bar{t} t)(Z \bar{t})$ & $(Wb)^4 Z$ \\ [0.1ex] 
      & $\tp \tpbar \rightarrow (\tphi_0 t)(W^+ W^- \bar{t}) \rightarrow (t \bar{t} t)(W^+ W^- \bar{t})$ & $(Wb)^4 W W$ \\ [0.1ex] 
      \hline
      $q \bar{q} \rightarrow \tphi_{++} \tphi_{--}$ & $\tphi_{++} \tphi_{--} \rightarrow (W^+ t \bar{b})(W^- \bar{t} b)$ & $(Wb)^4$ \\ [0.1ex] 
      \hline
    \end{tabular}
    \caption{Dominant production and decay modes for the multi-top final state.}
    \label{tab:lldecaymodes2}
  \end{table}
\end{center}

\newpage
\clearpage
\newpage

\bibliography{test}

\end{document}